\documentclass[12pt,a4paper,final]{iopart}
\eqnobysec

\usepackage{iopams,bbm}
\usepackage{graphicx}
\usepackage{float}
\usepackage{subcaption}
\usepackage{booktabs}
\usepackage{longtable}
\usepackage{multirow}
\usepackage[braket, qm]{qcircuit}
\usepackage{rotating}
\usepackage{hyperref}
\hypersetup{breaklinks=true,colorlinks=true,linkcolor=blue,urlcolor=blue,citecolor=blue}
\begin{document}
\title[]{Optimisation of diamond quantum processors}

\author{YunHeng Chen, Sophie Stearn, Scott Vella, Andrew Horsley, Marcus W. Doherty}
\address{Laser Physics Centre, Research School of Physics, Australian National University, Canberra, ACT 2601, Australia
}
\ead{marcus.doherty@anu.edu.au}

\begin{abstract}
Diamond quantum processors consisting of a nitrogen-vacancy (NV) centre and surrounding nuclear spins have been the key to significant advancements in room-temperature quantum computing, quantum sensing and microscopy. The optimisation of these processors is crucial for the development of large-scale diamond quantum computers and the next generation of enhanced quantum sensors and microscopes. Here, we present a full model of multi-qubit diamond quantum processors and develop a semi-analytical method for designing gate pulses. This method optimises gate speed and fidelity in the presence of random control errors and is readily compatible with feedback optimisation routines. We theoretically demonstrate infidelities approaching $\sim 10^{-5}$ for single-qubit gates and established evidence that this can also be achieved for a two-qubit CZ gate. Consequently, our method reduces the effects of control errors below the errors introduced by hyperfine field misalignment and the unavoidable decoherence that is intrinsic to the processors. Having developed this optimal control, we simulated the performance of a diamond quantum processor by computing quantum Fourier transforms. We find that the simulated diamond quantum processor is able to achieve fast operations with low error probability.

\end{abstract}

\vspace{2pc}
\noindent{\it Keywords}: nitrogen-vacancy centre, quantum control, quantum computing, quantum sensing
%\submitto{\NJP}
%\maketitle

\section{Introduction}
\label{sec:introduction}

\par Diamond is a promising architecture for quantum information processing \cite{Neumann2010a, Dolde2014,Taminiau2014, Waldherr2014,Wang2015,Kong2016,Bradley2019,Hou2019} and quantum sensing/ microscopy \cite{Neumann2013,Dolde2014a,Zaiser2016,Aslam2017} at both cryogenic and room temperautres. Optimised diamond quantum processors are crucial building blocks for large-scale diamond quantum computers and the next generation of quantum sensors and microscopes that are enhanced by embedded quantum memories and signal processing \cite{Unden2016, Haeberle2017,Pfender2017}. To date, diamond quantum processors have been used to implement quantum error correction codes \cite{Taminiau2014, Waldherr2014, Cramer2016}, quantum algorithms \cite{Shi2010, Xu2017}, detection of metallo-protein molecules \cite{Ermakova2013} as well as quantum simulation of the helium hydride cation \cite{Wang2015} and topological phase transition of a quantum wire \cite{Kong2016}. For technology applications, diamond quantum processors are distinguished from other quantum architectures due to their ability to operate in ambient conditions and with relatively simple microwave, radio-frequency and off-resonant optical control systems \cite{Doherty2017}. The resulting improvements in complexity, robustness and cost make diamond one of the most flexible and widely applicable quantum technology platforms. 

\par  Diamond quantum processors consist of a nitrogen-vacancy (NV) centre with a local cluster of hyperfine-coupled nuclear spins. These coupled nuclear spins include the intrinsic N nuclear spin of the NV centre and isotopic $^{13}$C lattice impurities. Quantum computations are realised by using the electron spin of the NV centre as a quantum bus that initialises, mediates interactions between, and reads-out the coupled nuclear spins, which act as the physical qubits. Scaling of the diamond quantum architecture requires a mechanism to couple multiple NV centres and their qubit clusters. Coupling of separate NV centres has been demonstrated at cryogenic temperatures using photons \cite{Bernien2013,Hensen2015,Kalb2017}, and at room temperature using magnetic dipole coupling between proximate NV centres \cite{Dolde2014,Dolde2013}. Spin chains \cite{Yao2012} and coherent spin transport \cite{Oberg2019} have also been proposed as coupling mechanisms. 

\par Previous work in optimising the quantum control of either the NV centre's electron spin or coupled nuclear spin qubits has already demonstrated excellent control fidelities, using techniques including dynamical decoupling \cite{Bradley2019,Hou2019,Viola1999,Khodjasteh2005,Lange2010,Wang2012,Zhao2012,Sar2012,Naydenov2011,Pham2012,Taminiau2012,Liu2013,Zhang2014,Abobeih2018} as well as numerical gate pulse-shaping techniques, such as the Chopped RAndom Basis (CRAB) quantum optimisation algorithm \cite{Doria2011, Caneva2011} and the GRadient Ascent Pulse Engineering (GRAPE) algorithm \cite{Khaneja2005}. Experimental application of the CRAB algorithm to the NV electron spin has yielded ultra-fast single-qubit gates (ie beyond the rotating wave approximation) with fidelities of $0.95\pm 0.01$ and $0.99\pm 0.016$ for $\pi/2$ and $\pi$ pulses, respectively \cite{Scheuer2014}. The GRAPE algorithm has been used to demonstrate single electron spin operations with fidelity $ F\approx 0.99$ \cite{Dolde2014} and generate entangled states of three nuclear spins with fidelities exceeding 85\% \cite{Waldherr2014}. Moreover, an average single-qubit gate fidelity of 0.999952 and two-qubit gate fidelity of 0.992 has been reported using composite pulses and a modified GRAPE algorithm, respectively \cite{Rong2015}. While some of the demonstrated gate fidelities are impressive, even with a gate fidelity of 0.999, for a fault-tolerant logical qubit to achieve logical error rates comparable to classical computers, effective surface code error correction is anticipated to require up to $10^4$ physical qubits \cite{Fowler2012,Campbell2017}. Therefore, there is a strong motivation to push for further reduction in gate errors. 

\par There are two different optimisation problems to address to improve the performance of diamond quantum processors: improvement of initialisation/ readout fidelities and improvement of gate fidelities and speeds. The initialisation/ readout fidelities are optimised by selecting the $^{13}$C nuclear spin lattice sites that have the longest nuclear spin relaxation time during the projective single-shot optical readout process employed in diamond quantum processors \cite{Waldherr2014}. Broadly speaking, the best lattice sites are those whose hyperfine field is well aligned to NV centre's axis. As will be discussed later, overcoming the effects of hyperfine field misalignment and decoherence are crucial to achieve high gate fidelities. As such, there is some correlation between the choices made to optimise initialisation/ readout fidelities and gate fidelities. Consequently, for simplicity, in the following we will not discuss the optimisation of initialisation/ readout fidelities. Instead, we assume a particular selection of nuclear spin qubits with well-aligned hyperfine fields, and focus on the problem of optimising the gate fidelities due to control errors and speeds.

\par Optimisation of the processor gate operations requires simultaneous maximisation of gate speed and minimisation of: (1) spurious effects of control fields on qubits other than the target qubit(s) (ie cross-talk), (2) the effects of random control field errors (ie those caused by fluctuations in amplitude, frequency and phase), (3) the effects of decoherence, and (4) the effects of hyperfine field misalignment. Additional practical requirements are that the gate design: (A) complies with the physical constraints of the control systems, (B) is readily incorporated into a feedback-based optimisation routine that uses measurements to optimise the actual physical processor (and not just models of the processor) and supports updating of the optimisation during operation to adjust for system drifts, and (C) is parameterised so that the degree of convergence to optimisation limits can be deterministically and systematically assessed (in order to support design decisions concerning the costs and benefits of further optimisation) and the dominant error modes can be diagnosed (to improve processor design). As GRAPE involves direct numerical solving of pulses using model systems, it does not readily achieve the practical requirements (B) and (C) and relies heavily on the accuracy of its models \cite{Wu2018}. On the other hand, the nature of CRAB allows the integration of (B) but not (C) directly due to its inherent reliance on random numbers, and the number of free parameters required to optimise the control field \cite{Caneva2011}.

\par We propose a different approach to this optimisation problem. Our approach has three steps. The first step is to generate a complete semi-analytical basis of pulses that comply with (A) and minimises (1) in the approximation where time-ordering is neglected in the simulated quantum evolution. The second step is to find the linear combination of these basis functions that minimise (2). The final step is to further optimise the pulses to account for the effects of time-ordering in the quantum evolution via closed-loop optimisations. We have neglected time-ordering in the quantum evolution to efficiently estimate the initial coefficients of the linear combinations of basis functions required for closed-loop optimisations. Without an initial estimate, optimising the linear combinations of basis functions via closed-loop optimisations is computationally time consuming. This linear minimisation is fast and can include measurements of the processor and therefore readily complies with (B). Furthermore, since the basis is complete, the dimension of the non-trivial basis functions provides a clear parameter to analyse convergence to optimisation limits and interpretation of different error modes, and thus also complies with (C). The principal strategy for both minimising the effects of decoherence (3) and the effects of hyperfine field misalignment (4) is to maximise the gate speed. This minimises the time over which decoherence accumulates and the time under which the nuclear spin experience the hyperfine misalignment during two-qubit gates, whilst ensuring that the infidelity introduced by control errors is less than that introduced by decoherence and hyperfine field misalignment. We have adopted this strategy to minimise the errors due to decoherence rather than directly targeting the primary source of qubit decoherence (eg via dynamical decoupling) because the primary source is the NV centre's electron spin, whose strong interaction with the nuclear spins is required for the implementation of both one- and two-qubit gates. Indeed, owing to the strong interaction, the qubit coherence time $1/T_{2,n}$ is limited by the electron spin relaxation time $1/T_{1,e}$ \cite{Maurer2012,Shim2013}, which is approximately 2.4~ms at room temperature \cite{Herbschleb2019}. Note that our discussion here, and the results of this paper, are in the context of room-temperature operation of diamond quantum processors. At cryogenic temperatures, weakly coupled nuclear spins can instead be employed as qubits and as a result, their decoherence is influenced by other mechanisms \cite{Kalb2018}, at the cost of slower gate speeds.

\par In this paper, we report (i) a demonstration and analysis of our optimal design approach and (ii) simulation of an optimised diamond quantum processor. In \sref{sec:operatingprinciples}, we first discuss the operating principles of a diamond quantum processor before presenting a complete model of the processor, its gate operations and control system errors. \Sref{sec:idealgates} demonstrates the generation of gate basis functions, while \sref{sec:nonidealgates} demonstrates the optimisation of gate fidelities in the presence of control errors. In \sref{sec:gatewithdecoherence}, the effects of decoherence on a diamond quantum computer are investigated via master equation simulations and in \sref{sec:simulation}, we simulated the performance of 3 and 5 qubit quantum Fourier transforms (QFTs) on a diamond quantum processor. Fidelities of QFTs were chosen as a simple performance metric because QFTs are the foundation of many quantum algorithms. Thus, the fidelities of QFTs are basic indicators for the system's performance with more sophisticated algorithms.
\hypersetup{pageanchor=false}
\section{Quantum control model of diamond quantum processors}\label{sec:operatingprinciples}
\hypersetup{pageanchor=false}
\subsection{Operating principles of diamond quantum processors}
\par The NV centre is a point defect in diamond consisting of a substitutional nitrogen and an adjacent carbon vacancy \cite{Doherty2013}. Its electronic structure consists of a ground state spin triplet $(^3A_2)$ and an excited state spin triplet $(^3E)$ with two intermediate singlet levels ($^1E$ and $^1A_1$) . There exist spin-selective non-radiative intersystem crossings between the triplet and singlet levels, which lead to initialisation of the electronic spin state upon optical excitation of the centre's $^3A_2\rightarrow ^3E$ transition as well as read out via the differing fluorescence intensities of the spin states (see Ref.~\cite{Doherty2013} for further details). In addition to high fidelity optical spin initialisation and readout, the NV centre also has the longest electron spin coherence time of any solid state spin at room temperature ($T_2\approx 2.4$ ms) \cite{Herbschleb2019}. 

\par
\begin{figure}[h]
	\centering
	\begin{subfigure}[b]{0.45\textwidth}
		\includegraphics[width=1\textwidth]{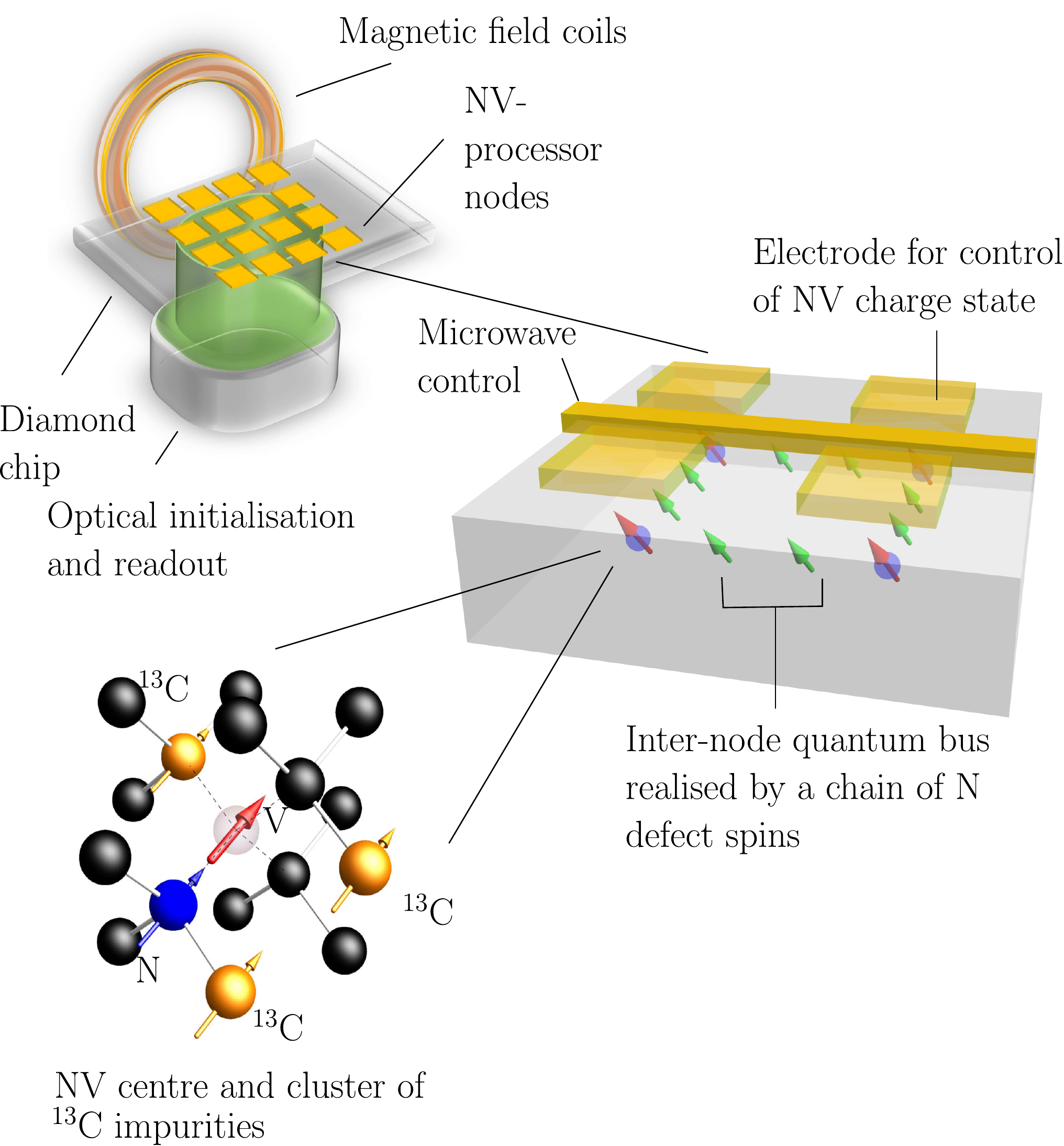}
	\caption[]{}
	\label{fig:conceptdesign}
	\end{subfigure}
	\begin{subfigure}[b]{0.45\textwidth}
	\includegraphics[width=1\textwidth]{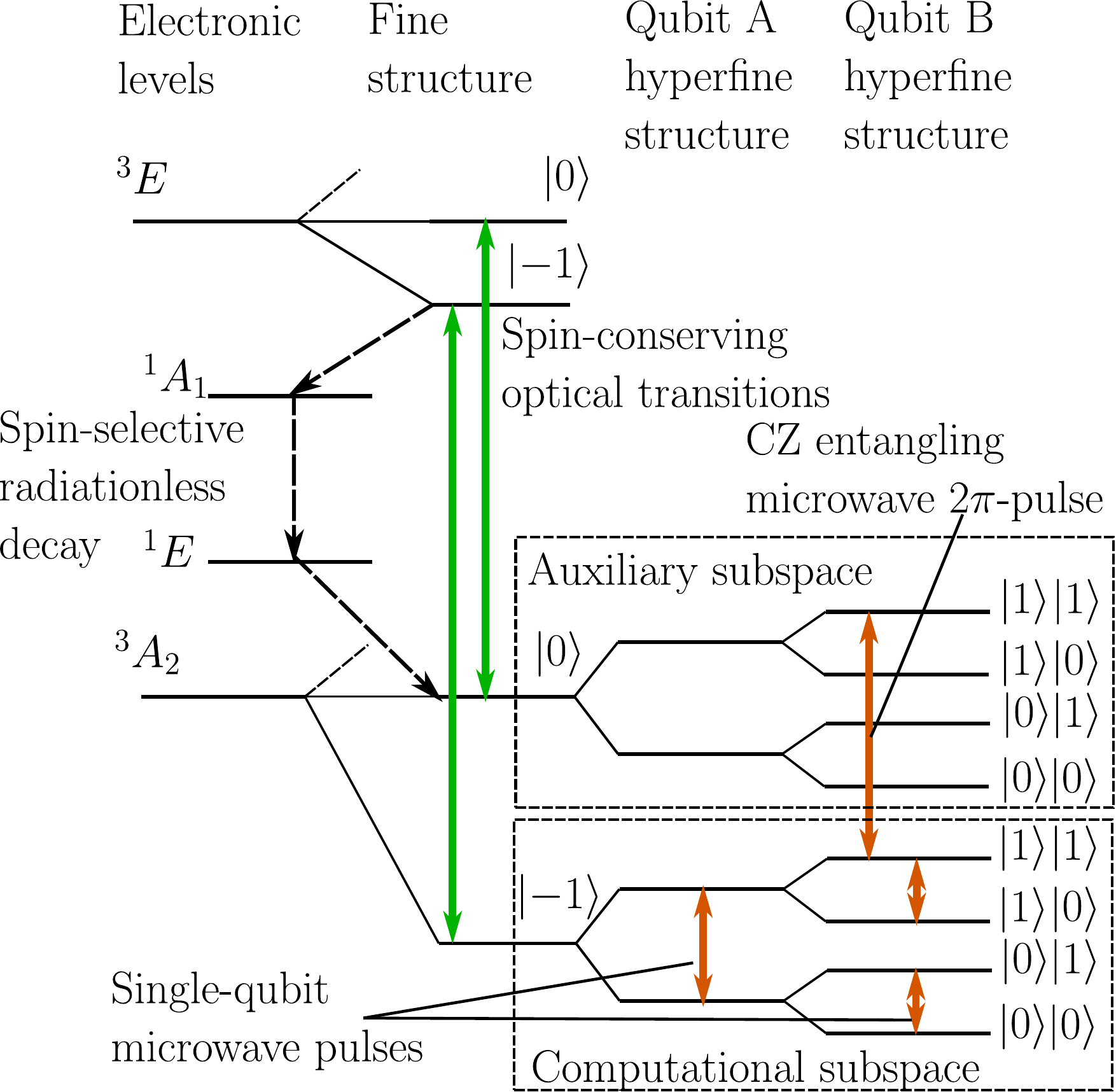}
	\caption[]{}
	\label{fig:estructure}
	\end{subfigure}
	\caption[]{ (a) Conceptual design of a diamond quantum computer (Adapted with permission from \cite{Doherty2017}. \copyright Australian Institute of Physics). At the device scale (top), the quantum computer contains a diamond chip with an array of quantum processing nodes. Each node is formed by magnetically coupling a surrounding cluster of $^{13}$C nuclear spin qubits to the NV centre. Optical initialisations and readout of these quantum processing nodes via their NV centres are done using an optical system placed below the diamond chip. At the scale of a single node (middle), surface mirowave structures are used to realise single and two qubit gate operations. Internode two-qubit gate operation is mediated using spin quantum buses which are realised through chains of substitutional N defects. At the cluster scale (bottom), the NV centre consists of a substitutional N defect (blue) adjacent to a carbon vacancy (transparent). The nuclear spins of the N defect and cluster of nearby $^{13}$C atoms are depicted in blue and orange respectively, while the NV centre's electronic spin is depicted in red. (b) The hyperfine structure of the NV centre which arises from the interaction of two nearby $^{13}$C nuclear spin qubits. Optical initialisation and readout of the NV centre's electron spin are realised via a combination of spin-conserving optical transitions and spin-selective radiationless decay \cite{Doherty2013}. Using microwave pulses, this capability can be extended to the nuclear spin qubits by selectively swapping the electronic and nuclear spin states \cite{Waldherr2014,Morton2006,Filidou2012}. The computational and auxiliary subspaces are defined to be the $\left|-1\right>$ and $\left|0\right>$ electronic spin projections, respectively. Single qubit gate operations are realised in the computational subspace using spectrally-selective microwave pulses. A conditional-z (CZ) two-qubit gate operation is realised via selective $2\pi$ microwave pulses that involves, but does not occupy, the auxiliary subspace \cite{Waldherr2014,Morton2006,Filidou2012}.}
	\label{fig:conceptdesign_estructure}
\end{figure}

 Each NV centre is coupled to a register of one or more nuclear spins, which we use as qubits \cite{ Taminiau2014,Waldherr2014}. The quantum register consists of the NV centre's intrinsic nitrogen nuclear spin and nearby $^{13}$C nuclear spins. Hyperfine coupling between the NV electron spin and the nuclear spins results in a splitting of the electronic and nuclear energy levels. This splitting depends on the particular hyperfine coupling strength between each nucleus and the NV electron spin, and also on the respective electron and nuclear spin states \cite{Felton2009,Smeltzer2011} (see \fref{fig:estructure}). We choose a register with non-overlapping hyperfine couplings, allowing the use of frequency selectivity to individually address each nuclear spin qubit in the register.

\par Key requirements for universal quantum computation are the initialisation and readout of the qubits, as well as the ability to apply single and two-qubit gate operations. In diamond quantum computing, each of these processes relies on high-fidelity quantum gates on the electron and nuclear spins. Initialisation and readout of a diamond quantum register is performed via a projective, single-shot readout of the nuclear spin qubits. This measurement scheme involves initialising the electron spin, entangling the nuclear spin qubits with the electron spin using a ${\rm C_nNOT_e}$ gate and then readout of the electron spin \cite{Neumann2010}. Single-qubit gate operations are realised using radiofrequency (RF) pulses. These pulses correspond to the $R_x$ and $R_y$ gates where they are the rotations about the $x$ and $y$ axes respectively. Other single-qubit gates can be constructed from combinations of these rotations. The intrinsic properties of the NV-nuclear spins system allows direct application of a CZ gate via microwave (MW) pulses. This CZ gate is achieved by performing a selective $2\pi$ pulse conditional on the nuclear spin register being in a particular state \cite{Waldherr2014,Morton2006,Filidou2012}. The CZ gate can be combined with single-qubit gates to realise any other two-qubit gate.

\par The splitting in the $^{3}A_2$ triplet ground state results in two types of subspaces which we identify as computational subspace and auxiliary subspace (\fref{fig:estructure}). The natural computational subspace is either of the $m_s=\pm 1$ states as they have non-zero hyperfine interactions, thus allowing the nuclear spin qubits to be individually addressed through frequency selectivity \cite{Dutt2007}.  Whilst the choice of either the $m_s=\pm 1$ state as the computational subspace is arbitrary, the $m_s=-1$ state is more often selected as it requires lower microwave frequencies for qubit gate operations. Single-qubit gates are realised in the computational subspace while a two-qubit CZ gate utilises, but does not occupy, the auxiliary subspace \cite{Waldherr2014,Morton2006,Filidou2012}. 

\hypersetup{pageanchor=false}
\subsection{Model Hamiltonian}
\label{section:modelhamiltonian}
\par The Hamiltonian $H_I$ of the nuclear spins coupled to the NV centre is
\begin{equation}
H_I=\sum_i  \vec{S}\cdot \boldsymbol{A}_i \cdot \vec{I}_i -\sum_i \gamma_i \Big[I_{i,z}B_0+I_{i,x}B_1(t)\Big]\label{eq:Hamiltonian1}
\end{equation}
where $\gamma_i$ is the gyromagnetic ratio of the $i^{th}$ nucleus, $B_0$ is the background static magnetic field aligned with the NV axis, $B_1(t)$ is the applied radio frequency field, $\boldsymbol{A}_i$ is the hyperfine tensor of the $i^{th}$ nucleus with $\vec{S}$ being the dimensionless electron spin operator and $\vec{I}_i$ is the dimensionless nuclear spin operator of the $i^{th}$ nucleus.
For this model, we apply the secular approximation as a very strong magnetic field is applied along the $z$-axis during the operation of this quantum computer. Therefore, the nuclear spin Hamiltonian in the computational subspace simplifies to
\begin{equation}
H_{I,m_s=-1}=-\sum_i\vec{A}_{i,z}\cdot\vec{I}_i- \sum_i\gamma_i\Big[I_{i,z}B_0+I_{i,x}B_1(t)\Big]\label{eq:Hamiltonian2}
\end{equation}
$\vec{A}_{i,z}$ is now a vector instead of a tensor where
\begin{equation}
\vec{A}_{i,z}=A_{i,xz}\hat{x}+A_{i,yz}\hat{y}+A_{i,zz}\hat{z}
\end{equation}
 We simplify the expression by diagonalising the nuclear spin Hamiltonian in the computational subspace via a rotation of the spin operators about the angles defined by their hyperfine interactions. Assuming that only nuclei with hyperfine fields nearly aligned with the NV axis are chosen, we perform small angle approximations and by undoing a rotation about $z$-axis for further simplifications, this yields
\begin{equation}
H_{I,m_s=-1}=- \sum_i \omega_i I_{i,z}-\sum_i I_{i,x}\gamma_i B_1(t)\label{eq:Hamiltonian3}
\end{equation}
where $\omega_i$ is the transition frequency of the $i^{th}$ nucleus. We also transform the Hamiltonian into the interaction picture using the following transformation operator
 \begin{equation}
 H_{single}= \mathcal{T} H_{I,m_s=-1}\mathcal{T}^{-1} -\mathbbm{i}\mathcal{T}\frac{d}{dt} \mathcal{T}^{-1}\label{eq:HamiltonianTransform}
 \end{equation}
 where
\begin{equation}
\mathcal{T}=\exp\Big[-\mathbbm{i}t\sum_i\omega_i I_{i,z}\Big]\label{eq:HamiltonianTransformOperator}
\end{equation}
 Thus, for single-qubit gate operations, the Hamiltonian for the computational subspace in the interaction picture is given by
\begin{equation}
H_{\rm{single}}=- \sum_i \Bigg[I_{i,x} \cos\omega_it + I_{i,y} \sin\omega_i t\Bigg]\gamma_i B_1(t)
\end{equation}
\par The auxiliary subspace is involved to perform a CZ gate and this is enabled via the NV electron spin. The effective Hamiltonian for two-qubit gate operations is given by
\begin{equation}
H'=\frac{\Delta}{2}\sigma_z +\frac{\Omega}{2}\sigma_x B_1(t)+\sum_i \left(\alpha_i + \beta_i\sigma_z\right)I_{i,z}\label{eq:twoqubit}
\end{equation}
with $\Delta=D-\gamma_eB_0$ where $D$ denotes the zero-field splitting, $\Omega=\sqrt{2}\gamma_e$, $\alpha_i=-\gamma_iB_0/2-\omega_i/2$ and $\beta_i=-\gamma_iB_0/2+\omega_i/2$. When defining the Hamiltonian above, we ignore the interactions with the $m_s=+1$ state of the electron spin and the direct interaction between the nuclear spins and the microwave field $B_1(t)$. It is possible to do this because the microwaves are far detuned from these transitions. We have also ignored the interactions of the static magnetic field acting on the nuclear spin when the electron spin is in $m_s=0$ state. This interaction term arises from the hyperfine field misalignment and is discussed further in \ref{appendix:misalignmenteffects}. Likewise to the Hamiltonian for single-qubit gate operations, we transform $H'$ into the interaction picture with the transformation operator $\mathcal{T}'$ is given by
\begin{equation}
\mathcal{T}'=\exp\left[\mathbbm{i}t\left(\frac{\Delta}{2}\sigma_z+\sum_i\left(\alpha_i+\beta_i\sigma_z\right)I_{i,z}\right)\right]
\end{equation}
The transformed Hamiltonian for two-qubit gate operations is then given by
\begin{eqnarray}
\fl H_{\rm{multi}}&=\frac{\Omega}{2}B_1(t)\Bigg[\sigma_x\otimes\Big(\left|11\right>\left<11\right|\cos\left(\left[\Delta+\beta_1+\beta_2\right]t\right)+\left|10\right>\left<10\right|\cos\left(\left[\Delta+\beta_1-\beta_2\right]t\right)\nonumber\\
\fl &+\left|01\right>\left<01\right|\cos\left(\left[\Delta-\beta_1+\beta_2\right]t\right)+\left|00\right>\left<00\right|\cos\left(\left[\Delta-\beta_1-\beta_2\right]t\right)\Big)\nonumber \\
\fl &-\sigma_y\otimes\Big(\left|11\right>\left<11\right|\sin\left(\left[\Delta+\beta_1+\beta_2\right]t\right)+\left|10\right>\left<10\right|\sin\left(\left[\Delta+\beta_1-\beta_2\right]t\right)\nonumber\\
\fl &+\left|01\right>\left<01\right|\sin\left(\left[\Delta-\beta_1+\beta_2\right]t\right)+\left|00\right>\left<00\right|\sin\left(\left[\Delta-\beta_1-\beta_2\right]t\right)\Big)\Bigg]
\end{eqnarray}
where 0 and 1 are the $m_I=-1/2$ and $+1/2$ nuclear spin projections, respectively. We use the notation where the most left entry of a tensor product corresponds to the first qubit, i.e $\left|q_1, q_2,\dots\right>$

\subsection{Control Pulses and Gate Operations}
\label{sec:controlpulses}
\par Focusing on single-qubit gate operations within the computational subspace, the applied radio frequency field $B_1(t)$ can be parametrised as a linear combination of oscillating components where
\begin{equation}
\gamma_iB_1(t)=a(t)\cos\omega t+b(t)\sin \omega t \label{eq:3_gammai}
\label{eq:gammaB}
\end{equation}
By neglecting the time-ordering in the quantum evolution, we can write the evolution operator for the $i^{th}$ nucleus as
\begin{equation}
U_i=e^{-\mathbbm{i} I_{i,x} X_i } \qquad  {\rm{or}} \qquad U_i=e^{-\mathbbm{i} I_{i,y} Y_i} \label{eq:3_uoperator}
\end{equation}
with
\begin{eqnarray}
X_i&=-\int\limits_{-\tau/2}^{\tau/2} a(t)\cos\omega_it \cos\omega t+b(t)\cos\omega_it \sin \omega t\ dt \label{eq:Xi1}\\ 
Y_i&=-\int\limits_{-\tau/2}^{\tau/2} a(t)\sin\omega_i t \cos \omega t + b(t) \sin \omega_i t \sin \omega t\ dt  \label{eq:Yi1}
\end{eqnarray}
where $\tau$ denotes the gate time while $X_i$ and $Y_i$ parametrises the rotations about the $x$ and $y$ axes respectively, which realises the gate operation. We have neglected time-ordering in the quantum evolution to efficiently estimate the initial coefficients of the linear combinations of basis functions required for closed-loop optimisations. The effects due to time-ordering in the quantum evolution is discussed further in \ref{appendix:timeordering}. 
\par If the $j^{th}$ nucleus is the intended target, then the operations on all other nuclei $(i\neq j)$ are simply identity operations. Thus we have : for $i\neq j$,
\begin{eqnarray}
X_i&=-\int\limits_{-\tau/2}^{\tau/2} a(t)\cos\omega_it \cos\omega_j t+b(t)\cos\omega_it \sin \omega_j t\ dt=0 \label{eq:Xi2}\\ 
Y_i&=-\int\limits_{-\tau/2}^{\tau/2} a(t)\sin\omega_i t \cos \omega_j t + b(t) \sin \omega_i t \sin \omega_j t\ dt=0 \label{eq:Yi2}
\end{eqnarray}
and for $i= j$
\begin{eqnarray}
X_j&=-\int\limits_{-\tau/2}^{\tau/2} a(t)\cos\omega_jt \cos\omega_j t+b(t)\cos\omega_jt \sin \omega_j t\ dt=X_T \label{eq:Xj1}\\
Y_j&=-\int\limits_{-\tau/2}^{\tau/2} a(t)\sin\omega_j t \cos \omega_j t + b(t) \sin \omega_j t \sin \omega_j t\ dt =Y_T \label{eq:Yj1}
\end{eqnarray}
We impose the restriction that $\omega=\omega_j$ where $\omega_j$ is the transition frequency of the targeted qubit. Using these gate parametrisations, we introduce the following inverse Fourier transforms, where
\begin{eqnarray}
a(t)&=\frac{1}{\sqrt{2\pi}}\int\limits_{-\infty}^{\infty} a(\omega)e^{\mathbbm{i}\omega t}\ d\omega \label{eq:fourierat}\\
b(t)&=\frac{1}{\sqrt{2\pi}}\int\limits_{-\infty}^{\infty} b(\omega)e^{\mathbbm{i}\omega t}\ d\omega \label{eq:fourierbt}
\end{eqnarray}

As the signal is finite in time domain, we can enforce that $a(t)$ and $b(t)$ are zero outside $t\in \left[-\tau/2,\tau/2\right]$. This enables us to change the limits of the time integral to $\pm \infty$ and pass the time integral through the frequency integral. We also use the following identity
\begin{equation}
\frac{1}{\sqrt{2\pi}}\int\limits_{-\infty}^{\infty} e^{-\mathbbm{i}\omega t}dt =\sqrt{2\pi}\delta\left(\omega\right)
\end{equation}
for further simplifications. For $a(t)$ and $b(t)$ to be real functions with well defined phases, we enforce
\begin{eqnarray}
a^*(\omega)&=a(\omega)=a(-\omega)\\
b^*(\omega)&=b(\omega)=b(-\omega)
\end{eqnarray}
Using the above conditions, we arrived at the expressions where : for $i\neq j$
\begin{eqnarray}
X_i&=-\frac{\sqrt{2\pi}}{2} \Bigg[a\left(\omega_i+\omega_j\right)+a\left(\omega_i-\omega_j\right)\Bigg]=0 
\label{eq:xi}\\
Y_i&=\frac{\sqrt{2\pi}}{2} \Bigg[b\left(\omega_i+\omega_j\right)-b\left(\omega_i-\omega_j\right)\Bigg]=0
\label{eq:yi}
\end{eqnarray}
and for $i=j$
\begin{eqnarray}
X_j&=-\frac{\sqrt{2\pi}}{2} \Bigg[a\left(2\omega_j\right)+a\left(0\right)\Bigg]=X_T
\label{eq:xj}\\
Y_j&=\frac{\sqrt{2\pi}}{2} \Bigg[b\left(2\omega_j\right)-b\left(0\right)\Bigg]=Y_T
\label{eq:yj}
\end{eqnarray}
where $X_T$ and $Y_T$ are the intended angles of rotation on the $x$-axis and $y$-axis respectively.
\hypersetup{pageanchor=false}
\subsection{Statistical Model of Gate Errors}
\label{sec:statisticalmodel}
In reality, the control fields have noises and the frequencies of the qubits fluctuate slowly between the computational shots. Thus, the real control field can be written as
\begin{equation}
\fl \gamma_iB_1(t)=\left(1+\epsilon\right)\Bigg[a(t)\cos\left(\left(\omega+\delta \right)t+\phi\right)+b(t)\sin\left(\left(\omega+\delta\right) t+\phi\right)\Bigg]
\end{equation}
where $\epsilon,\phi$ and $\delta$ are free parameters representing the amplitude, phase and frequency noises respectively. Let the target evolution operator for a single qubit gate on the $j^{th}$ nucleus within an $N$ qubit cluster to be
\begin{eqnarray}
G_{T,x}&=I_1\otimes \dots \otimes I_{j-1}\otimes e^{-\mathbbm{i}\left(I_{j,x}X_T\right)}\otimes I_{j+1}\otimes \dots \otimes I_N\\
G_{T,y}&=I_1\otimes \dots \otimes I_{j-1}\otimes e^{-\mathbbm{i}\left(I_{j,y}Y_T\right)}\otimes I_{j+1}\otimes \dots \otimes I_N
\end{eqnarray}
The actual evolution operator is defined as
\begin{equation}
G_{A,x}=\bigotimes_{i=1}^{N} e^{-\mathbbm{i}\left(I_{i,x}X_i\right)} \qquad {\rm{or}}\qquad  G_{A,y}=\bigotimes_{i=1}^{N} e^{-\mathbbm{i}\left(I_{i,y}Y_i\right)}
\end{equation}
The analytical expression for infidelity can be written as
\begin{equation}
I=1-\frac{\Tr\Big[G_T^\dagger G_A\Big]}{\Tr\Big[G_T^\dagger G_T\Big]} \label{eq:infidelity}
\end{equation}
This expression for infidelity compares an ideal gate to an experimental gate \cite{White2007}. It is used as it requires minimal computation when compared to other figures of merit, and is equivalent to the usual performance function implemented in the GRAPE algorithm \cite{Khaneja2005}. Assuming the gates are rotating about the x-axis, the infidelity expression can be written as
\begin{eqnarray}
\fl I &=1-\frac{1}{2^N}\Bigg[\prod_{i\neq j}^{N} \Tr\left(e^{-\mathbbm{i}\left(I_{i,x}X_i\right)}\right)\Bigg]\Tr\Bigg[\left(e^{-\mathbbm{i}\left(I_{j,x}X_T\right)}\right)^\dagger e^{-\mathbbm{i}\left(I_{j,x}X_j\right)}\Bigg]\nonumber\\
\fl &=1-\Bigg[\prod_{i\neq j}^{N}\cos\left(\frac{1}{2}X_i\right)\Bigg]\Bigg[\cos\left(\frac{X_j-X_T}{2}\right)\Bigg] \label{eq:intrinsicinfone}
\end{eqnarray}
Assuming small errors in the regime where $X_i\ll 1$ for $i\neq j$ and $\left(X_j-X_T\right)\ll 1$, we can then expand the above expression, and keep only the terms up to second order, which gives
\begin{equation}
I\approx \frac{1}{8}\sum_{i\neq j}^{N} X_i^2 +\frac{1}{8}\delta_X^2 
\end{equation}
 Repeating the same procedure for the rotations about the y-axis gives
\begin{equation}
I\approx \frac{1}{8}\sum_{i\neq j}^{N} Y_i^2 +\frac{1}{8}\delta_Y^2
\end{equation}
where $\delta_X=X_j-X_T$ and $\delta_Y=Y_j-Y_T$
\par The control signal has a general form of 
\begin{eqnarray}
a(\omega)&=\sum_nc_n a^{(n)}(\omega) \label{eq:linearcombX}\\
b(\omega)&=\sum_nd_n b^{(n)}(\omega) \label{eq:linearcombY}
\end{eqnarray}
As such, the set of linear coefficients $c_n$ and $d_n$ that minimise the infidelity caused by the amplitude, phase and frequency noises can be solved.
\par We assumed the noise in the qubit frequencies $\delta$, pulse amplitudes $\left(1+\epsilon\right)$ where $\epsilon$ is the fractional error of the pulse amplitudes and phase $\phi$ are described by Gaussian distributions centred at zero with their respective standard deviations of $\sigma_\delta,\sigma_\epsilon$ and $\sigma_\phi$. The Gaussian distribution for frequency error is justified by experiments \cite{Rong2015,Balasubramanian2009}. Whilst there is no clear evidence of either Gaussian or Lorentzian distribution for the amplitude error, we assumed Gaussian distribution and noting that Ref.~\cite{Rong2015} observed a Lorentzian distribution for the convolution of power fluctuations and pulse shape errors. The analytical expression for the average infidelity can be written as
\begin{eqnarray}
\fl \left<I\right>&=\int\limits_{-\infty}^{\infty} P(\delta;\sigma_\delta)\int\limits_{-\infty}^{\infty}P(\epsilon;\sigma_\epsilon)\int\limits_{-\infty}^{\infty}P(\phi;\sigma_\phi) I(\omega_1,\dots,\omega_N) d\delta d\epsilon d\phi\nonumber\\
\fl&=\frac{1}{8}\Bigg[\int\limits_{-\infty}^{\infty} p(\delta;\sigma_\delta)\int\limits_{-\infty}^{\infty}p(\epsilon;\sigma_\epsilon)\int\limits_{-\infty}^{\infty}p(\phi;\sigma_\phi)\delta_X^2  d\delta d\epsilon d\phi \nonumber\\
\fl&+\sum_{i\neq j}^{N} \int\limits_{-\infty}^{\infty} p(\delta;\sigma_\delta)\int\limits_{-\infty}^{\infty}p(\epsilon;\sigma_\epsilon)\int\limits_{-\infty}^{\infty}p(\phi;\sigma_\phi) X_i^2  d\delta d\epsilon d\phi\Bigg] \label{eq:avginf}
\end{eqnarray}
A similar infidelity expression can be obtained for rotations about the y-axis. The minimum average gate infidelity is found when
\begin{eqnarray}
\frac{\partial \left<I\right>}{\partial c_n}&=0\\
\frac{\partial \left<I\right>}{\partial d_n}&=0
\end{eqnarray}
are satisfied for each function in the expansion.

\section{Generation of Basis Functions}
\label{sec:idealgates}
In this paper, we use frequency-shifted sinc functions as an ansatz for our control pulses in the frequency domain. Sinc functions were chosen as they represent pulses of finite duration in the time domain. As previously demonstrated in equation~\ref{eq:linearcombX} and \ref{eq:linearcombY}, the key property is the amplitude at very specific frequencies $\left(\omega_i,\omega_j\right)$. The interference of frequency-shifted sinc functions allow us to cancel the pulse amplitude at certain frequencies whilst at the same time amplifying other frequencies. The parametrisation of $a(\omega)$ and $b(\omega)$ is given by
\begin{eqnarray}
a^{(n)}\left(\omega\right)&=f^{(n)} \tau\Bigg[{\rm sinc}\Big(\frac{\tau}{2}\left(\omega-\mu^{(n)}\right)\Big)+{\rm sinc}\Big(\frac{\tau}{2}\left(\omega+\mu^{(n)}\right)\Big)\Bigg]
\label{eq:basisfunctiona}\\
b^{(n)}\left(\omega\right)&=g^{(n)} \tau \Bigg[{\rm sinc}\Big(\frac{\tau}{2}\left(\omega-\nu^{(n)}\right)\Big)+{\rm sinc}\Big(\frac{\tau}{2}\left(\omega+\nu^{(n)}\right)\Big)\Bigg]
\label{eq:basisfunctionb}
\end{eqnarray}
where $f^{(n)}$, $g^{(n)}$ are the pulse amplitudes in the frequency domain, $\tau$ is the gate time, $\mu^{(n)}$ and $\nu^{(n)}$ are the frequency shifts and $(n+1)$ is the total number of basis functions used in the optimisation procedure.
\par  If the frequency shifts for the $n^{\rm th}$ solution are defined as
\begin{eqnarray}
\mu^{(n)}=n\left(\frac{2\pi}{\tau}\right)\\
\nu^{(n)}=n\left(\frac{2\pi}{\tau}\right)
\end{eqnarray}
then the above simply becomes a Fourier series, which is known to be a complete basis function. However, as seen in equation \ref{eq:basisfunctiona} and \ref{eq:basisfunctionb}, the $a^{(n)}(0)$ and $b^{(n)}(0)$ terms would then be
\begin{eqnarray}
a^{(n)}(0)&=  2f^{(n)}{\rm sinc}\Bigg(n\pi\Bigg)\\
b^{(n)}(0)&=  2g^{(n)} {\rm sinc}\Bigg(n\pi\Bigg)
\end{eqnarray}
 where for $n\geq1$, $a^{(n)}(0)$ and $b^{(n)}(0)$ terms will always be 0. Hence, without any contributions from the  $a^{(n)}(0)$ and $b^{(n)}(0)$ terms, generating an optimal pulse from only the $a^{(n)}\left(2\omega_j\right)$ and $b^{(n)}\left(2\omega_j\right)$ terms would require extremely large amplitudes.
 \par
\begin{figure}[H]
\centering
\includegraphics[width=0.7\textwidth]{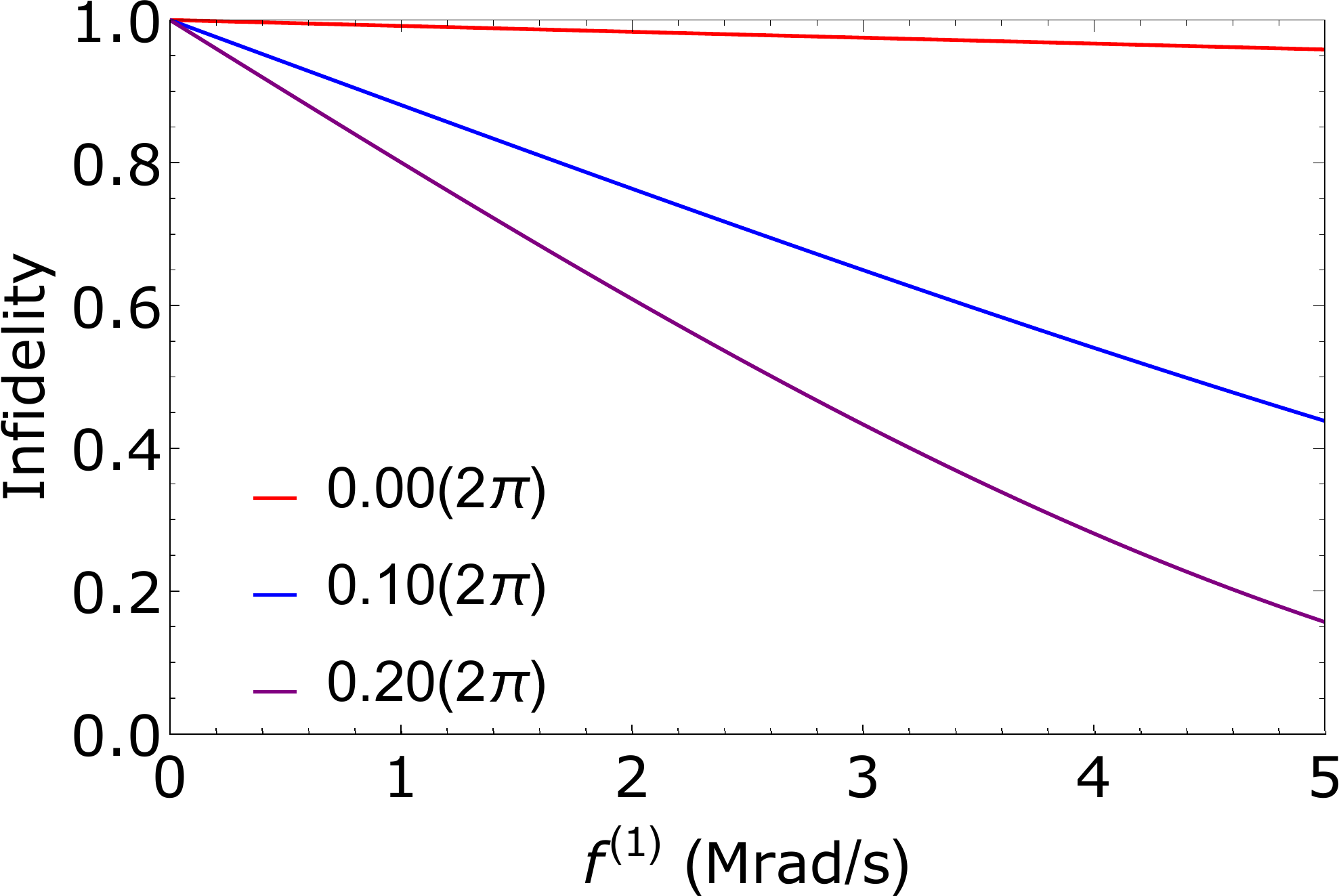}
\caption[]{Plots of infidelity (equation~\ref{eq:intrinsicinfone}) versus pulse amplitude $\left(f^{(1)}\right)$ for a $\pi$ rotation about the x-axis with $\tau=1~\mu$s and $n=1$. This is the first step of the optimisation procedure where we generate optimal solutions for the pulse amplitudes in the absence of control errors. Due to the experimental hardware constraints, the maximum pulse amplitude in the time domain is set to be approximately 25~Mrad/s. We arbitrarily set the corresponding search range for $f^{(n)}$ and $g^{(n)}$ to be $f^{(n)},g^{(n)} \in \left[-5,5\right]$ and noting that the amplitude in the time domain has an additional factor of $2\sqrt{2\pi}$ from the inverse Fourier transform of the basis functions. An additional shift of $0.2\left(2\pi\right)$ allows us to resolve better solutions with lower infidelity compared to no additional shift and a 0.1$\left(2\pi\right)$ shift within a bounded search range for the amplitudes. See text for more discussions regarding the optimisation procedures.} 
\label{fig:f1shifts}
\end{figure}

 One possible method to overcome this complication is to use Kadec's 1/4 Theorem, where an additional small frequency shift is introduced to the Fourier series whilst retaining the completeness of the Fourier series. It was shown that when the additional frequency shift is bounded by a maximum value of $0.25\left(2\pi/\tau\right)$, the inequality will still be able to generate a continuous set of sinc basis functions \cite{Avantaggiati2016}. In this paper, we have chosen the upper bound to be $0.2\left(2\pi/\tau\right)$ as it enables us to resolve better solutions for the pulse amplitudes (see figure~\ref{fig:f1shifts}).

The frequency shifts are then redefined as
\begin{eqnarray}
\mu^{(n)}&=\left(n+\frac{1}{5}\right)\left(\frac{2\pi}{\tau}\right)\\
\nu^{(n)}&=\left(n+\frac{1}{5}\right)\left(\frac{2\pi}{\tau}\right)
\end{eqnarray}
As a result, there are many solutions in $a^{(n)}\left(\omega\right)$ and $b^{(n)}\left(\omega\right)$ that we can consider and these solutions form a linear basis for the construction of optimal pulse functions. These optimal pulse functions are found by determining the linear coefficients that minimise the effects of the pulse errors on average.
\par As an example, consider a two-qubit system which consists of $^{15}$N and $^{13}$C nuclear spins. Their respective hyperfine interactions are given by $A_N\approx  3$~MHz \cite{Felton2009} and $A_C\approx 0.413$~MHz \cite{Waldherr2014}. The background static magnetic field is chosen to be $B_0=0.62$~T \cite{Waldherr2014}. Using these parameters, we demonstrate the optimisation procedure for an X gate targeted at the $^{13}$C nuclear spin $\left(X_2\right)$. Since we are performing rotations only in the x-axis/ y-axis, we assumed there are no mixed signals in the pulse and thus, there are no contributions from the $b(\omega)$/ $a(\omega)$ components. Hence, the equations that are satisfied by the solutions are given by 
\begin{eqnarray}
X_1&=-\frac{\sqrt{2\pi}}{2} \Bigg[a\left(\omega_N+\omega_C\right)+a\left(\omega_N-\omega_C\right)\Bigg]=0
\label{eq:onequbitx1}\\
X_2&=-\frac{\sqrt{2\pi}}{2} \Bigg[a\left(2\omega_C\right)+a\left(0\right)\Bigg]=\pi
\label{eq:onequbitx2}
\end{eqnarray}
where $\omega_N=A_N+\gamma_NB_0$ and $\omega_C=A_C+\gamma_CB_0$. The first step of the optimisation routine is to generate a set of basis functions that minimise the infidelity landscape in the absence of control errors against the pulse amplitude, $f^{(n)}$ and $g^{(n)}$ in the frequency domain as described by equation \ref{eq:intrinsicinfone}. This step minimises crosstalk between the qubits.

\par The search range for the pulse amplitudes is constrained by the design of our MW/RF system and this corresponds to the shortest gate time that we can perform for our qubit gate operations. In the time domain, the pulse amplitude, $a(t)$ is determined by $\gamma_iB_1(t)$ where $\gamma_i$ is the gyromagnetic ratio of the nuclear spin qubits and $B_1(t)$ is the maximum amplitude of the oscillating MW/RF magnetic field. Implementing the typical values used in an experiment, the pulse amplitude $a(t)$ is thus limited to a maximum value of approximately 25~Mrad/s. The corresponding search range for $f^{(n)}$ and $g^{(n)}$ is then arbitrarily set to $f^{(n)},g^{(n)} \in \left[-5,5\right]$ and noting that the amplitude in the time domain has an additional factor of $2\sqrt{2\pi}$ from the inverse Fourier transform of the sinc basis functions. 

\par Similarly, the set of basis functions for a two-qubit CZ gate is generated using the same procedure. In this case, the pulse amplitude is limited to approximately 80 Mrad/s. The search range for $f^{(n)}$ is bounded in the region of $f^{(n)}\in \left[0.5,15\right]$ as the intrinsic infidelity expression used to describe two-qubit gate operations, equation \ref{eq:intrinsicinftwo} is symmetrical. Using $f^{(n)}=0$, the optimisation will be stuck in a local minimum and clearly, $f^{(n)}=0$ corresponds to no physical pulse. Thus, in order to find sensible solutions, we shifted the initial search boundary by 0.5 to stimulate the optimisation to find another local minimum. The equations that are satisfied by the solutions are given by equations \ref{eq:twoqubitx1}, \ref{eq:twoqubitx2}, \ref{eq:twoqubitx3} and \ref{eq:twoqubitx4}.

\par As seen in table~\ref{tab:1qdiffrotation}, the pulse amplitudes are dependent on the angle of rotations and gate time. This is consistent with the formulation of our sinc basis functions where larger amplitudes are expected to generate greater angle of rotations for a given frequency shift and gate time. We are also able to resolve solutions with lower infidelity for smaller angle of rotations at shorter gate times as the pulse amplitudes are smaller. We observed a trend, shown in table~\ref{tab:1qdiffrotation} and \ref{tab:diffrotationCZ}, where at short gate times, higher order basis functions have solutions which correspond to the lower or upper bound of the allowed values. These pulse amplitudes minimise the infidelities within the search range imposed on them as set by physics. While some of the pulse amplitudes correspond to the maximum or minimum allowed values, the critical factor in this optimisation routine is the linear combinations of these generated basis functions as discussed in the following section.

\section{Optimal Gates for Non-Ideal Control System}
\label{sec:nonidealgates}
\par Applying this optimal control method, we optimise the infidelities of an X gate, Hadamard gate and a CZ gate due to the control errors. We approximated the standard deviations of the Gaussian phase, amplitude and frequency noises to be $\sigma_\phi=1/2\pi$~kHz, $\sigma_\epsilon=10^{-3}$, $\sigma_\delta=  1$~kHz for nuclear spins and $\sigma_\delta= 27.5$~kHz for the electron spin \cite{Balasubramanian2009}. Phase noises are excluded from the calculation of average infidelities for a CZ gate as we are performing a $2\pi$ pulse. The first order effects due to the phase errors are negligible.

\begin{figure}[ht]
\centering
	\begin{subfigure}[b]{0.45\textwidth}
		\includegraphics[width=1\textwidth]{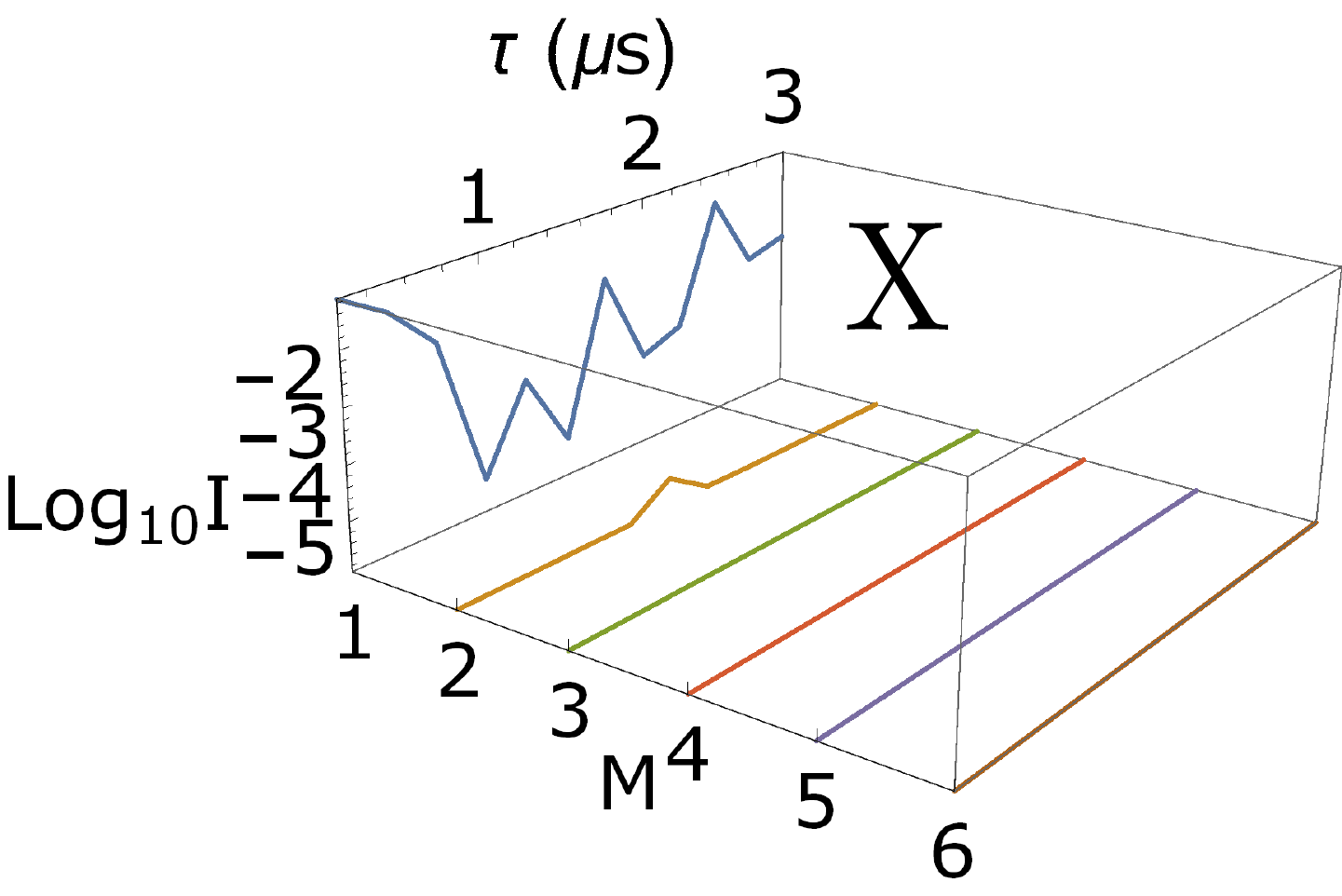}
	\caption[]{}
	\label{fig:xgate}
	\end{subfigure}
	\begin{subfigure}[b]{0.45\textwidth}
	\includegraphics[width=1\textwidth]{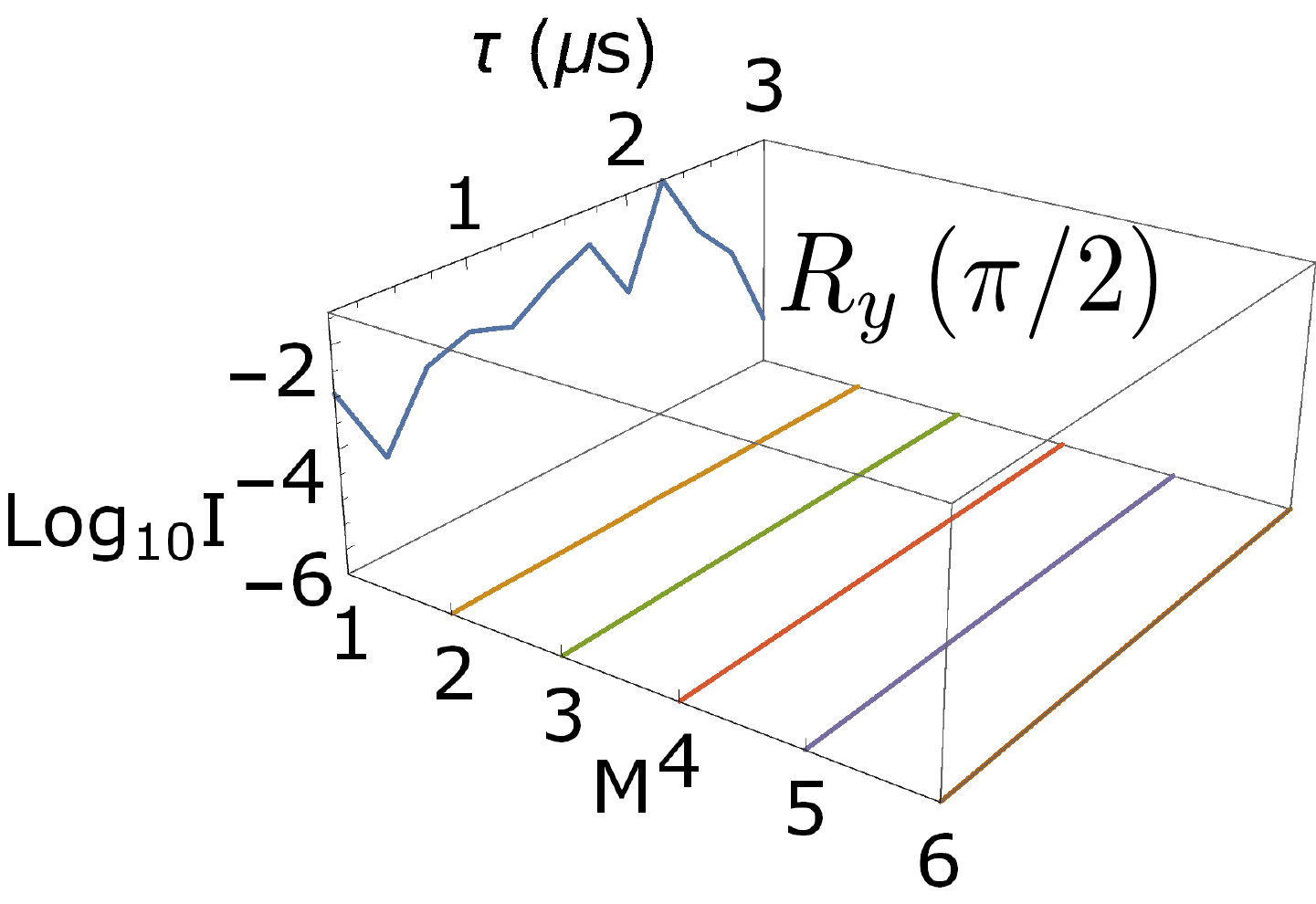}
	\caption[]{}
	\label{fig:ygate}
	\end{subfigure}
	\begin{subfigure}[b]{0.45\textwidth}
	\includegraphics[width=1\textwidth]{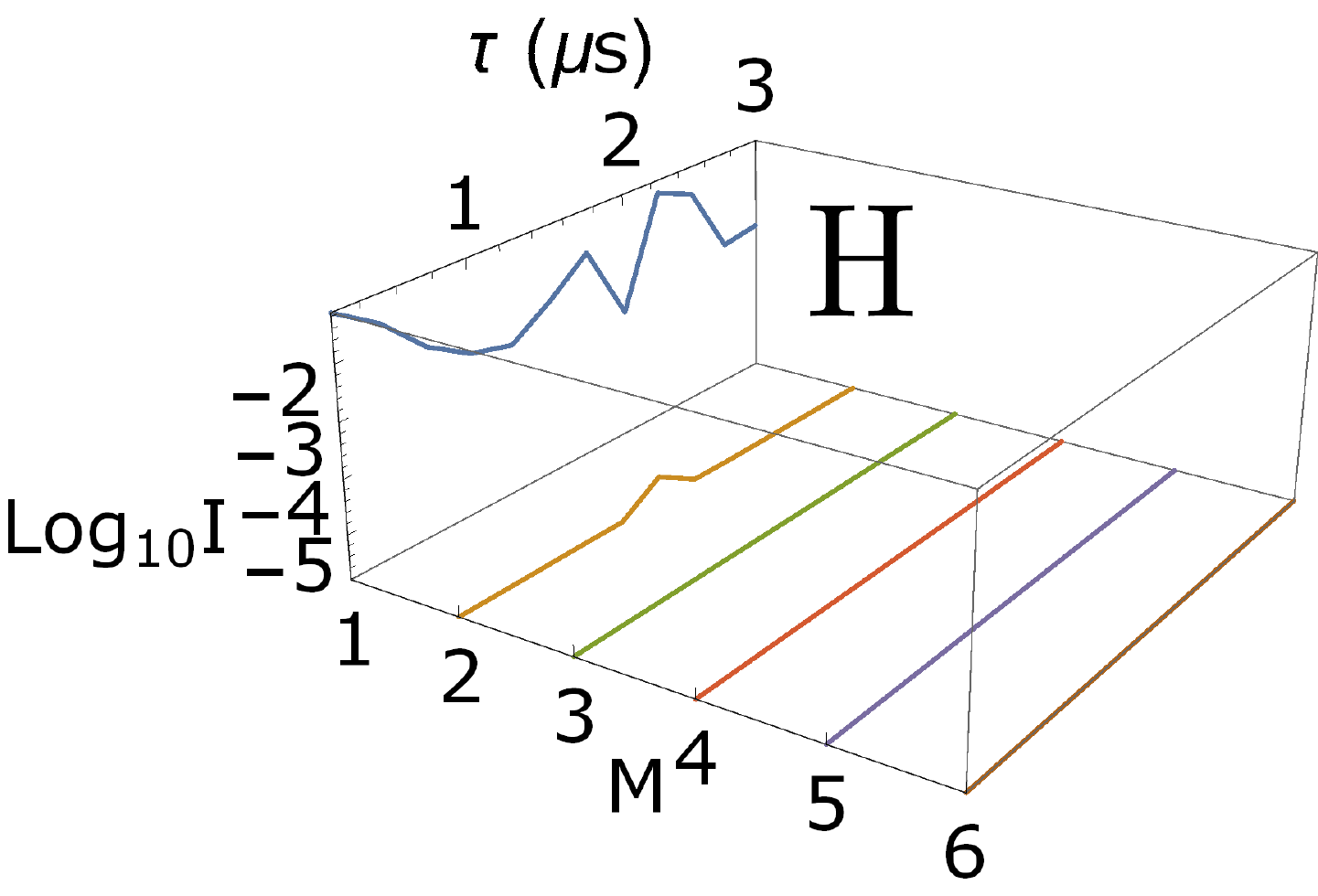}
	\caption[]{}
	\label{fig:Hgate}
	\end{subfigure}
	\caption[]{Calculated infidelities of various gates for different gate times. (a) The performance of an X gate, (b) a $\pi/2$ rotation about the y-axis and (c), a Hadamard gate in the presence of frequency, amplitude and phase noises. The infidelities are plotted as a function of total number of basis functions, $M$ and gate time, $\tau$. Gate infidelities of $\sim 10^{-6}$ can be achieved by using linear combinations of two or more basis functions. The infidelities are decreasing with an increasing number of basis functions used in the optimisation procedure (not shown on scale). In general, 3 basis functions are needed for the infidelities to converge to the optimisation limits. For a system with two qubits, using more than 3 basis functions does not have significant improvements on achieving lower infidelities.}
	\label{fig:1qubitgateinf}
\end{figure}
\newpage
\par As shown in figure~\ref{fig:1qubitgateinf}, the average infidelities for an X gate, a $\pi/2$ rotation about the y-axis and a Hadamard gate fluctuate with a single basis function, since the function parameters depend on the local infidelity landscape during the initial basis function computation. However,  the optimised linear combinations of two or more basis functions yield infidelities of approximately $10^{-6}$ for an X gate and a Hadamard gate. Infidelities up to $10^{-7}$ can be achieved for a $\pi/2$ rotation about the y-axis. The infidelities are monotonically decreasing with increasing number of basis functions and only 3 basis functions are required for the infidelities to converge to the optimisation limits. Thus, this demonstrates the capability of this optimal control method, and allow us to systematically assess the degree of convergence to the optimisation limits. For a system with two qubits, using more than 3 basis functions does not significantly lower the infidelities. In general, we expect to achieve infidelities on this order (of $10^{-6}$) for any single-qubit operation. 

\par Given that this method can achieve low infidelities, the main objective is to shorten the gate time without significantly affect the infdelity, in order to minimise the effects due to decoherence. We analyse the overall amplitudes of the respective linear combinations of the sinc basis functions in order to find the minimum gate time for which the pulse amplitude can still be practically generated. Using estimated constraints described in the previous section, the maximum threshold for the amplitude is set to be $\log_{10}$(25)$\approx 1.40$. 

\par Figure~\ref{fig:singleabsamp} depicts the overall amplitudes of the basis functions in the linear combinations for different basis size and gate times. At some gate times, the linear combinations have larger amplitudes when more basis functions are used. Since the optimal control method optimises the linear combinations of the basis functions such that the infidelities are monotonically decreasing, therefore, if the amplitude in the initial basis function is large, then it may result in larger amplitudes for the subsequent basis functions when the basis size is increased. Based on the generated data, we determined that a $0.5~\mu$s X gate with an infidelity of $\sim 10^{-6}$ can be achieved with just two basis functions (figure \ref{fig:single2lc}). Despite being able to perform a $0.5~\mu$s X gate, we need to take into account various gate times required for different types of rotations where slower gates are required for bigger angle of rotations. Thus, on average, a conservative estimate for the fastest single-qubit gate that we are able to perform with $\sim 10^{-6}$ infidelity is approximately $1~\mu$s.

\begin{figure}[H]
\centering
	\begin{subfigure}[b]{0.45\textwidth}
		\includegraphics[width=1\textwidth]{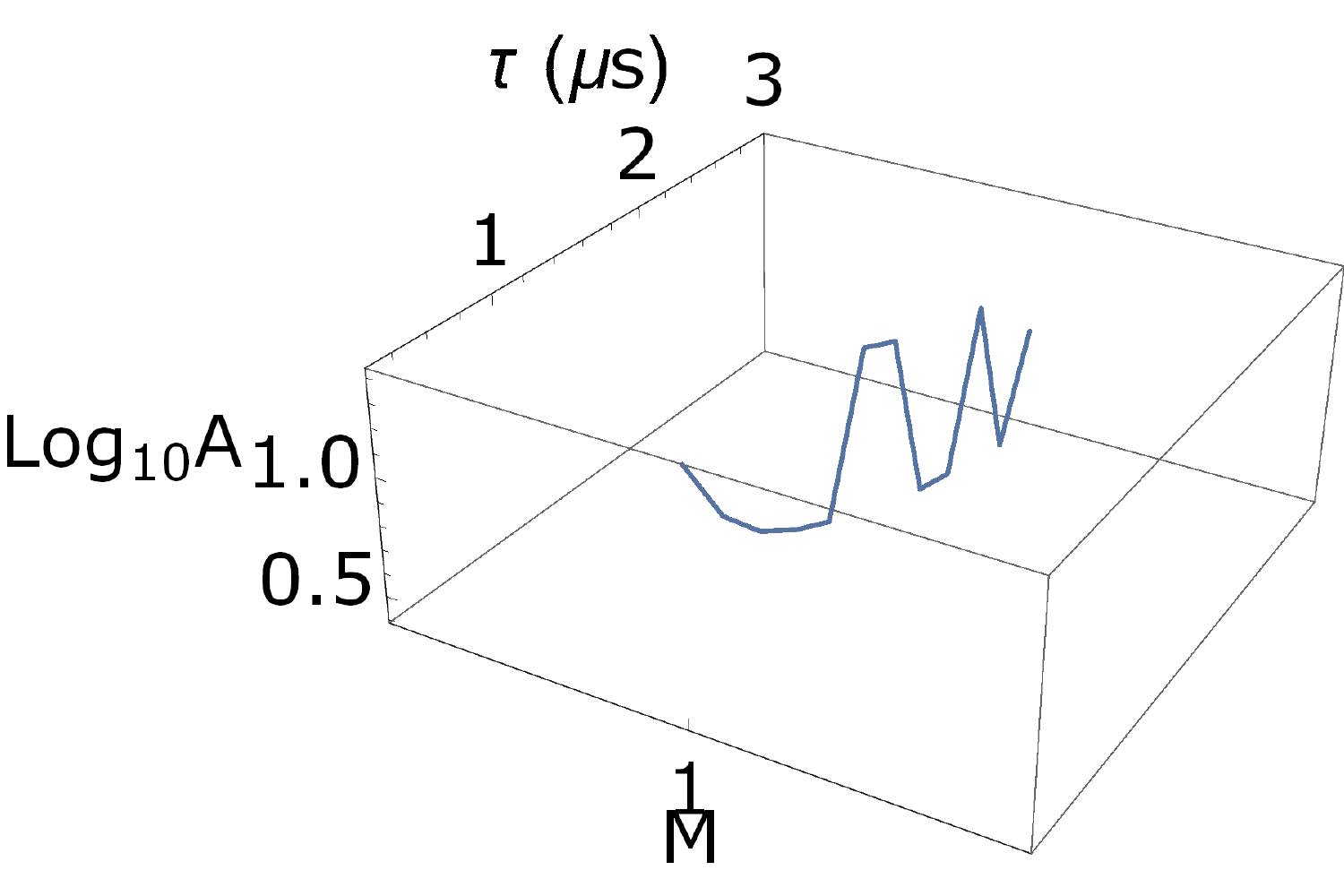}
	\caption[]{}
	\label{fig:single1lc}
	\end{subfigure}
	\begin{subfigure}[b]{0.45\textwidth}
	\includegraphics[width=1\textwidth]{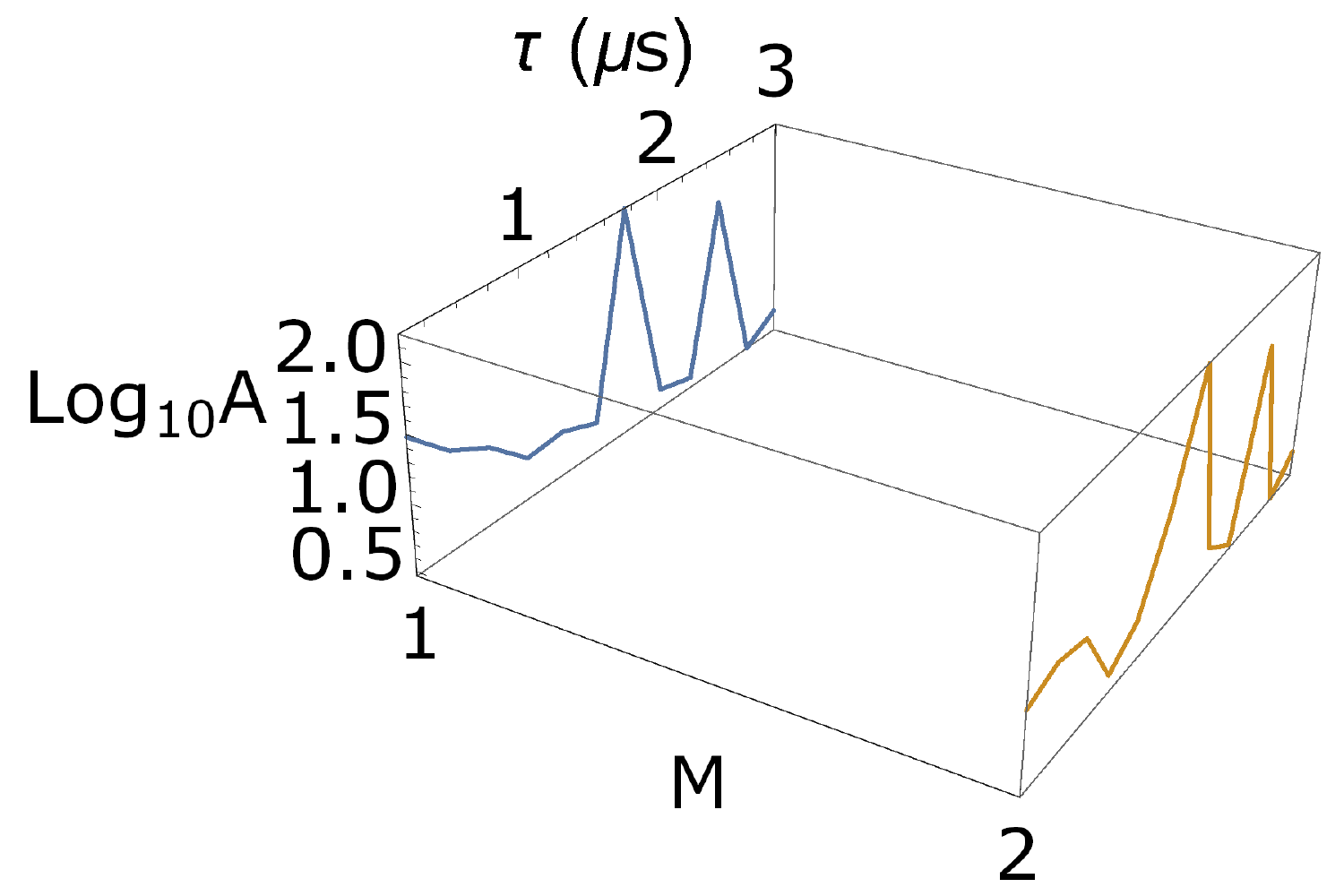}
	\caption[]{}
	\label{fig:single2lc}
	\end{subfigure}
	\begin{subfigure}[b]{0.45\textwidth}
		\includegraphics[width=1\textwidth]{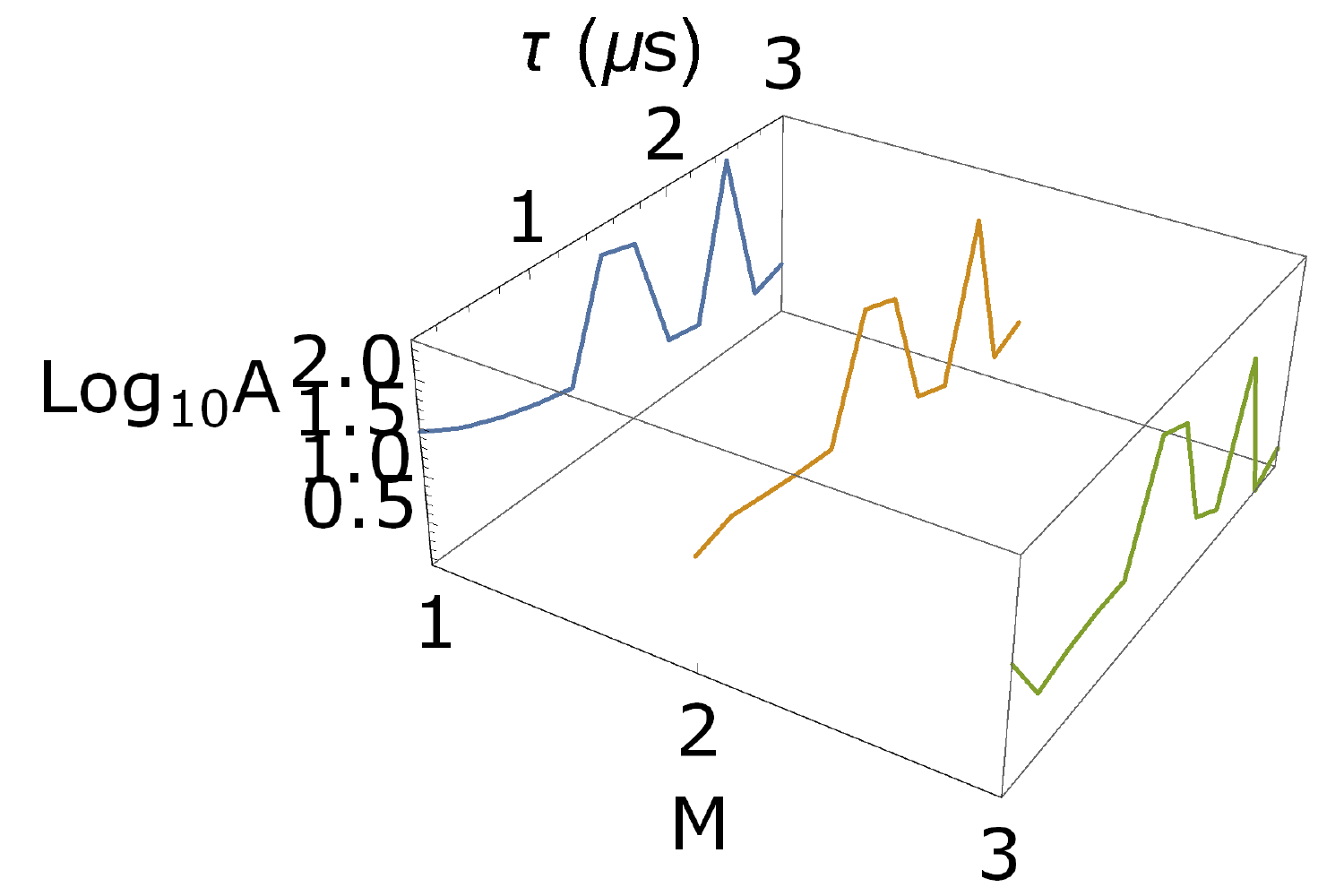}
	\caption[]{}
	\label{fig:single3lc}
	\end{subfigure}
	\begin{subfigure}[b]{0.45\textwidth}
	\includegraphics[width=1\textwidth]{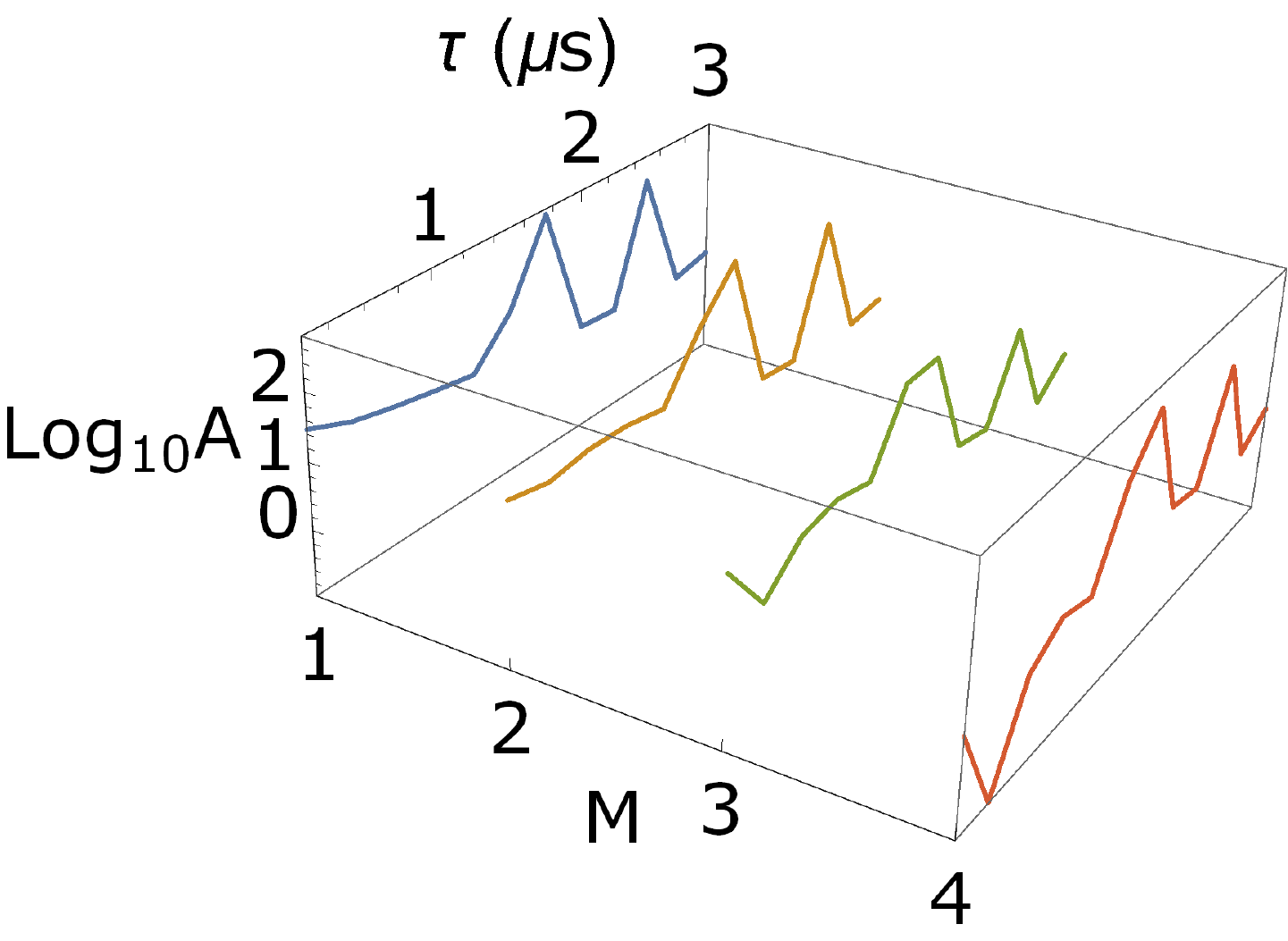}
	\caption[]{}
	\label{fig:single4lc}
	\end{subfigure}
\end{figure}
\begin{figure}[H]
\ContinuedFloat
\centering
	\begin{subfigure}[b]{0.45\textwidth}
		\includegraphics[width=1\textwidth]{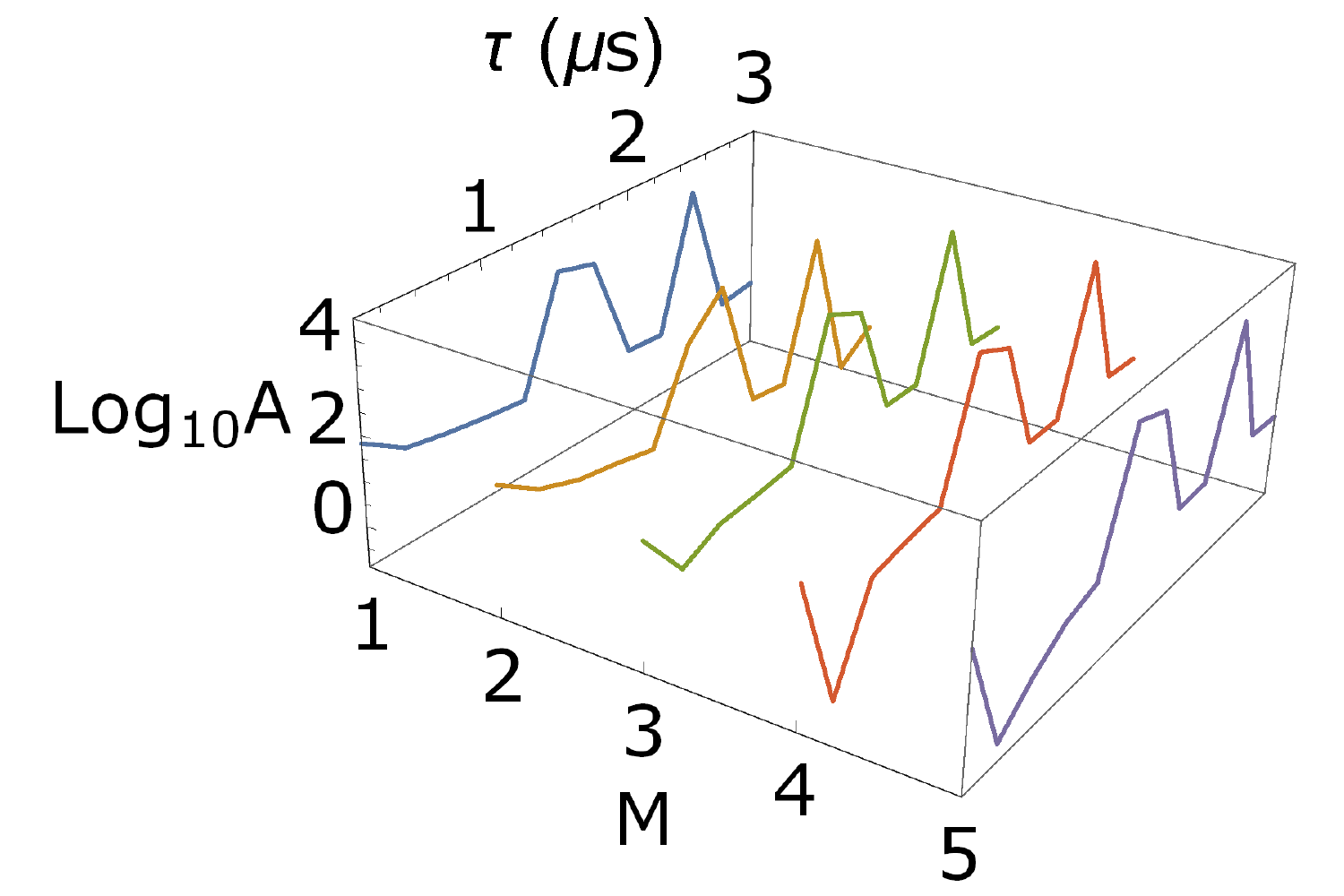}
	\caption[]{}
	\label{fig:single5lc}
	\end{subfigure}
	\begin{subfigure}[b]{0.45\textwidth}
	\includegraphics[width=1\textwidth]{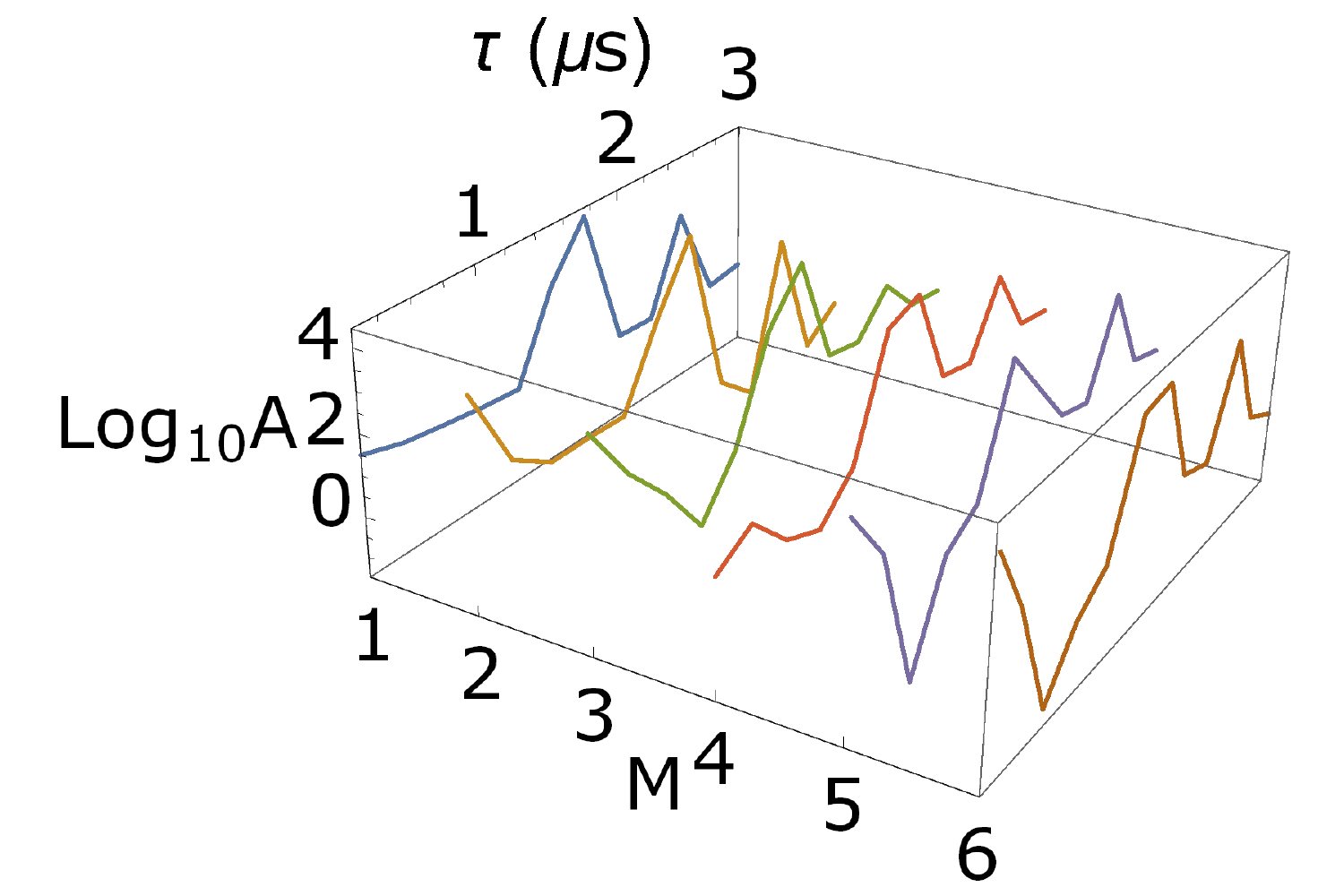}
	\caption[]{}
	\label{fig:single6lc}
	\end{subfigure}
	\caption[]{Plots of the overall amplitudes, $A$, which realises an X gate with the minimum average infidelity ranging from 1 basis function (a), to 6 basis functions (f). At certain gate times, the overall amplitudes become very large when the number of basis functions are increased. The overall amplitudes are dependent on the initial solutions generated in the minimisation of the infidelity expression (equation \ref{eq:intrinsicinfone}). These amplitudes are optimised to generate lower infidelities with an increasing number of basis functions. Thus, a large amplitude in the first basis function may result in linear combinations with larger amplitudes when the basis size is increased.}
	\label{fig:singleabsamp}
\end{figure}
\par
\begin{figure}[H]
\centering
	\includegraphics[width=0.7\textwidth]{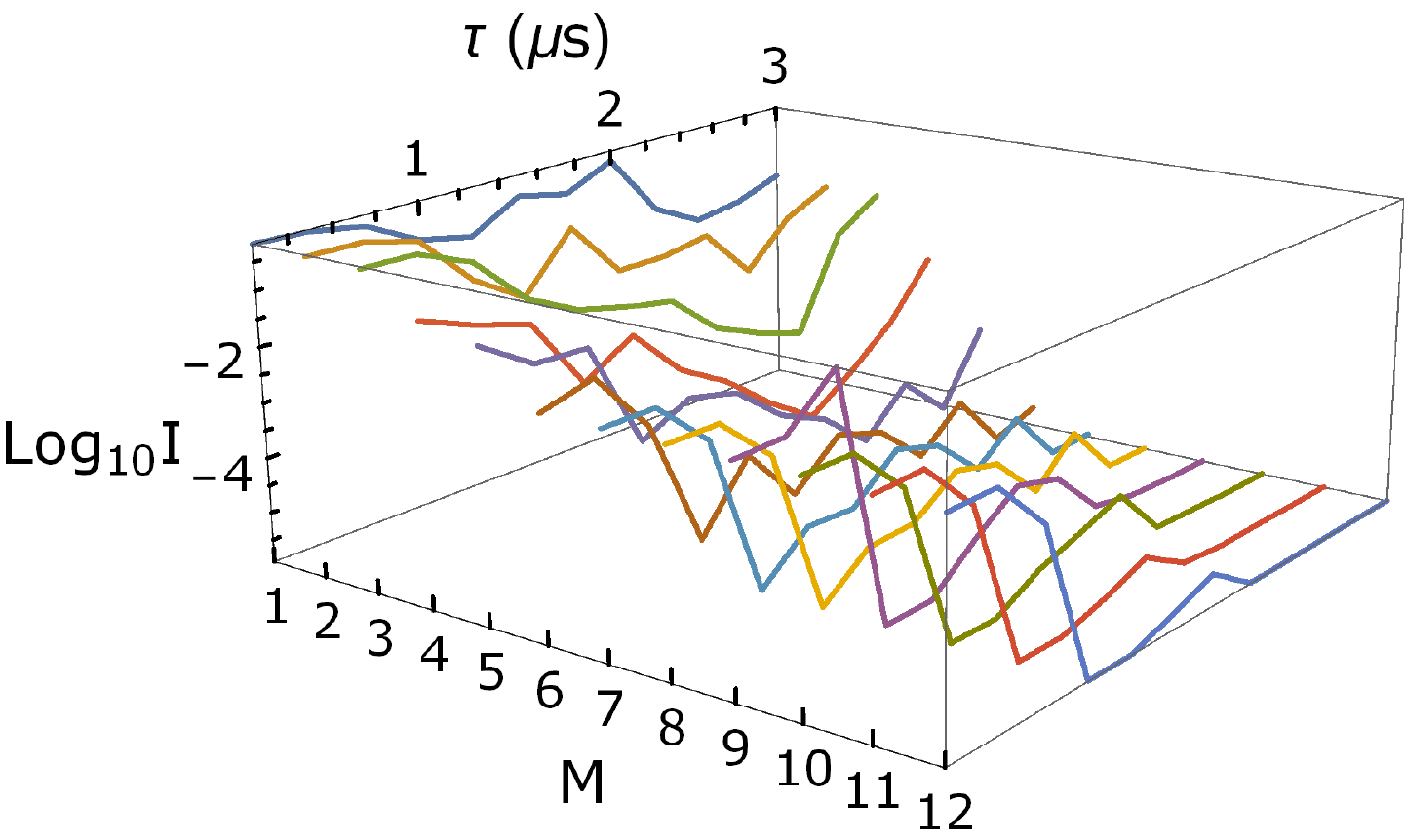}
	\caption[]{The performance of a two-qubit CZ gate in the presence of frequency and amplitude noise. The infidelities are plotted as a function of total number of basis functions $M$ and gate time $\tau$. On average, gate infidelities of $10^{-4}\sim 10^{-6}$ can be achieved by using linear combinations of at least 4 basis functions for gate times $\tau \geq 1~\mu$s. Monotonicity was not fully demonstrated for $\tau<1~\mu$s due to numerical integration errors. This is caused by many linear combinations of basis functions with extremely large amplitudes which results in a high oscillatory integrand. We observed convergence of these infidelities to their respective optimisation limits and it requires at least 8 basis functions on average.}
	\label{fig:czgate}
\end{figure}

 Applying our method now to two-qubit gates, the maximum threshold for the amplitude of a two-qubit CZ gate is set to be $\log_{10}$(80)$\approx 1.9$. From figure~\ref{fig:czgate}, on average, gate infidelities of $10^{-4}\sim 10^{-6}$ can be achieved with 4 or more basis functions for gate times $\tau \geq 1~\mu$s. The calculated infidelities are monotonically decreasing with increasing basis size for $\tau \geq 1~\mu$s. Monotonicity was not fully demonstrated for $\tau<1~\mu$s due to numerical integration errors. These errors are exacerbated for $\tau<1~\mu$s because of many linear combinations of basis functions with extremely large amplitudes at short gate times, resulting in a high oscillatory integrand.  On average, more than 8 basis functions are required for the convergence of these infidelities to their respective optimisation limits. Based on the optimisations shown in figures \ref{fig:czgate} and \ref{fig:CZabsamp}, the fastest two-qubit CZ gate that we can perform with an infidelity of $\sim 10^{-6}$ is $1~\mu$s and it requires 6 linear combinations of basis functions.

\begin{figure}[H]
\centering
	\begin{subfigure}[b]{0.45\textwidth}
		\includegraphics[width=1\textwidth]{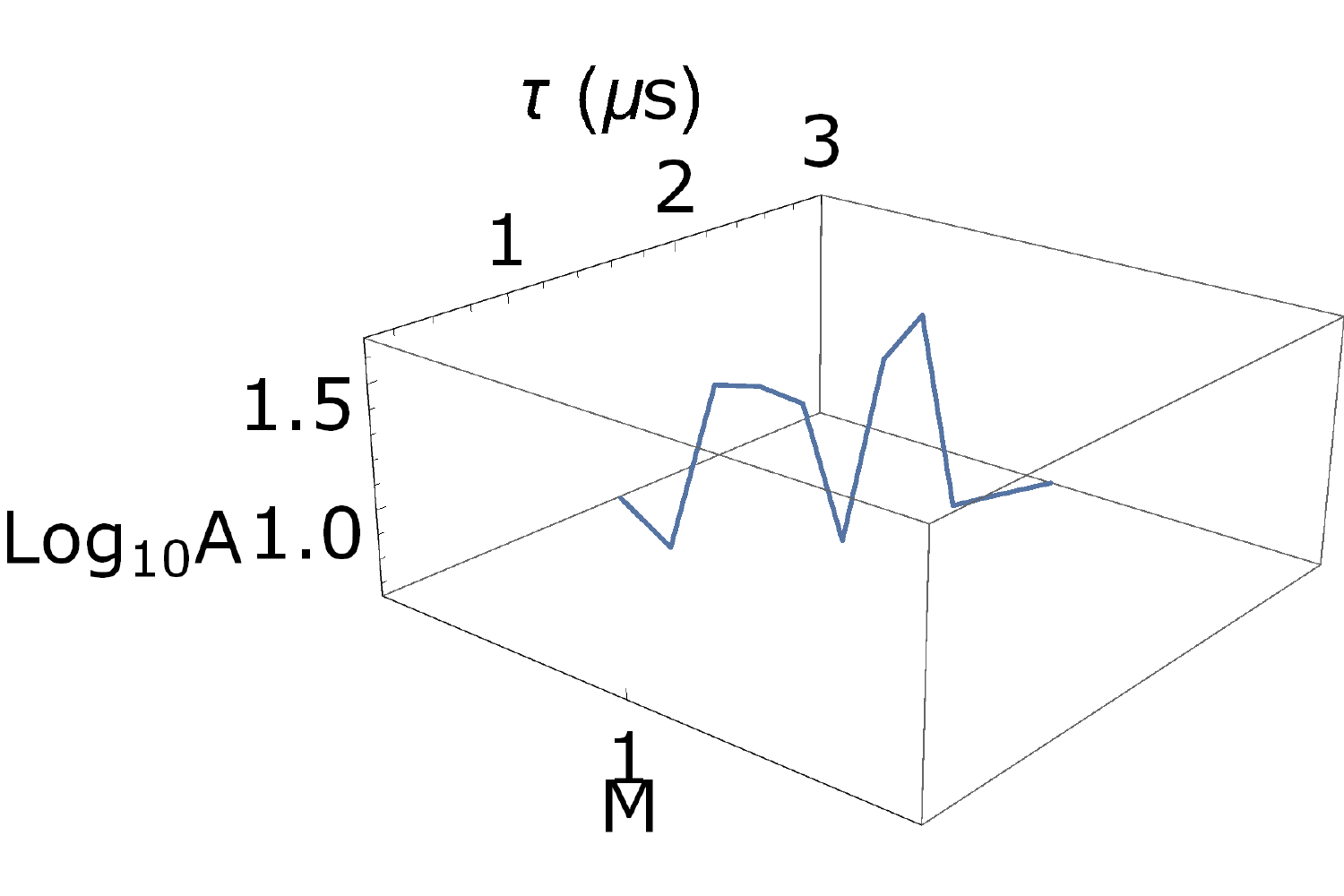}
	\caption[]{}
	\label{fig:CZ1lc}
	\end{subfigure}
	\begin{subfigure}[b]{0.45\textwidth}
	\includegraphics[width=1\textwidth]{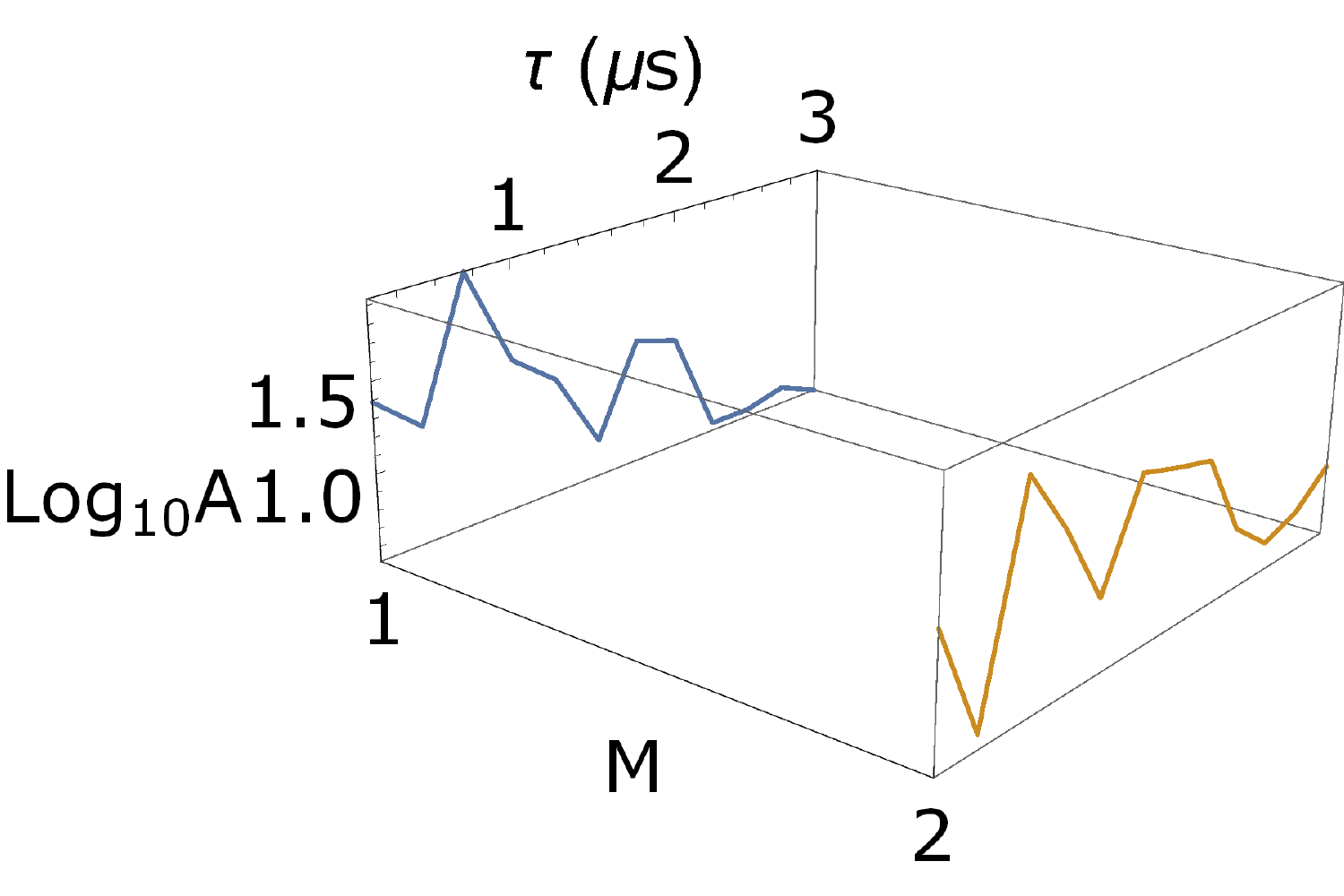}
	\caption[]{}
	\label{fig:CZ2lc}
	\end{subfigure}
	\begin{subfigure}[b]{0.45\textwidth}
		\includegraphics[width=1\textwidth]{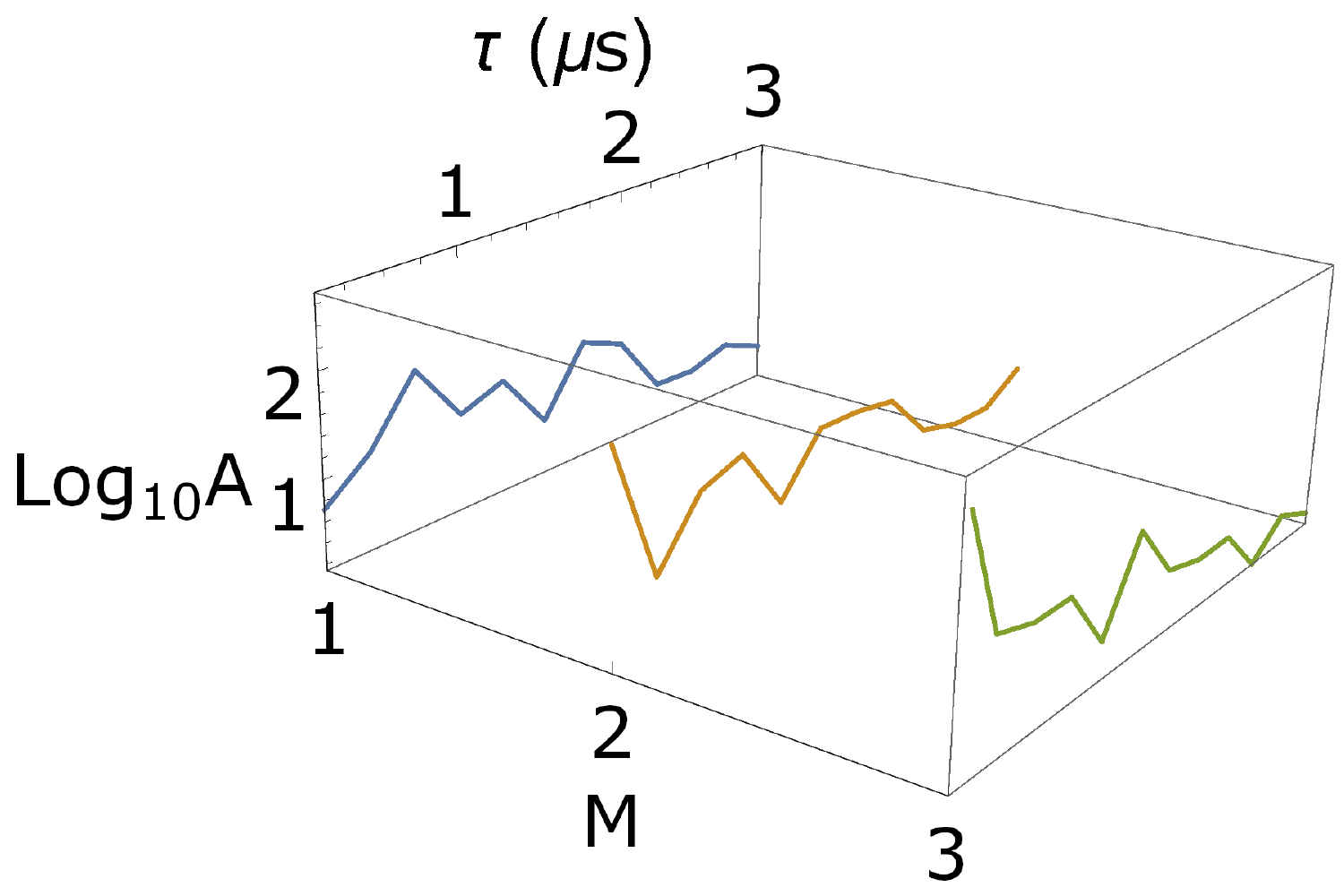}
	\caption[]{}
	\label{fig:CZ3lc}
	\end{subfigure}
	\begin{subfigure}[b]{0.45\textwidth}
	\includegraphics[width=1\textwidth]{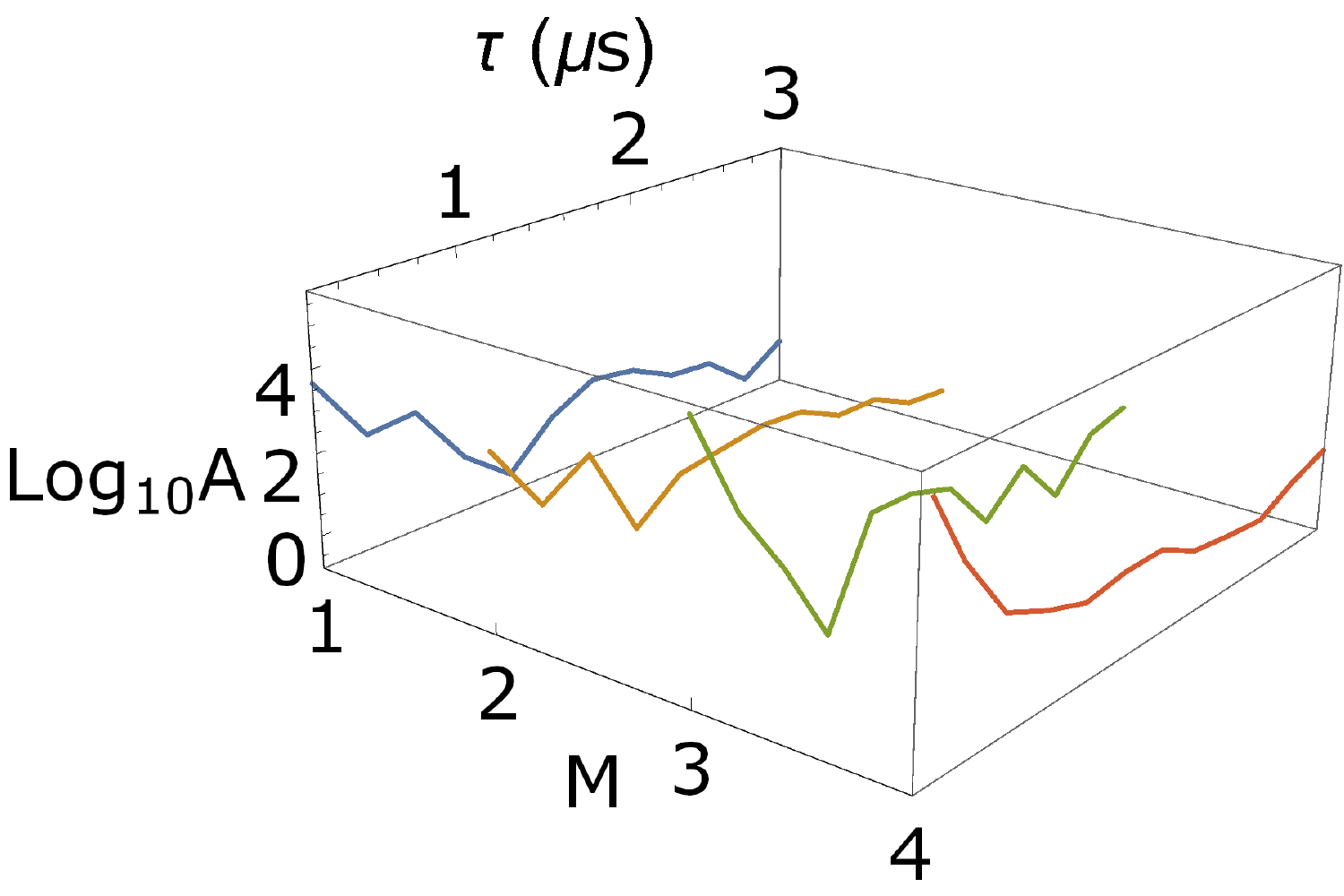}
	\caption[]{}
	\label{fig:CZ4lc}
	\end{subfigure}
\end{figure}
\begin{figure}[H]
\centering
\ContinuedFloat
	\begin{subfigure}[b]{0.45\textwidth}
		\includegraphics[width=1\textwidth]{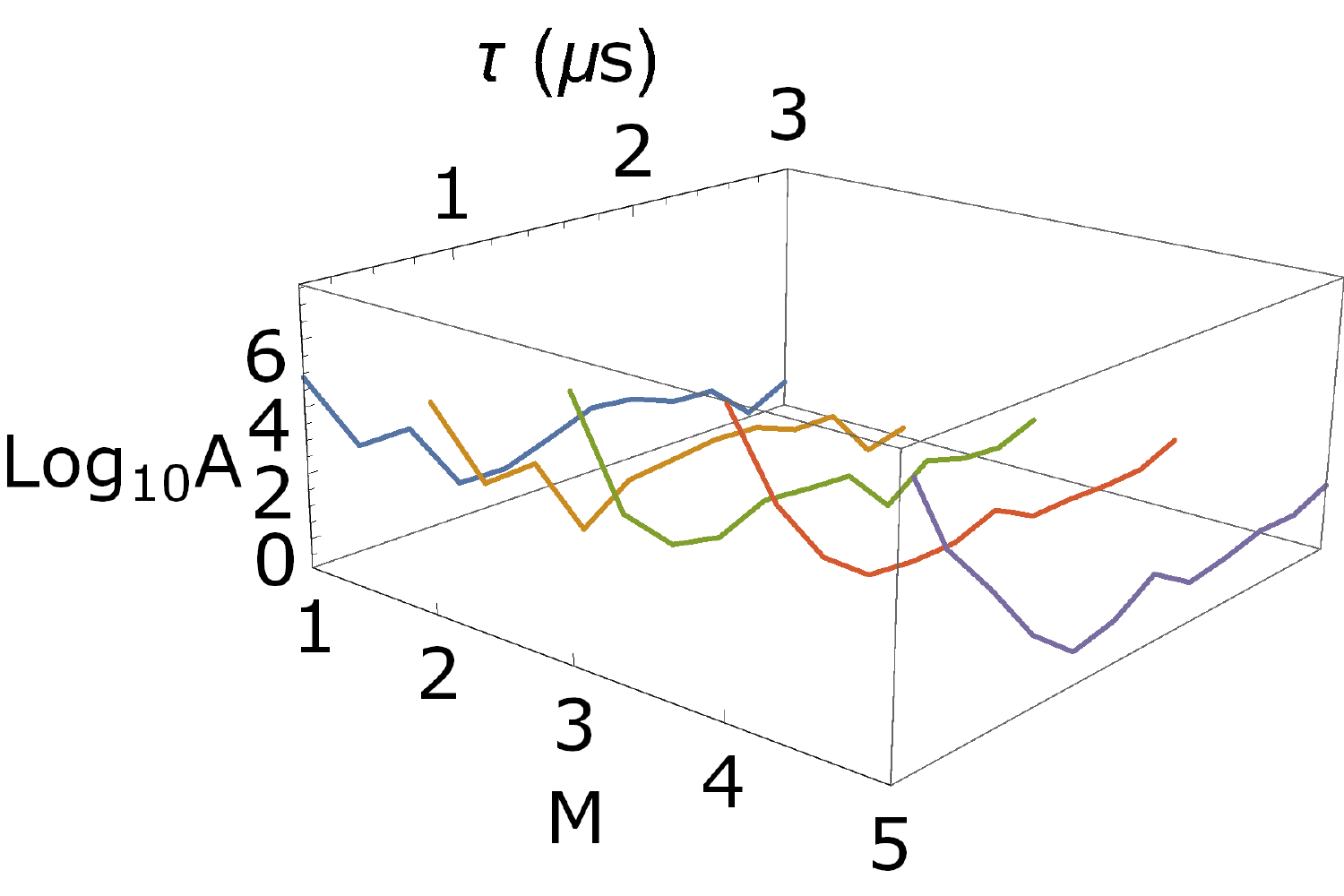}
	\caption[]{}
	\label{fig:CZ5lc}
	\end{subfigure}
	\begin{subfigure}[b]{0.45\textwidth}
	\includegraphics[width=1\textwidth]{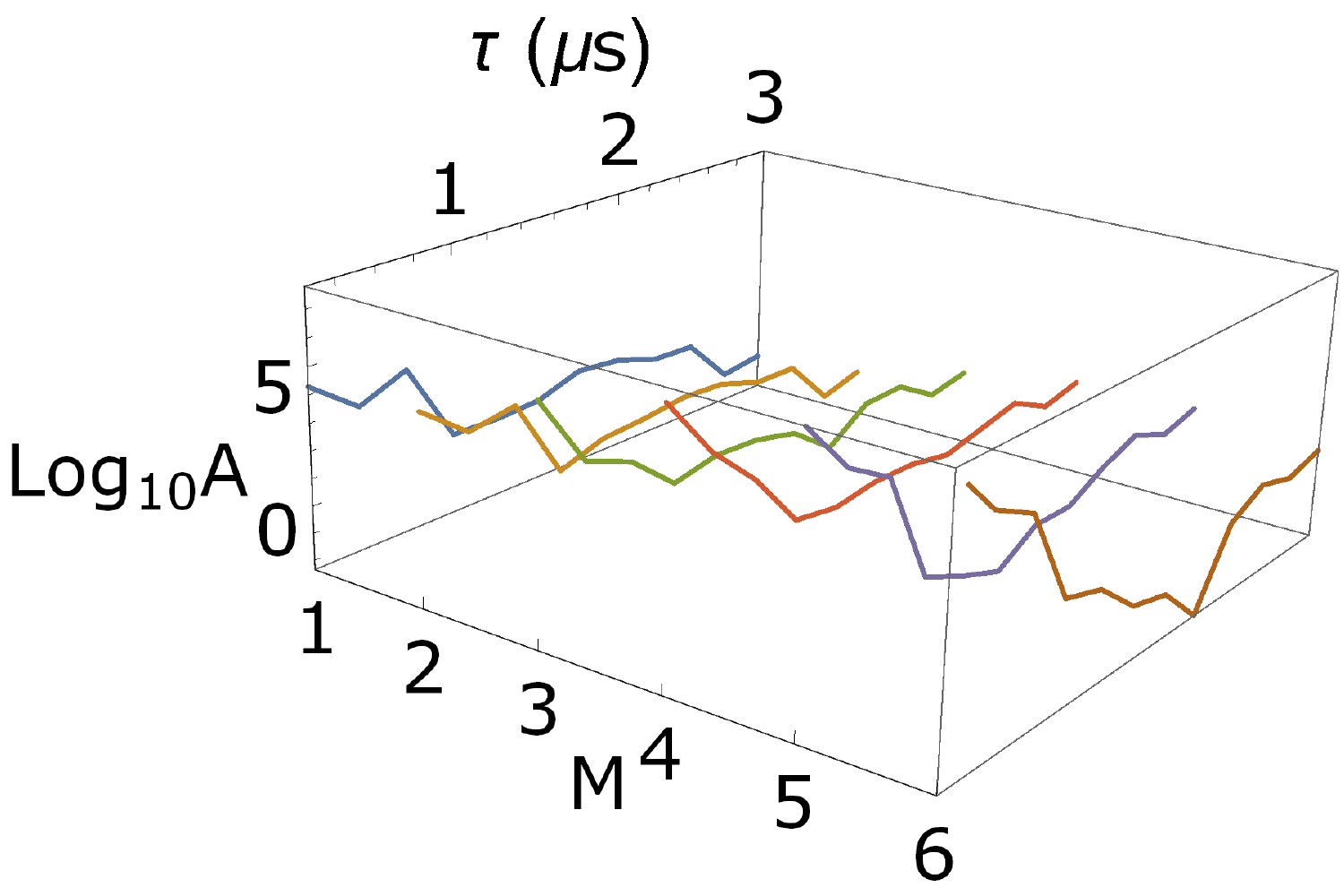}
	\caption[]{}
	\label{fig:CZ6lc}
	\end{subfigure}
	\begin{subfigure}[b]{0.45\textwidth}
		\includegraphics[width=1\textwidth]{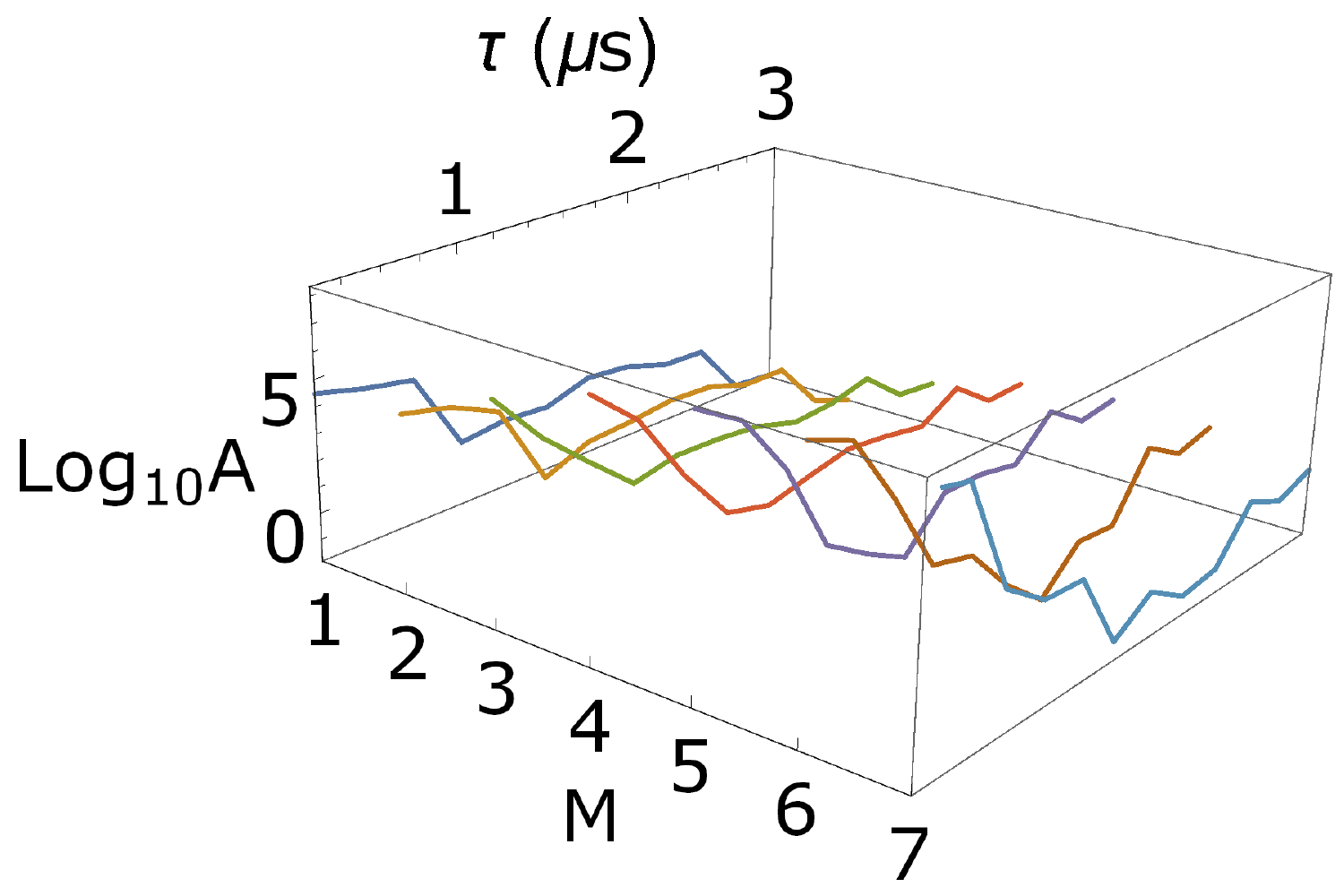}
	\caption[]{}
	\label{fig:CZ7lc}
	\end{subfigure}
	\begin{subfigure}[b]{0.45\textwidth}
	\includegraphics[width=1\textwidth]{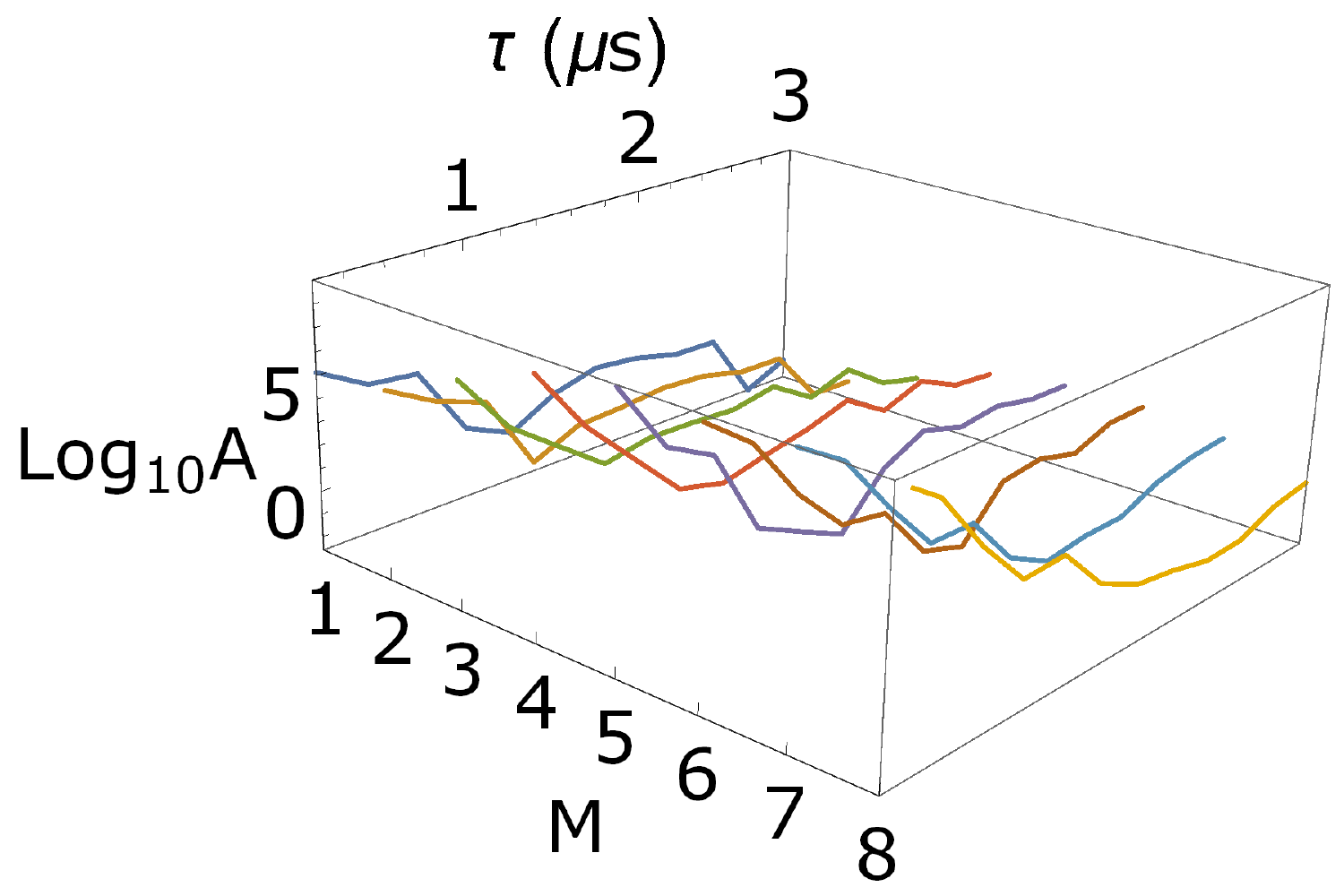}
	\caption[]{}
	\label{fig:CZ8lc}
	\end{subfigure}
	\begin{subfigure}[b]{0.45\textwidth}
		\includegraphics[width=1\textwidth]{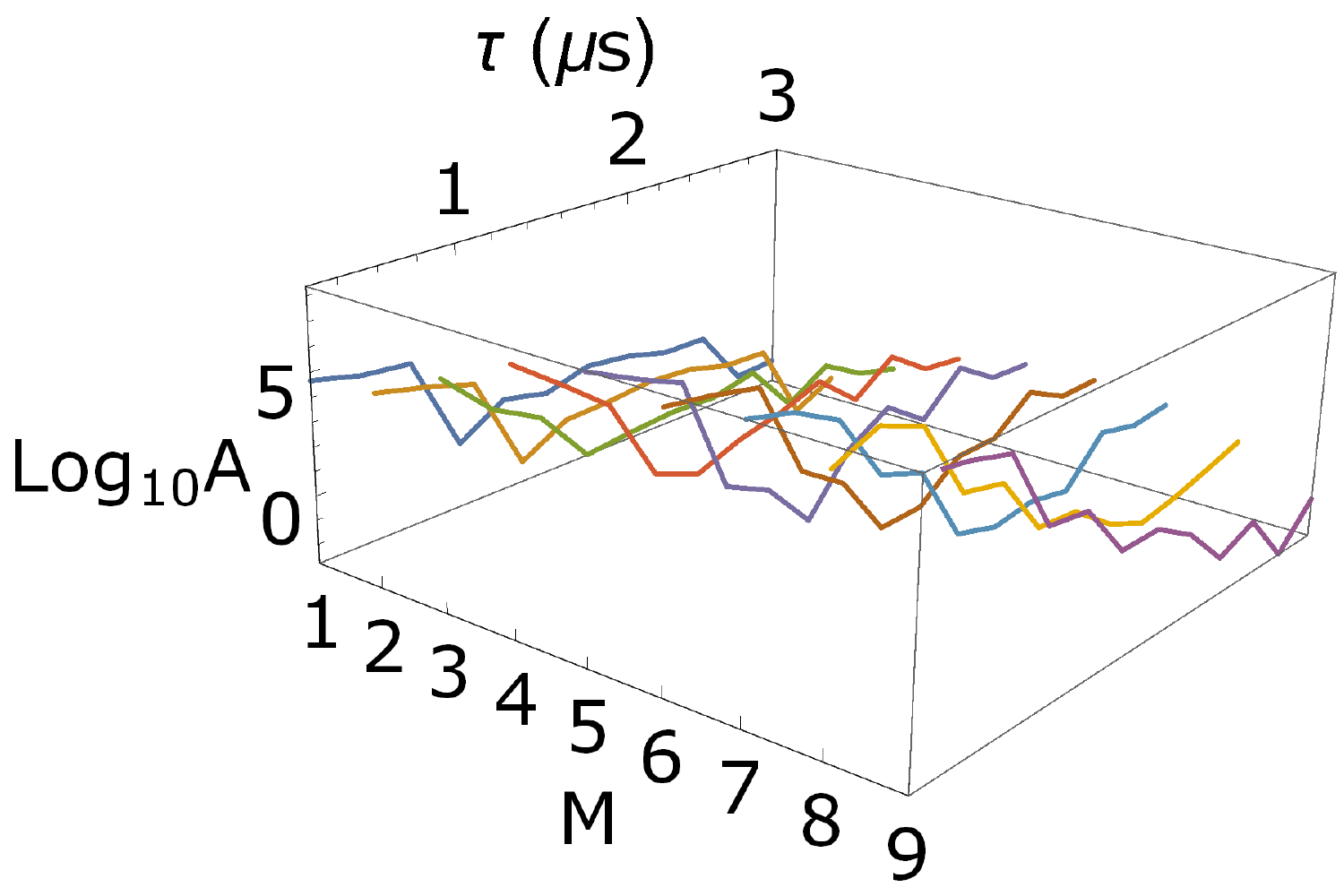}
	\caption[]{}
	\label{fig:CZ9lc}
	\end{subfigure}
	\begin{subfigure}[b]{0.45\textwidth}
	\includegraphics[width=1\textwidth]{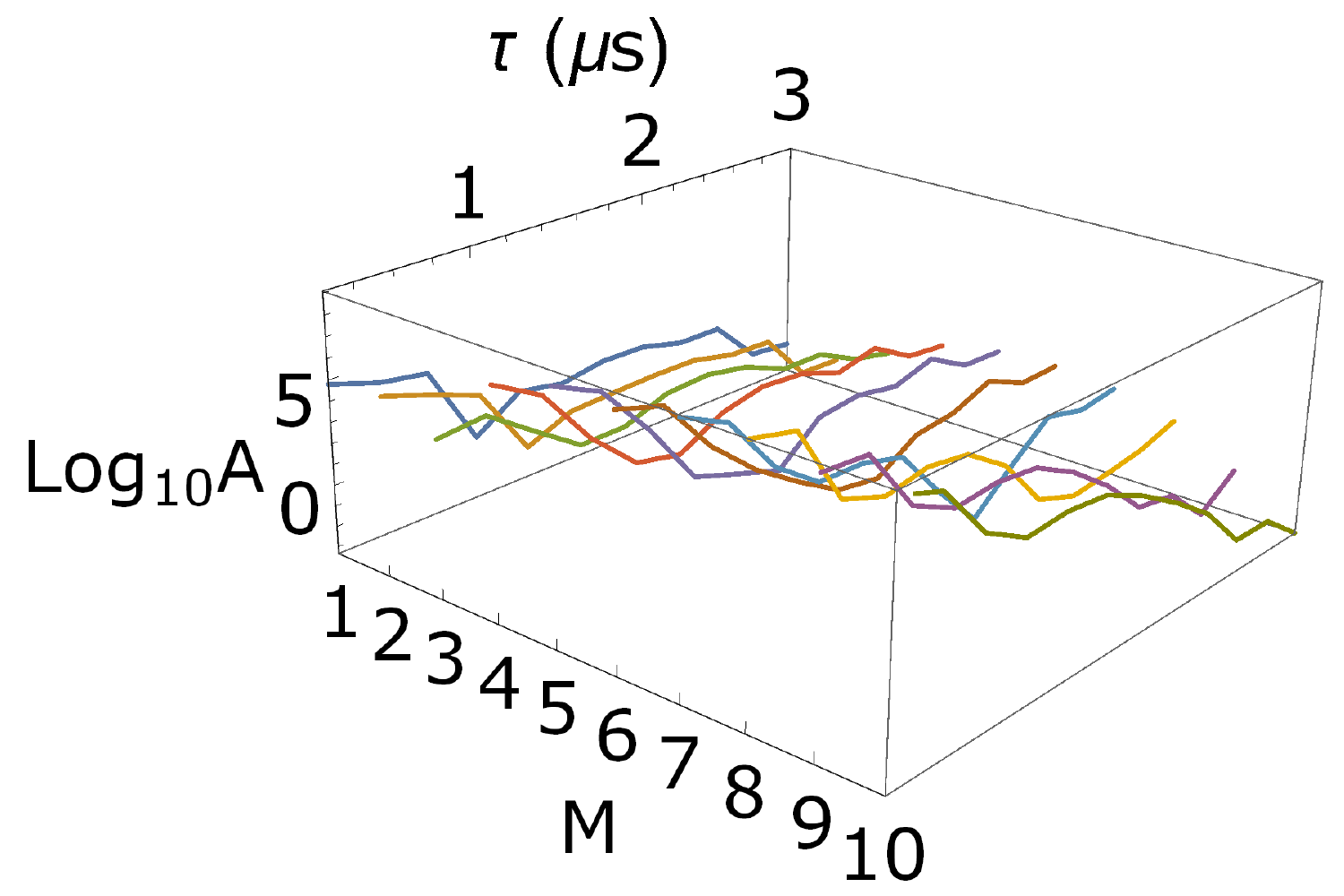}
	\caption[]{}
	\label{fig:CZ10lc}
	\end{subfigure}
\end{figure}
\begin{figure}[H]
\ContinuedFloat
\centering
	\begin{subfigure}[b]{0.45\textwidth}
		\includegraphics[width=1\textwidth]{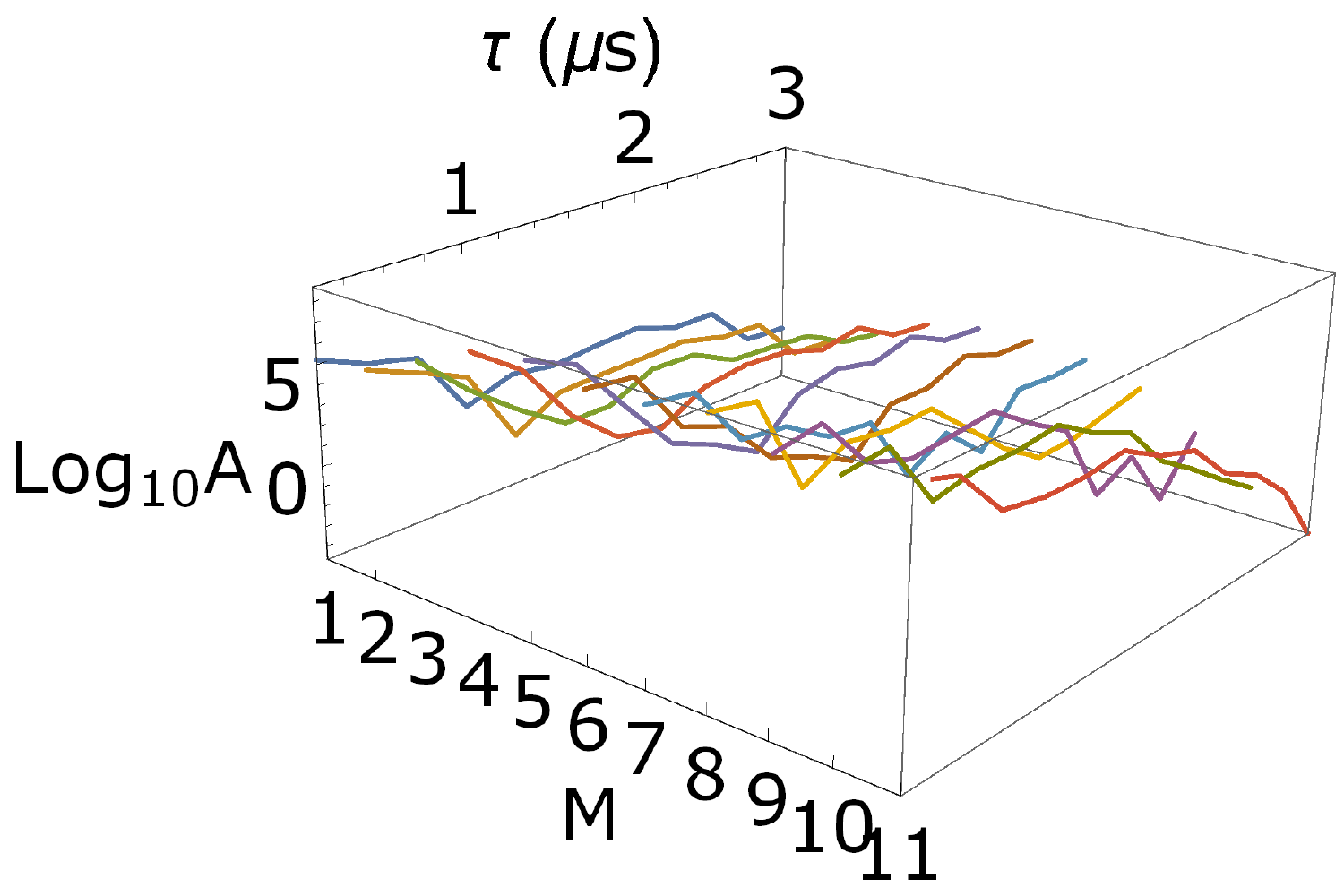}
	\caption[]{}
	\label{fig:CZ11lc}
	\end{subfigure}
	\begin{subfigure}[b]{0.45\textwidth}
	\includegraphics[width=1\textwidth]{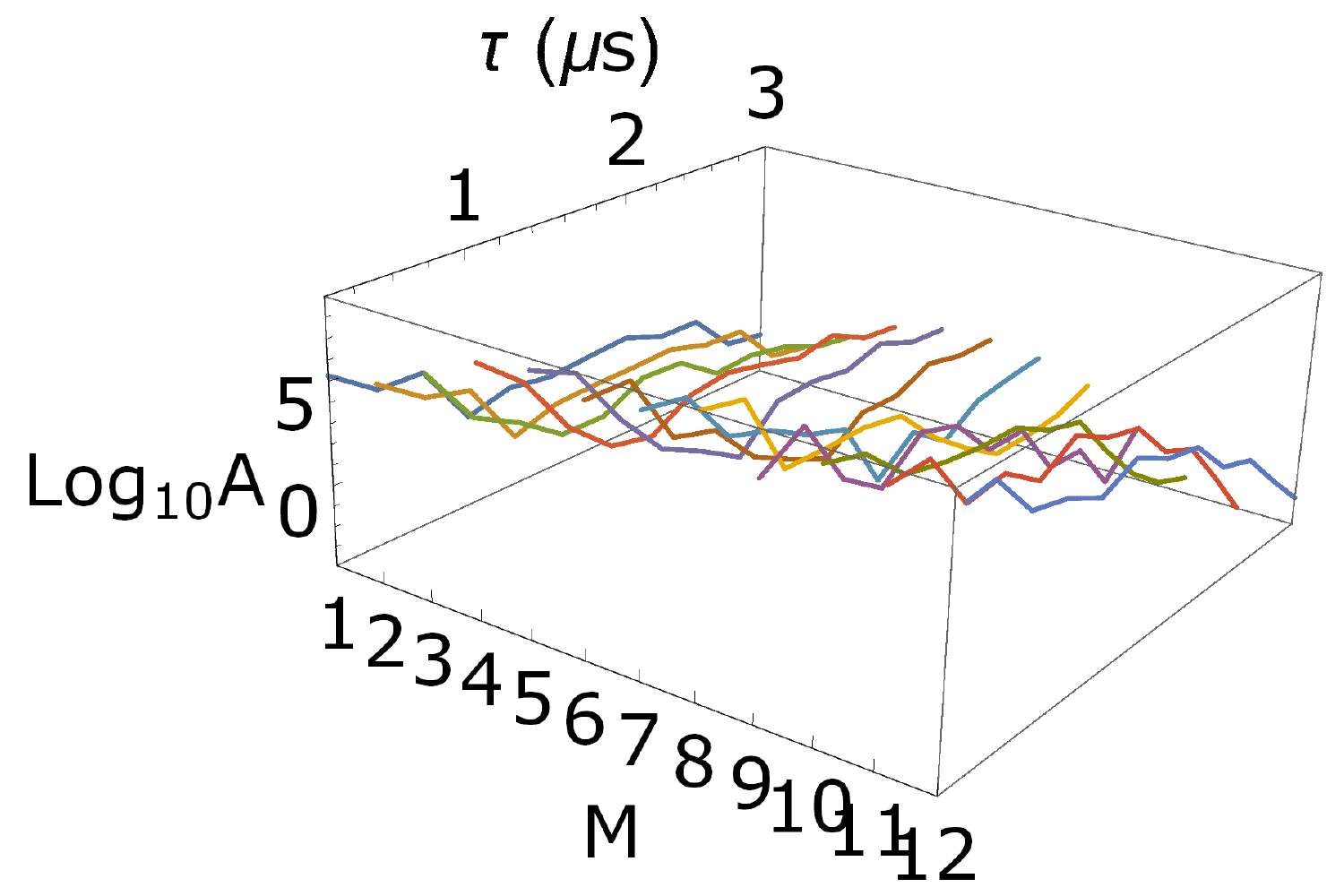}
	\caption[]{}
	\label{fig:CZ12lc}
	\end{subfigure}
	\caption[]{Plots of the overall amplitudes, $A$, which generate the minimum averaged infidelity of a CZ gate ranging from 1 basis function (a), to 12 basis functions (l). Similar to an X gate, at certain gate times, the overall amplitudes are much larger when the number of basis functions are increased to generate monotonically decreasing infidelities and they are dependent on the initial solutions generated in the minimisation of the intrinsic infidelity expression (equation \ref{eq:intrinsicinftwo}).}
	\label{fig:CZabsamp}
\end{figure}
\par Additionally, we have analysed the infidelities of a CZ gate for a range of electron spin linewidths that may occur in materials with different densities of $^{13}$C. Figure~\ref{fig:CZlinewidth2} and \ref{fig:CZlinewidth3} depicts the CZ gate infidelities with $\sigma_\delta=2\times27.5$~kHz and $4\times 27.5$~kHz , respectively. Meanwhile, Figure~\ref{fig:CZconvergence} demonstrates convergence of CZ gate infidelities with $\tau=2.25~\mu$s occuring at different number of basis functions for different electron spin linewidths, with larger linewidths requiring larger number of basis functions. Overall, we are still able to obtain infidelities less than $10^{-4}$ across the range of linewidths, but at a cost of longer gate time and larger number of basis functions.

\begin{figure}[H]
\centering
\begin{subfigure}[b]{0.45\textwidth}
		\includegraphics[width=1\textwidth]{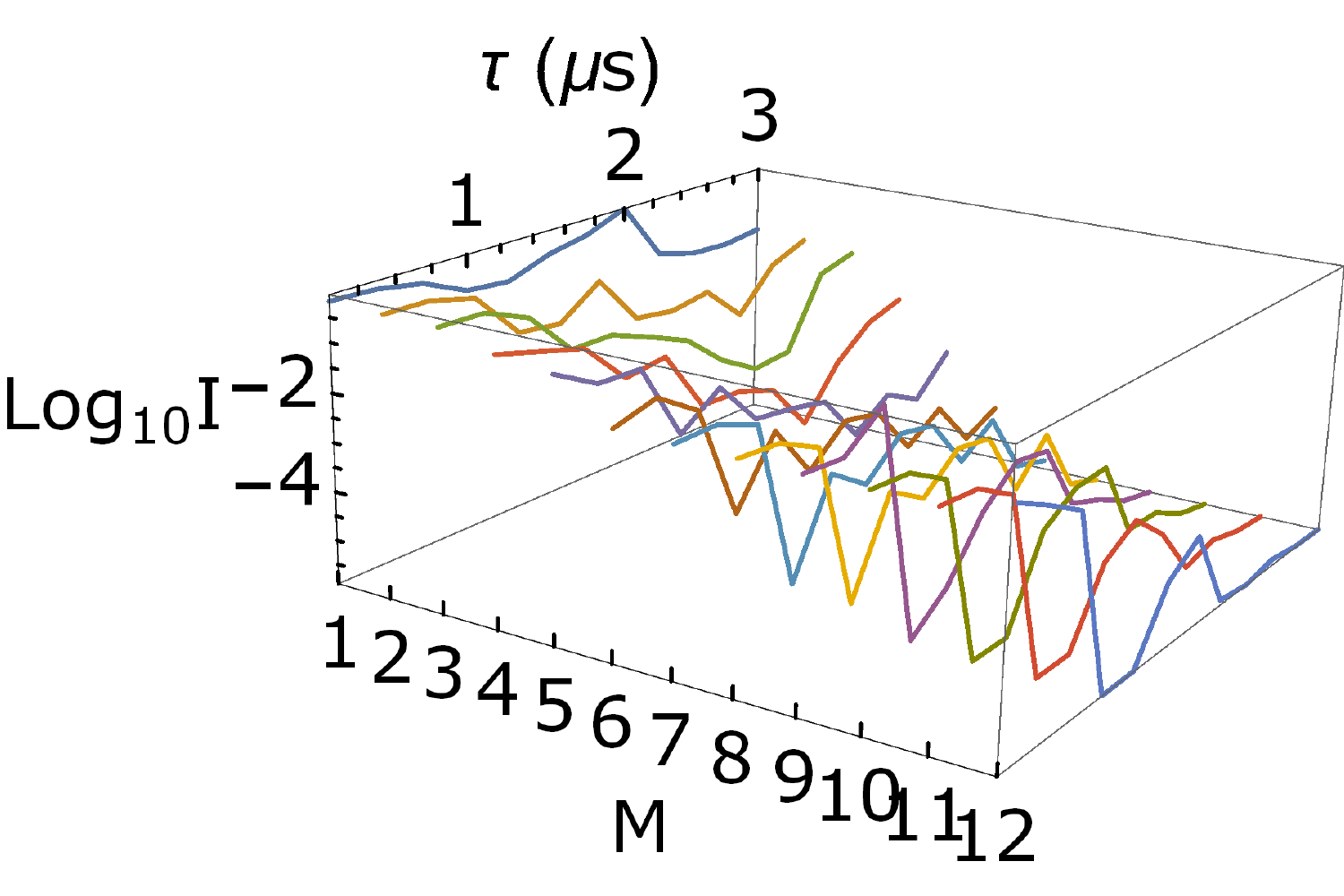}
	\caption[]{}
	\label{fig:CZlinewidth2}
	\end{subfigure}
\begin{subfigure}[b]{0.45\textwidth}
		\includegraphics[width=1\textwidth]{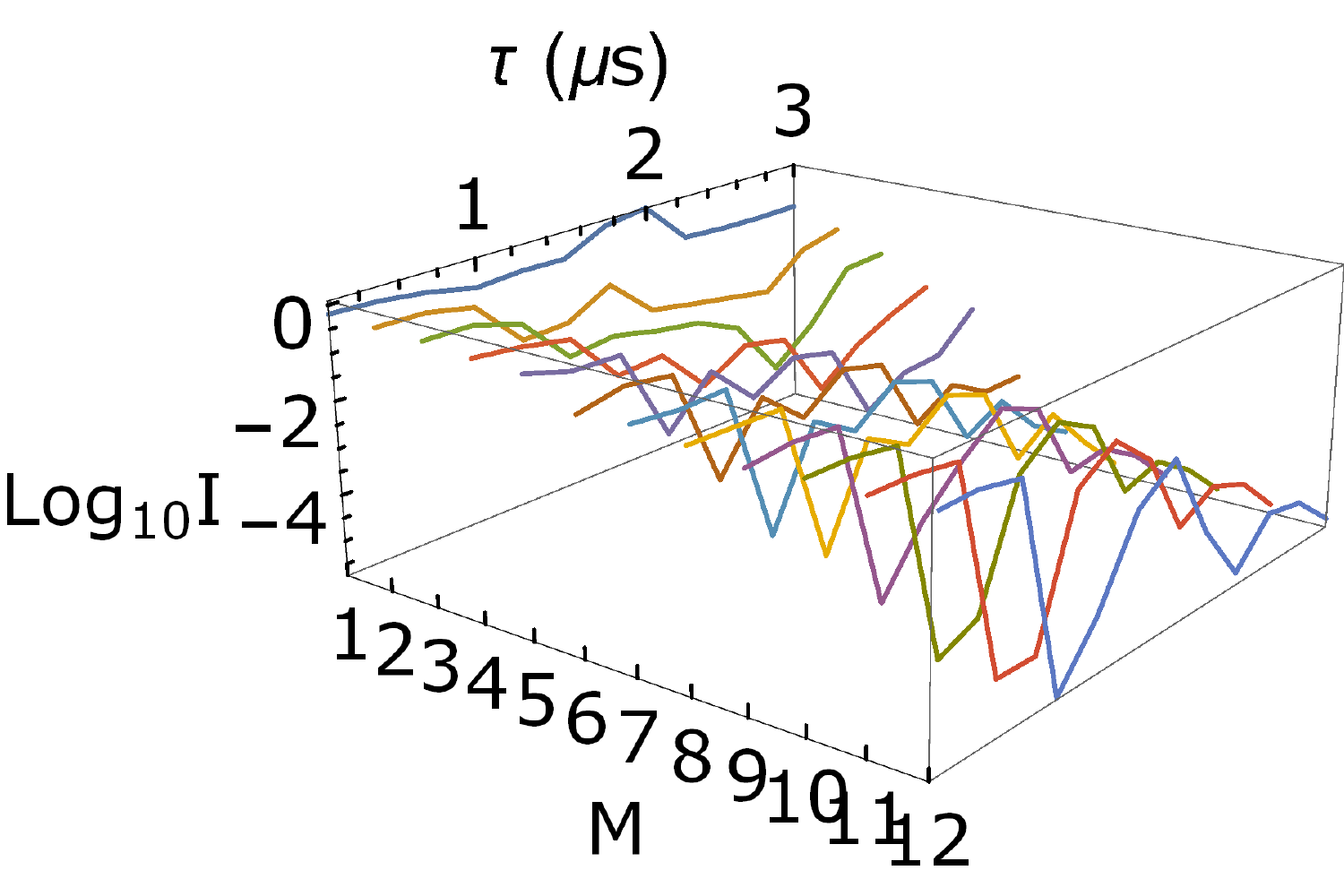}
	\caption[]{}
	\label{fig:CZlinewidth3}
	\end{subfigure}
\end{figure}
\begin{figure}[H]
\ContinuedFloat
\centering
	\begin{subfigure}[b]{0.45\textwidth}
		\includegraphics[width=1\textwidth]{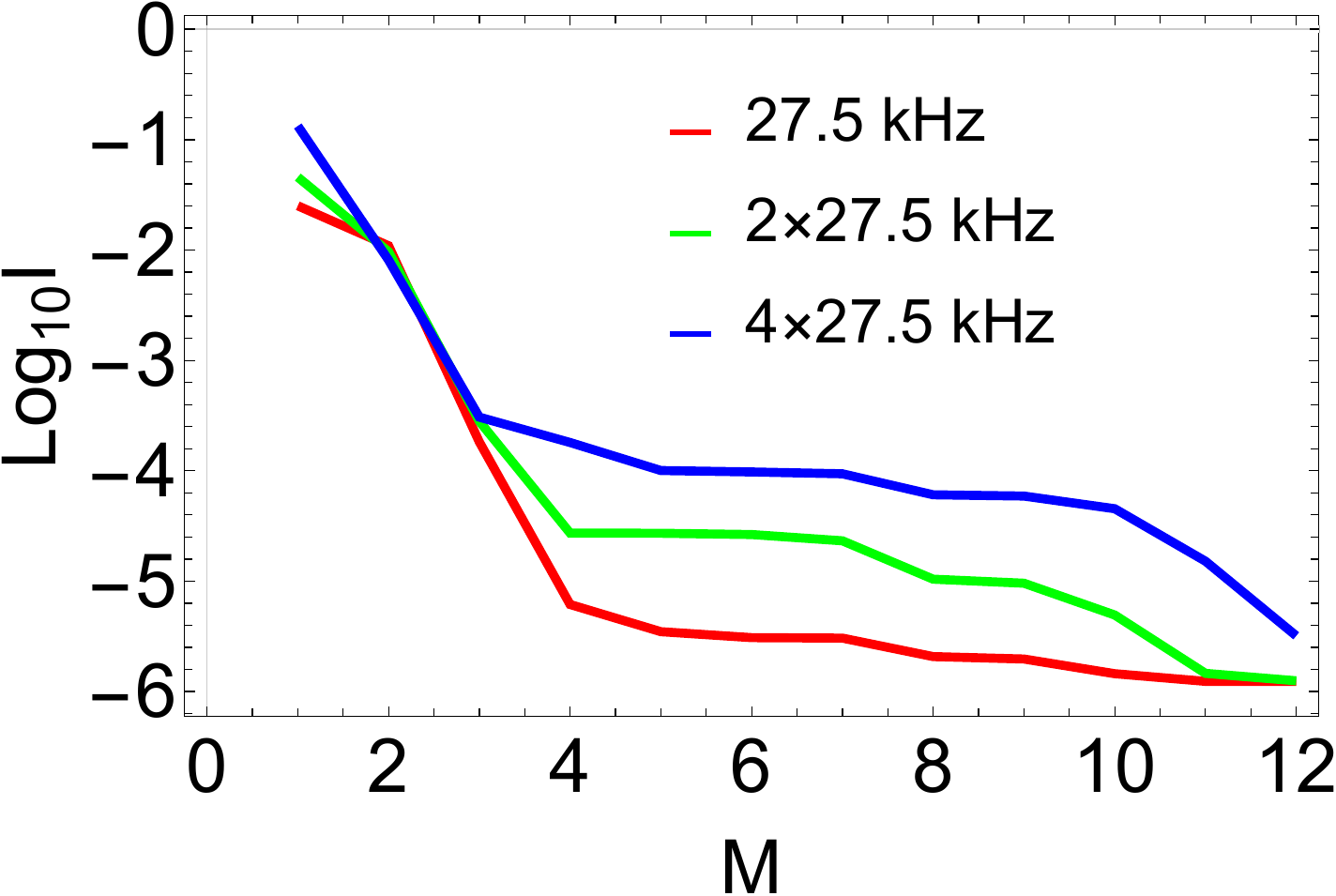}
	\caption[]{}
	\label{fig:CZconvergence}
	\end{subfigure}
	\caption{CZ gate infidelities for electron spin linewidth $(\sigma_\delta)$ of (a) $ 2\times 27.5$~kHz and (b) $4\times 27.5$~kHz. (c) shows the convergence of CZ gate infidelities for different electron spin linewidths with $\tau=2.25~\mu$s. Convergence of infidelities occurs at approximately 5 and 11 basis functions for $\sigma_\delta= 27.5$~kHz and $ 2 \times 27.5$~kHz, respectively. Convergence of infidelities was not observed for $\sigma_\delta=4\times 27.5$~kHz, implying that we would require more basis functions for convergence to occur. While increasing the linewidths would amount to increasing the gate infidelities, we are still able to obtain infidelities of less than $10^{-4}$, but at the cost of longer gate times and larger number of basis functions.}
	\label{fig:CZadditionallinewidth}
\end{figure}
\par We now turn our attention to the final step of the optimisation process, which is to incorporate time-ordering in the quantum evolution into our optimisation routine. As a proof of principle demonstration, we have performed this optimisation process on an X gate (see \ref{appendix:timeordering}). Through further optimisations where the quantum evolution is evaluated numerically using a finite-difference method, we found that an X gate can theoretically achieve infidelities of $10^{-5}$ or lower. We believe that with the addition of more advanced numerical optimisation techniques, then higher number of basis functions can be included as well as the treatment of two-qubit gates. Thus, we are confident that we can reduce the control errors to the order of $10^{-5}$ or better for all gates. However, it is not necessary to involve the numerical optimisation techniques if we are performing experiments on the diamond quantum processors as the same feedback-based optimisation routine will occur.
\par Now, we must assess the implications for neglecting the hyperfine field misalignments when attaining these infidelities. From \ref{appendix:misalignmenteffects}, the hyperfine field misalignment only affects the two-qubit gate. While we are able to minimise the gate infidelities due to control errors to the order of $10^{-5}$ or $10^{-6}$, the inclusion of the correction terms from the hyperfine field misalignment introduced an additional infidelity on the order of $10^{-3}$. This additional infidelity arises from the projection of the applied magnetic field due to the rotation of the nuclear spin coordinates in the computational subspace. One way to overcome this effective field is to include radiowave pulses which are resonant with the nuclear spins in the $m_s=0$ state in conjunction with the microwave pulses which drive the electron spin transitions. This dual application of radio and microwave pulses is beyond the scope of this work and should be pursued in future work. There is no physical reason why these errors due to hyperfine field misalignment cannot be reduced to the level of our control errors.

\section{Gate Performances in Presence of Decoherence}
\label{sec:gatewithdecoherence}
 \par As mentioned in \sref{sec:introduction}, the nuclear spin qubits undergo pure dephasing due to the relaxation of the electron spin. Thus, the coherence time of the qubits is bounded by the relaxation time of the electron spin. Here, we introduce the master equation which is also known as the Lindblad equation \cite{Gorini1976, Lindblad1976}
\begin{equation}
\fl\dot{\rho}(t)=\mathcal{L}(\rho) \equiv -\frac{i}{\hbar}\left[H,\rho(t)\right]+\sum_{m=1}^{N}\Bigg(L_m\rho(t)L_m^\dagger-\frac{1}{2}L_m^\dagger L_m\rho(t)-\frac{1}{2}\rho(t)L_m^\dagger L_m\Bigg)
\end{equation}
where the sum over $m$ is the summation of the decoherence mechanism over the individual nuclear spins. Instead of solving the differential equations, we can express the Lindblad equation as a vectorized density matrix \cite{Havel2003}
\begin{equation}
\dot{\rho}=\left(\mathbf{H}+\mathbf{G}\right)\rho
\end{equation}
where $\mathbf{G}$ is the decoherent part of the Lindblad equation with the form of
\begin{equation}
\mathbf{G}=\sum_{m=1}^N \Bigg(\overline{L}_m \otimes L_m -\frac{1}{2}\mathbbm{I}\otimes \left(L_m^\dagger L_m\right)-\frac{1}{2}\left(\overline{L}_m^\dagger \overline{L}_m\right)\otimes \mathbbm{I}\Bigg)
\end{equation}
and
\begin{equation}
\mathbf{H}=-\frac{i}{\hbar}\Bigg(\overline{H}\otimes \mathbbm{I}-\mathbbm{I}\otimes H\Bigg)
\end{equation}
The overline denotes the complex conjugate, $\dagger$ is the adjoint, $H$ is the Hamiltonian of the qubit system and $\mathbbm{I}$ is the $2^N\times 2^N$ identity matrix where $N$ is the number of qubits in the system. For a time dependent Hamiltonian, the Linblad equation is given by 
\begin{equation}
\rho(t)=\exp\left[\int\limits_{-\tau/2}^{\tau/2}\mathbf{H}~dt+\mathbf{G}t\right]\rho(0)
\label{eq:density}
\end{equation}
The Lindblad operator $L$ describing the dephasing of the nuclear spin qubits induced by a random electron spin flip can be written as
\begin{equation}
L_m=\sqrt{\frac{1}{2T_2}} \sigma_{z,m}
\end{equation}
where the nuclear spin $T_2$ is defined by the relaxation time $T_1$ of the electron spin (1.8~ms \cite{Balasubramanian2009}) and $\sigma_{z,m}$ is the Pauli matrix for the $z$ component of each nuclear spin.
 \par To assess the effects of decoherence due to dephasing, we consider an example with a perfect X gate. The aim now is to solve for $\rho(t)$ and calculate its state fidelity defined as
 \begin{equation}
 F=\frac{\Tr\Big[\rho_I^\dagger \rho\Big]}{\Tr\Big[\rho_I^\dagger \rho_I\Big]}
 \label{eq:statefid}
 \end{equation}
where $\rho_I$ is the ideal density matrix without the effects of decoherence and $\rho$ is the simulated density matrix of the system.

\begin{figure}[H]
	\centering
	\includegraphics[width=0.6\textwidth]{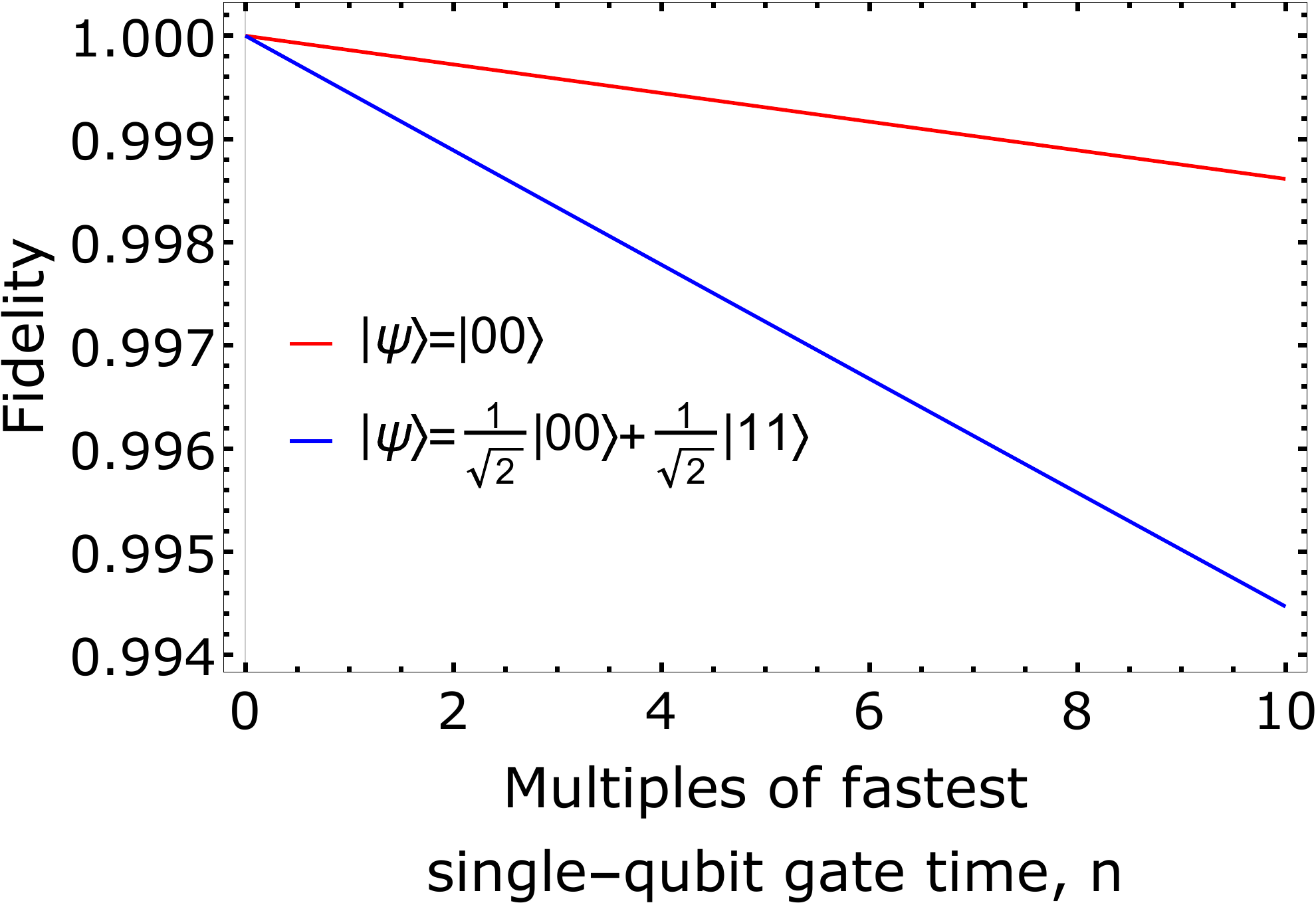}
	\caption[]{Effects of decoherence on two different different initial states by applying a perfect X gate. The fidelity is simulated with respective of multiples of the fastest single-qubit gate time of $1~\mu$s. The errors are found to be on the scale of $10^{-3}\sim10^{-4}$.}
	\label{fig:decoherenceexample}
\end{figure}
\par As shown in figure \ref{fig:decoherenceexample}, the errors in both cases have magnitudes of approximately $10^{-3}\sim 10^{-4}$. These errors are much larger than the errors caused by the effects of frequency, phase and amplitude noises as demonstrated in section \ref{sec:nonidealgates} and the decoherence errors increase with longer gate times. The decoherence errors also have the same order of magnitude $\left(\sim 10^{-3}\right)$ as the errors introduced by hyperfine field misalignment on CZ gates (see \ref{appendix:misalignmenteffects}). Since we cannot remove the electron-nuclear spin coupling as it is required for the selective operations, this leads to the conclusion that for a given electron relaxation time $T_{1,e}$, the only solution to improve the gate fidelity is to make the gate operations faster.

\section{Simulation of QFT On A Diamond Quantum Processor}
\label{sec:simulation}

In this section, we set a benchmark for the optimal performance of a diamond quantum processor by simulating quantum algorithms. The benchmark will provide us with insights into the limits of the processor, which can be used to aid the design and comparison of the device in the near future. 
\par Owing to our expectations that with modifications to the control system, errors due to the hyperfine field misalignment can be reduced to the level of control errors, which are small compared to the effects of decoherence. Hence, building on the results from \sref{sec:gatewithdecoherence} and \ref{appendix:misalignmenteffects}, we can simulate quantum algorithms on the diamond quantum computer by only considering the effects of decoherence. The key metrics for simulation will be the error probability and the total computational time (ignoring initial loading time of the computational control systems). These two metrics are chosen since computational time is the primary resource and the error rate is the key quality of a quantum computer. We have chosen to compute quantum Fourier transforms (QFT) for simulation as QFT is widely used in quantum algorithms \cite{NielsenChuang2010}. A further motivation is the similarity to the algorithms used for enhanced quantum sensing using a register of nuclear spin qubits. The initial state was chosen in a way such that an output state of $\left|001\right>$ (3 qubit QFT) and $\left|00001\right>$ (5 qubit QFT) will be the only outcome with a probability of 1.

\begin{figure}[h]
	\centering
	\includegraphics[width=1\textwidth]{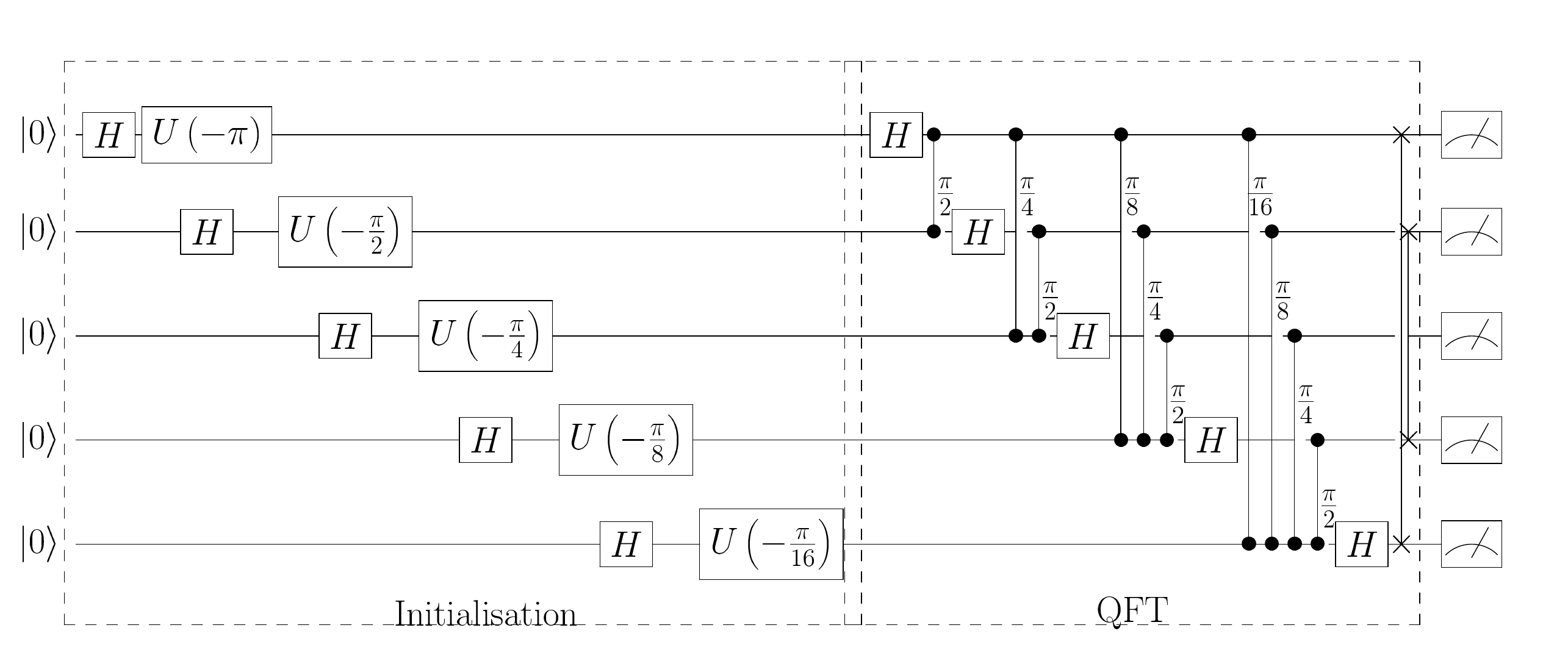}	
	\caption[]{The full circuit used for the simulations of 5 qubit quantum Fourier transform. $U\left(\theta\right)$ denotes the phase gate of $\theta$. The 2 dots joined with a line denotes a controlled-phase gate with their respective phases written in the circuit. We can write the controlled-phase gate in this notation as the matrix representation for this operation is the same regardless of which qubit is the control/target qubit, i.e ${\rm C}_1{\rm PHASE}_2={\rm C}_2{\rm PHASE}_1$. For a 3 qubit quantum Fourier transform, gate operations are performed on the first three qubits only (starting from the top).}
	\label{qftcircuit}
\end{figure}

\par The simulations of QFT on the diamond quantum computer are done using equation~\ref{eq:density} with electron relaxation time of $T_{1,e}\approx 1.8$ ms. These simulations will be iteratively solved for multiples of the fastest single and two-qubit gate times by only considering the effects of decoherence as the effects due to hyperfine field misalignments have the same order of magnitude and only affect CZ gates. The total number of pulses required for 3 qubits (QFT3) and 5 qubits (QFT5) quantum Fourier transforms after the decomposition into rotations about the $x$ and $y$ axis and CZ gate are 75 and 195 for QFT3 and QFT5 respectively (\ref{appendix:gatedecomposition}).

\par The total computation time on a diamond quantum computer can be broken down into shot time and initialisation/readout time. Shot time can be regarded as the total duration of the pulses required for an experiment. Assuming single-qubit gate times of 1~$\mu$s and CZ gate time of $1~\mu$s, the optimal pulse duration for QFT3 and QFT5 are approximately $75~\mu$s and $195~\mu$s, respectively. For the initialisation/readout time (single-shot readout), we are going to apply $M$ number of readout cycles per qubit. Thus, the total time for initialisation/readout is given by
\begin{equation}
{\rm Initialisation/Readout\ Time}=n\times M\times t_c
\end{equation}
where $n$ is the number of qubits, $M$ is the number of readout cycles applied per qubit and $t_c$ is the time per cycle. Time per cycle is based on the time of the optical pulse required for readout $(t_{\rm opt})$ and the time of the microwave pulse $(t_{\rm mw})$ required to perform the CNOT gate for repetitive measurements. These two time quantities are approximately 1~$\mu$s each. $M$ is chosen to be 500 as it has the same magnitude as other numbers of repetition which achieve an initialisation fidelity of 0.99 given a specific relative shift of the initialisation threshold \cite{Waldherr2014}. The time required for a single-shot readout is given by
\begin{eqnarray}
T_{\rm QFT3}&=75\times 10^{-6} + 3\times 500\times 2\times 10^{-6}\approx 0.0031~{\rm s}\\
T_{\rm QFT5}&=195\times 10^{-6} + 5\times 500\times 2\times 10^{-6}\approx 0.0052~{\rm s}
\end{eqnarray}

\begin{figure}[H]
	\centering
	\begin{subfigure}[b]{0.45\textwidth}
		\centering
		\includegraphics[width=1\textwidth]{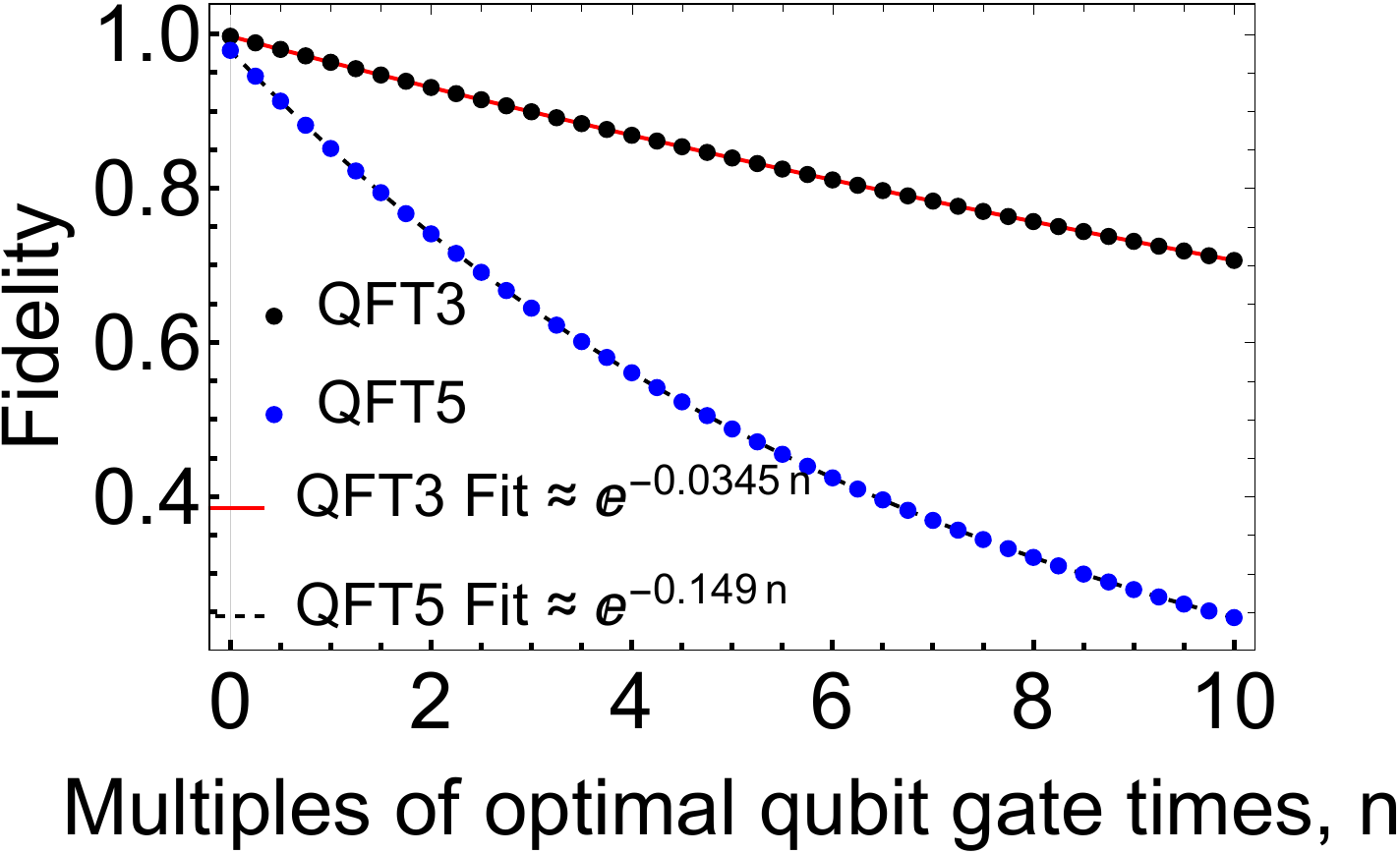}
		\caption[]{}
		\label{fig:nvfidelity}
	\end{subfigure}
	\begin{subfigure}[b]{0.45\textwidth}
		\centering
		\includegraphics[width=1\textwidth]{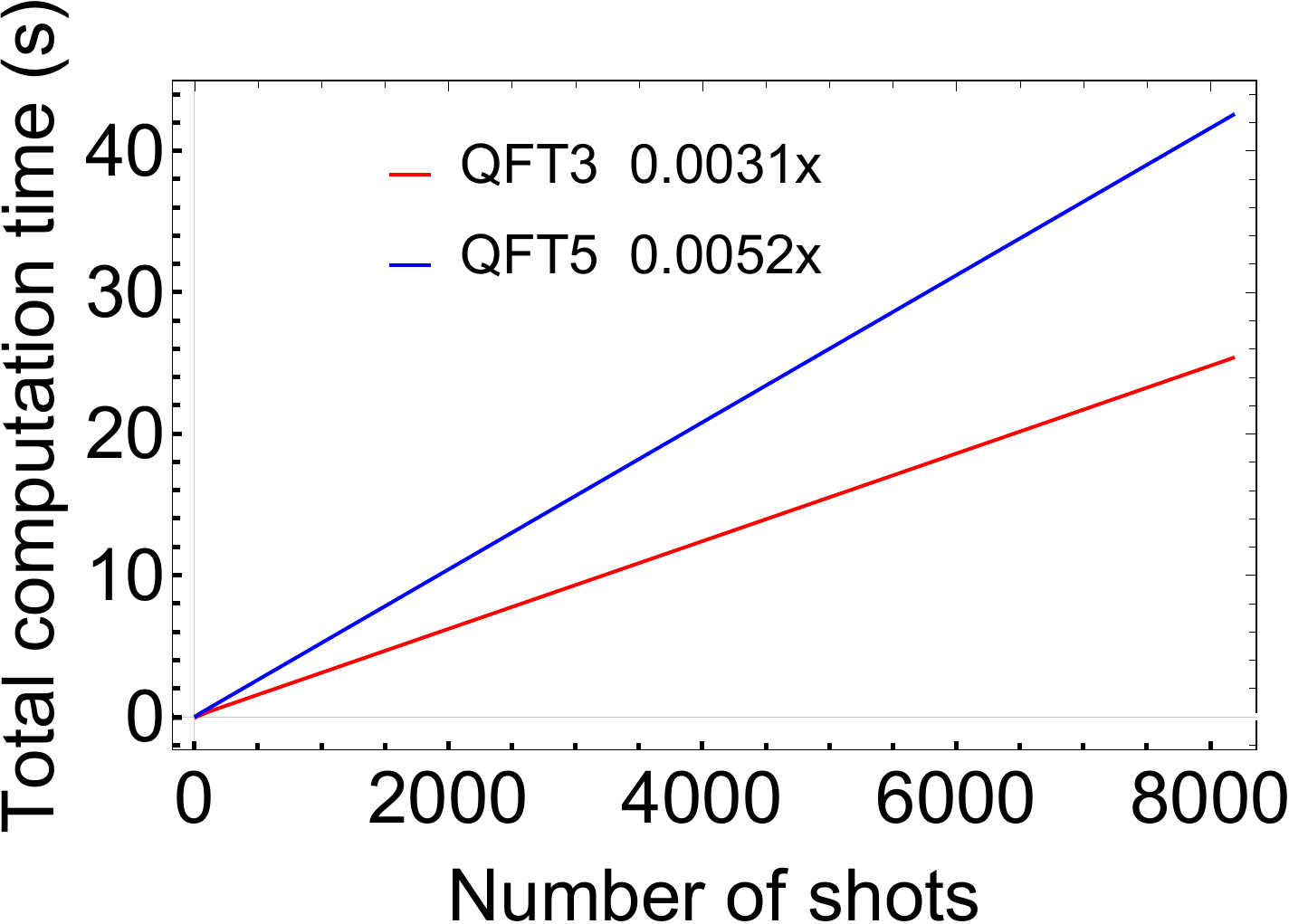}
		\caption[]{}
		\label{fig:nvtime}
	\end{subfigure}
	\caption[]{(a) Single-shot simulated fidelity of 3 qubits and 5 qubits quantum Fourier transform. We assumed perfect initialisation/readout fidelity. (b) Simulated total computation time of QFT3 and QFT5 assuming optimal gate time of 1~$\mu$s and $1~\mu$s for single-qubit gates and CZ gate, respectively. }
	\label{fig:nvperformance}
\end{figure}
As depicted in figure \ref{fig:nvfidelity}, simulation of QFT3 is able to achieve higher fidelity ($\approx 0.964$) than QFT5 ($\approx 0.855$) when they are simulated using the optimal gate time of $1~\mu$s for single-qubit gates and $1~\mu$s for CZ gate as QFT3 has smaller circuit size. Implementing a lesser number of gates will introduce less errors during the evolution of the quantum states, thus an algorithm with a smaller circuit size will achieve greater output state fidelity. The larger decay constant in the fitted model of QFT5 indicates that it is important to perform optimal control on the pulses to obtain the fastest gate possible with the lowest infidelity, as well as optimising our circuit size. Consequently, this will give us the best result when an experiment is performed using an actual diamond quantum computer.

\begin{figure}[H]
	\centering
	\includegraphics[width=0.6\textwidth]{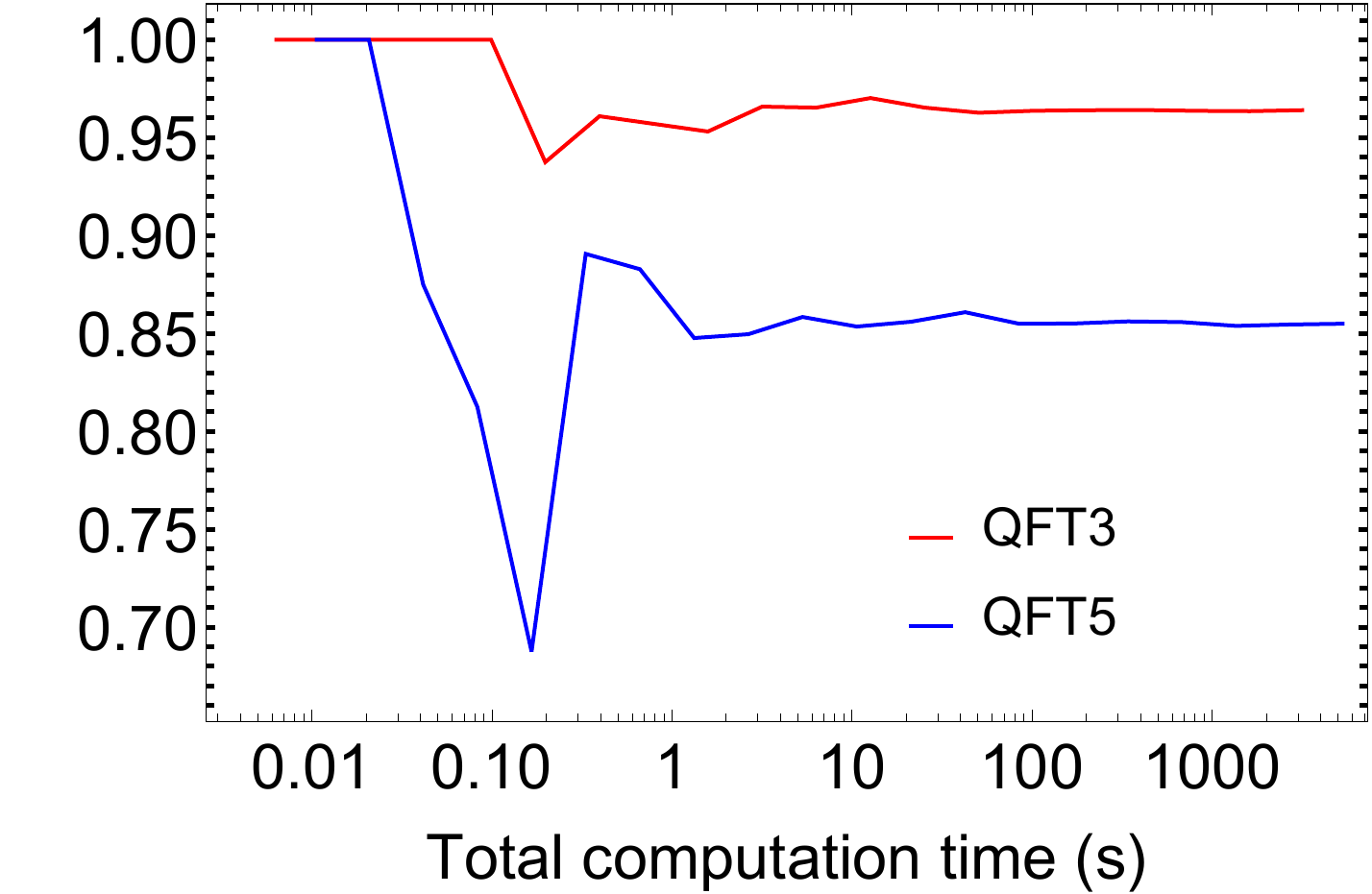}
	\caption[]{Total computation time and the fidelity of performing QFT3 and QFT5 on a simulated diamond quantum computer. The simulation on a diamond quantum computer assumed perfect initialisation/readout fidelity.}
	\label{ibmvsnv}
\end{figure}
\par To simulate the performance of a diamond quantum computer over time, we first create a binomial distribution of probabilities with the respective optimal values for QFT3 and QFT5 simulated on a diamond quantum computer. Then, we simulate the probability distribution for a range of shots using Markov chain Monte Carlo method. Markov chain Monte Carlo allows us to approximate the probability distribution of the fidelities through random sampling. With short computation time per shot for QFT3 and QFT5, the simulations of QFT3 and QFT5 on a diamond quantum computer are able to converge to their respective optimal values of 0.964 and 0.855 in less than 10~s. These results can be used as a benchmark for comparison with other quantum computing architectures in the future.

\section{Conclusion}
\par In summary, we have presented a complete model of a diamond quantum processor, the gate operations and their implementations. We have developed a semi-analytical optimal control method which theoretically produces fast, high fidelity gates with control errors that are less than those introduced by decoherence and hyperfine field misalignment. This optimal control method has three steps. First, we generate a complete semi-analytical basis of pulses that minimises cross-talk between qubits in absence of time-ordering in the quantum evolution. Second, we generate linear combination of these basis functions to minimise the control field errors. Third, we further optimise the pulses to account for the effects of time-ordering in the quantum evolution. We used frequency-shifted sinc functions as an ansatz for the control pulses. We have demonstrated the first two steps of this optimal control method on a Hadamard gate and a CZ gate with an additional third step on an X gate.  The simulated performance of a diamond quantum computer shows promising results where it can perform fast computations with low error probability. Our results will aid the design and the development of diamond quantum computers and enhanced quantum sensors. 
\par A future extension should include more advanced numerical optimisation techniques to optimise the pulses in the presence of time-ordering in the quantum evolution. However, if the optimisation is being performed experimentally, this step may not be required. Instead, we can implement a feedback control system as an optimal control method which tunes the pulse parameters based upon the output of a physical diamond quantum computer. Subsequent work should also include the introduction of a more complex optimal control technique to mitigate the errors of the hyperfine field misalignment through the application of simultaneous radiofrequency and microwave pulses.

\ack
We acknowledge support from the Australian Research Council (DP170103098 and DE170100169) as well as the Australian National University, University of Canberra and Charles Sturt University through the Discovery Translation Fund 2.0, managed by ANU Connect Ventures.

\appendix
\section{Time-Ordering In The Quantum Evolution}
\label{appendix:timeordering}
\par In section~\ref{sec:controlpulses}, we ignored the time-ordering in the quantum evolution to simplify the unitary operators in terms of rotation about the $x$-axis or $y$-axis. In this appendix, we examine the effects due to time-ordering in the quantum evolution on the gate fidelity.
\par  We consider the case where $b(t)$ of equation~\ref{eq:gammaB} is zero. The nuclear spin Hamiltonian parameterised with the control error parameters $\delta,\epsilon$ and $\phi$ as described in section~\ref{sec:nonidealgates} is given by
\begin{eqnarray}
 H_{\rm{single}}&=- \sum_i a(t)(1+\epsilon)\Bigg[I_{i,x} \cos\left(\omega_it\right)\cos\left(\left[\omega +\delta\right] t+\phi\right) \nonumber\\
&+ I_{i,y} \sin\left(\omega_i t\right)\cos\left(\left[\omega+\delta\right] t+\phi\right)\Bigg]
\label{eq:Htimeordering}
\end{eqnarray}
This Hamiltonian is the same as the Hamiltonian described in section~\ref{section:modelhamiltonian}. Using a finite-difference method, we approximate the Hamiltonian as being time-independent over the short period of $\Delta t$. The quantum evolution of the Hamiltonian can be written as
\begin{equation}
U_{\rm{fd}}(\tau, \delta,\epsilon,\phi)\approx \prod_{m=0}^{N-1} e^{-i H(t_{m})\Delta t}
\label{eq:Hquantumevolution}
\end{equation}
in which the product is a time-ordered product, $t_n=n\Delta t$ and $\Delta t=\tau/N$ where $N$ is the number of time steps and $\tau$ is the gate time.
\par We use the pulse, $a(t)$ generated from section~\ref{sec:nonidealgates}. The functional form of $a(t)$ can be obtained by performing an inverse Fourier transform on $a(\omega)$. It is given by
\begin{equation}
a(t)=\sum_{n=1} 2\sqrt{2\pi}c_n \cos\left(\frac{2\pi t\left[(n-1)\mu\right]}{\tau}\right)
\end{equation}
where $c_n$ is the overall amplitudes of the linear combinations for the $n^{th}$ basis functions as estimated by optimising over the control errors and $\mu=0.2$ is the shift of the basis functions. We further optimise $a(t)$ to account the effects of time-ordering by introducing a change $d_n$ to the coefficients where
\begin{equation}
a(t)=\sum_{n=1} 2\sqrt{2\pi}(c_n+d_n) \cos\left(\frac{2\pi t\left[(n-1)\mu\right]}{\tau}\right)
\end{equation}
The best parameters for $d_n$ are obtained by minimising the average infidelity as given by
\begin{equation}
\fl \left<I\right>=\int\limits_{-\infty}^{\infty} P(\delta;\sigma_\delta)\int\limits_{-\infty}^{\infty}P(\epsilon;\sigma_\epsilon)\int\limits_{-\infty}^{\infty}P(\phi;\sigma_\phi)\left[1-\frac{\rm{Tr}\left(U_{\rm{fd}}^\dagger\cdot U_{\rm{perfect}}\right)}{\rm{Tr}\left(U_{\rm{fd}}^\dagger\cdot U_{\rm{fd}}\right)}\right] d\delta d\epsilon d\phi
\label{eq:averageinf}
\end{equation}
where the probability distributions $P$ for all control parameters are of Gaussian.

\par Applying this method to an X gate of $1~\mu$s with 2 basis functions where $c_1\approx -1.20$ and $c_2\approx 0.79$, we found that the infidelities have converged for $N>4000$. For the further optimisation of $a(t)$, we performed a simple grid search to obtain the best parameters for $d_n$. The parameters of $d_1\approx-0.20$ and $d_2\approx-1.17$ resulted in an averaged infidelity of approximately $2.90\times 10^{-5}$, which is comparable to the infidelities obtained in section~\ref{sec:nonidealgates} and is still below the errors introduced by decoherence.

\begin{figure}[h]
	\centering
	\includegraphics[width=0.8\textwidth]{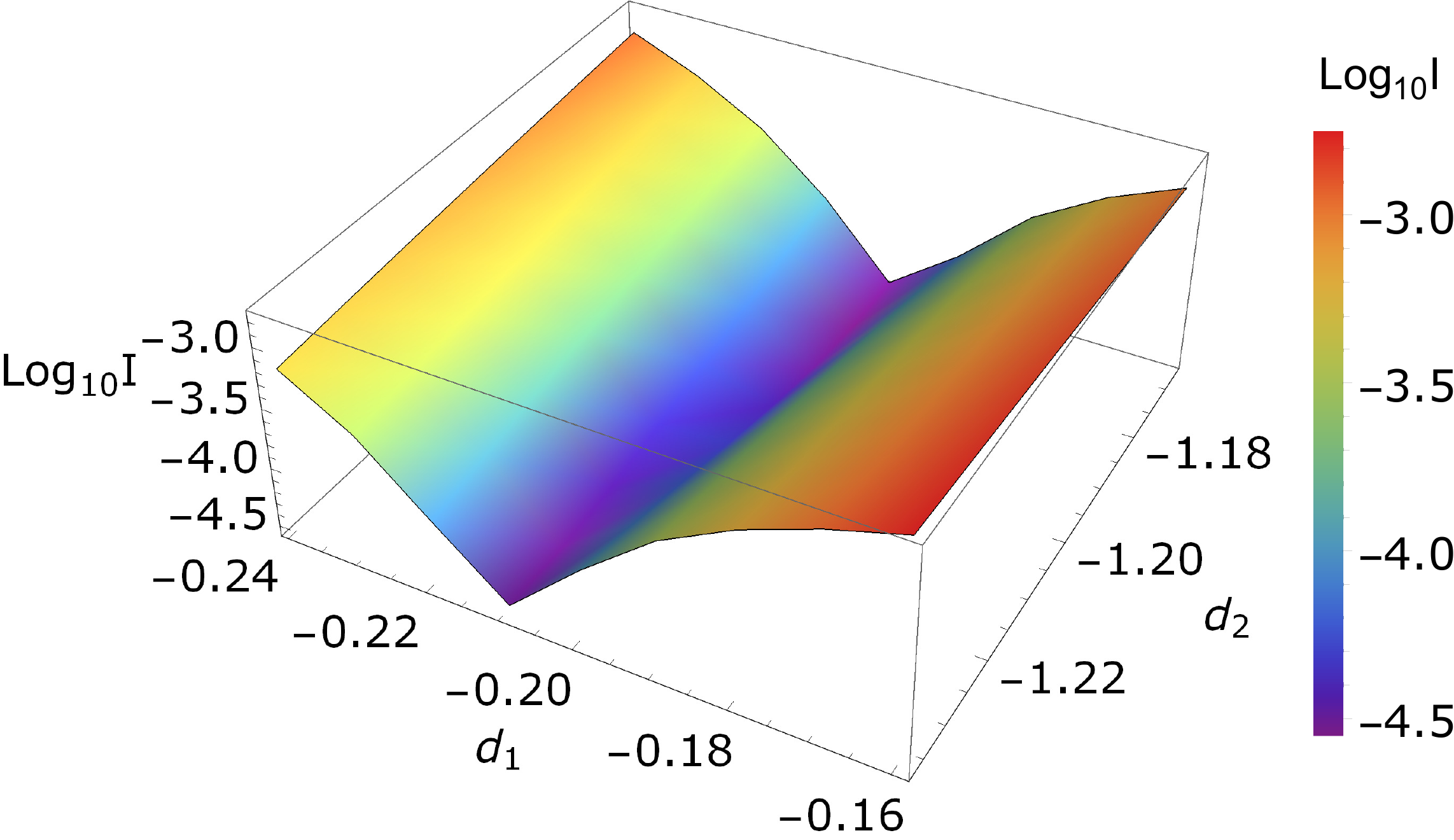}
	\caption[]{Optimisation of an X gate against the effects of time-ordering in the quantum evolution. We use a simple grid search to obtain the optimal parameters for $d_1$ and $d_2$ which are found by minimising equation~\ref{eq:averageinf}.}
	\label{fig:fd}
\end{figure}

\par Calculations for a two-qubit CZ gate are not demonstrated here due to sheer size of the search space considering it requires at least 6 basis functions to achieve an infidelity of approximately $10^{-6}$. It is computationally expensive to perform these calculations without implementing more advanced algorithms for closed-loop optimisations. These calculations will be considered in future extensions.
\par Since we are able to optimise an X gate with two linear combinations with infidelities below the decoherence limit, there is no reason why we couldn't achieve comparable infidelities for more linear combinations. Thus, while the time-ordering in the quantum evolution introduces additional infidelities to the gate operations, they can still be optimised to a level below the errors introduced by decoherence.

\section{Approximations Of Model Hamiltonian}
\label{appendix:modelapproximations}
\par In section~\ref{section:modelhamiltonian}, we applied the secular approximation and ignored the hyperfine field misalignment to simplify the nuclear spin Hamiltonian of the quantum processor. In this appendix, we examine the validity of these two approximations, identified their leading order corrections and their effects on gate fidelity.
\subsection{Secular Approximation}
\label{appendix:secularapproximation}

\par In this subsection, we examine the validity of the secular approximation. The full electron-nuclear Hamiltonian can be divided into two terms
\begin{equation}
H=H^{(a)}+H^{(b)}
\end{equation}
where
\begin{eqnarray}
 H^{(a)}&=DS_z^2+\gamma_eB_0S_z+S_z\left(A_{zz}I_z+A_{xz}I_x+A_{yz}I_y\right)-\gamma_nB_0I_z\nonumber\\
 H^{(b)}&=S_x\left(A_{xx}I_x+A_{yx}I_y+A_{zx}I_z\right)+S_y\left(A_{xy}I_x+A_{yy}I_y+A_{zy}I_z\right)
\end{eqnarray}
$H^{(a)}$ is the secular component that contains terms with $S_z$ whereas $H^{(b)}$ is the non-secular component, containing $S_x$ and $S_y$. For large magnetic fields aligned with the NV axis, the non-secular terms are small compared to the secular terms. In this Hamiltonian, the nuclear spin coordinates are co-aligned with the NV coordinates.
\par Treating the non-secular terms as a perturbation, we can apply perturbation theory to collapse the full Hamiltonian into effective nuclear spin Hamiltonians corresponding to the different electron spin projections. The nuclear spin Hamiltonian of the $m_s=n$ electron spin projection, correct to second order in the non-secular terms can be written as
\begin{equation}
H_n=\left<n\left|H^{(a)}\right|n\right>+\sum_{m\neq n} \frac{\left<n\left|H^{(b)}\right|m\right>\left<m\left|H^{(b)}\right|n\right>}{E_n-E_m}
\end{equation}
It follows that
\begin{eqnarray}
\fl H_{I,m_s=0}&=-\gamma_nB_0I_z-\frac{\gamma_eB_0}{D^2-\gamma_e^2B_0^2}\Bigg[\left(A_{yy}A_{zx}-A_{yx}A_{zy}\right)I_x +\left(A_{xx}A_{zy}-A_{xy}A_{zx}\right)I_y \nonumber \\
\fl &+\left(A_{xy}^2-A_{xx}A_{yy}\right)I_z\Bigg]\label{eq:secular1}
\end{eqnarray}
and
\begin{eqnarray}
\fl H_{I,m_s=-1}&=D-\gamma_eB_0-\left(A_{zz}I_z+A_{xz}I_x+A_{yz}I_y\right)-\gamma_nB_0I_z-\frac{1}{2\left(D-\gamma_eB_0\right)}\Bigg[\nonumber \\
\fl &\left(A_{yy}A_{zx}-A_{yx}A_{zy}\right)I_x +\left(A_{xx}A_{zy}-A_{xy}A_{zx}\right)I_y+\left(A_{xy}^2-A_{xx}A_{yy}\right)I_z\Bigg]\label{eq:secular2}
\end{eqnarray}
In equation~\ref{eq:secular2}, the scalar terms $D-\gamma_eB_0$ do not affect the nuclear spin so they can be neglected from the nuclear spin Hamiltonian. However, they do perturb the electron spin transition frequencies. The terms in the square brackets of equations~\ref{eq:secular1} and \ref{eq:secular2} are the lowest order corrections to the nuclear spin Hamiltonians. For $m_s=-1$, the above can be rewritten in terms of $\vec{A}'\cdot \vec{I}$ where $\vec{A}'$ is the corrected hyperfine field. Thus, the correction terms add a quantitative difference to the values of $\vec{A}_{i,z}$, but not a functional difference that affects the validity of our model. However, the corrections do represent a change for the $m_s=0$ level as there is no equivalent terms used in the model. They are equivalent to an effective magnetic field misaligned with the NV axis that introduces error to our model. The error introduced by the neglect of these terms will be examined in the next subsection alongside with the analysis of hyperfine field misalignments.

\subsection{Effects of Hyperfine Field Misalignment}
\label{appendix:misalignmenteffects}
In this subsection, we examine the effects due to the hyperfine field misalignment. After applying the secular approximation, the nuclear spin Hamiltonian (equation.~\ref{eq:Hamiltonian1}) in the auxiliary and computational subspace are 
\begin{eqnarray}
H_{I,m_s=0}&=-\sum_i  \gamma_i\Big[I_{i,z}B_0+I_{i,x}B_1(t)\Big]\\
H_{I,m_s=-1}&=-\sum_i  \vec{A}_{i,z}\cdot\vec{I}_i-\gamma_i\Big[I_{i,z}B_0+I_{i,x}B_1(t)\Big]
\end{eqnarray}
To diagonalise the nuclear spin Hamiltonian of the computational subspace (i.e for $m_s=-1$), we performed a coordinate transform of the nuclear spin operators. We did this by a rotation defined by the angles
\begin{eqnarray}
\tan \phi_i&=\frac{A_{i,xz}}{A_{i,yz}}\\
\tan \theta_i&=\frac{\left(A_{i,xz}^2+A_{i,yz}^2\right)^{1/2}}{A_{i,zz}+\gamma_iB_0}
\end{eqnarray}
The transformed Hamiltonians for the auxiliary $m_s=0$ and computational $m_s=-1$ subspaces are then
\begin{eqnarray}
\fl H_{I,m_s=0}&=-\sum_i\gamma_iB_0\left(I'_{i,z}\cos\theta_i-I'_{i,y}\sin\theta_i\right)-\sum_i\gamma_iB_1(t)\Bigg[\Big(I'_{i,x}\cos\phi_i\nonumber\\
\fl&+I'_{i,y}\sin\phi_i\cos\theta_i+I'_{i,z}\sin\theta_i\sin\phi_i\Big) \Bigg]
\label{eq:auxiliaryrotated}\\
\fl H_{I,m_s=-1}&=-\sum_i\omega_iI'_{i,z}-\sum_i\gamma_iB_1(t)\Bigg[\Big(I'_{i,x}\cos\phi_i+I'_{i,y}\sin\phi_i\cos\theta_i\nonumber\\
\fl &+I'_{i,z}\sin\theta_i\sin\phi_i\Big) \Bigg]\label{eq:computationalrotated}
\end{eqnarray}
where $\omega_i=\left(\left[A_{i,zz}+\gamma_iB_0\right]^2+A_{i,xz}^2+A_{i,yz}^2\right)^{1/2}$ is the transition frequency of the $i^{th}$ nucleus in the $m_s=-1$ subspace. We assume that only nuclei with hyperfine fields nearly aligned to the NV axis are chosen. We then perform small angle approximations on $\theta_i$ and simplify the expressions by only including the terms which are zero-order in $\theta_i$. Finally, undoing rotations about the $z$-axis gives us the zero-order nuclear spin Hamiltonians 
\begin{eqnarray}
H_{I,m_s=0}^{(0)}\approx- \sum_i\gamma_iB_0 I''_{i,z}-\sum_i I''_{i,x}\gamma_iB_1(t)\label{eq:auxiliary0order}\\
H_{I,m_s=-1}^{(0)}\approx-\sum_i\omega_i I''_{i,z}-\sum_i I''_{i,x}\gamma_iB_1(t)\label{eq:computational0order}
\end{eqnarray}
In the main text, we have dropped the $''$ signs for the spin operators. The first-order corrections to equation~\ref{eq:auxiliary0order} and \ref{eq:computational0order} are obtained by considering the terms that are first-order in $\theta_i$ in equation~\ref{eq:auxiliaryrotated} and \ref{eq:computationalrotated}. They are given by
\begin{eqnarray}
H_{I,m_s=0}^{(1)}&=\sum_i\gamma_iB_0\theta_i\left(I'_{i,y}\right)\nonumber\\
&=\sum_i \gamma_iB_0\theta_i \left(I''_{i,x}\sin\phi_i+I''_{i,y}\cos\phi_i\right)\\
H_{I,m_s=-1}^{(1)}&=0
\end{eqnarray}
In both $m_s=0$ and $-1$ states, we have ignored the correction term due to the oscillating component of $\gamma_iB_1(t)I'_{i,z}\sin\theta_i\sin\phi_i$. While the correction term is of first order, within rotating wave approximation, the introduced errors have effects of order greater than one. Therefore, we find that the hyperfine field misalignments only influence the two-qubit gates and the first-order correction for the auxiliary subspace is effectively a static magnetic field in the $x$ and $y$ direction. 
\par Including this leading order correction and the leading order correction from the secular approximation, the new two-qubit gate Hamiltonian is
\begin{eqnarray}
\fl H'&=\frac{\Delta}{2}\sigma_z +\frac{\Omega}{2}\sigma_x B_1(t)+\sum_i \left(\alpha_i + \beta_i\sigma_z\right)I_{i,z}+\sum_i\Bigg[\left(\gamma_iB_0\theta_i\sin\phi_i-\chi_i+\xi_i \theta_i\sin\phi_i\right)I_{i,x}\nonumber\\
\fl&+\left(\gamma_iB_0\theta_i\cos\phi_i-\eta_i+\xi_i\theta_i\cos\phi_i\right)I_{i,y}-\left(\chi_i\theta_i\sin\phi_i+\eta_i\theta_i\cos\phi_i+\xi_i\right)I_{i,z}\Bigg]\label{eq:fulltwoqubitHamiltonian}
\end{eqnarray}
where
\begin{eqnarray}
\chi_i&=\frac{\gamma_eB_0}{D^2-\gamma_e^2B_0^2}\left(A_{i,yy}A_{i,zx}-A_{i,yx}A_{i,zy}\right)\\
\eta_i&=\frac{\gamma_eB_0}{D^2-\gamma_e^2B_0^2}\left(A_{i,xx}A_{i,zy}-A_{i,xy}A_{i,zx}\right)\\
\xi_i&=\frac{\gamma_eB_0}{D^2-\gamma_e^2B_0^2}\left(A_{i,xy}^2-A_{i,xx}A_{i,yy}\right)\
\end{eqnarray}
\par  We can estimate the effect of these corrections by considering the rotations that they generate over time when the auxiliary subspace is populated during the implementation of a CZ gate. Taking into account the evolution of electron spin and nuclear spins explicitly, we adopt a simple model as follows to estimate the infidelity introduced by these correction terms. For an ideal unitary operator $U_T$, we consider a perfect $2\pi$ rotation on the electron spin, conditional on the nuclear spin. The unitary operator with the presence of correction terms as described equation~\ref{eq:fulltwoqubitHamiltonian} are given by $U_A$. None of these corrections are applied to the $^{15}$N nuclear spin as it is perfectly aligned with the NV quantisation axis. Thus, the additional infidelity due to these first-order corrections is given by
\begin{equation}
{\rm Infidelity}=1-\frac{\Tr \left[U_T^\dagger U_A\right]}{\Tr\left[U_T^\dagger U_T\right]}
\end{equation} 
where
\begin{eqnarray}
\fl U_T&=\exp\Bigg[-\frac{\mathbbm{i}}{2}\Big[2\pi \left(\sigma_{e,x} \sigma_{n}\right)\Big]\Bigg]\\
\fl U_A&=\exp\Bigg[-\frac{\mathbbm{i}}{2}\Big[2\pi \left(\sigma_{e,x} \sigma_{n}\right)+\tau\left(\gamma_n B_0\theta\sin\phi-\chi+\xi\theta\sin\phi\right)\left(\sigma_n\sigma_x\right)\nonumber\\
\fl&+\tau\left(\gamma_n B_0\theta\cos\phi-\eta+\xi\theta\cos\phi\right)\left(\sigma_n\sigma_y\right)-\tau\left(\chi\theta\sin\phi+\eta\theta\cos\phi+\xi\right)\left(\sigma_n\sigma_z\right)\nonumber\\
\fl\sigma_n&= \left(\begin{array}{cc}
0&0\\
0&1
\end{array}\right)
\end{eqnarray}
$\sigma_{x,y,z}$ are the Pauli matrices, $\sigma_e$ and $\sigma_n$ are the matrices operating on the electron spin and $^{13}$C nuclear spin, respectively. 
\par For a distant nuclei, we can approximate the hyperfine terms to be $A_{i,xx}=A_{i,yy}=A_{i,zz}$ and they are the leading order terms. We can also approximate $A_{i,zx}=A_{i,zy}=A_{i,nd}$ where $A_{i,nd}^2=A_{i,zx}^2+A_{i,zy}^2$. The $^{13}$C that we have chosen with $A_C\approx 0.413~$MHz closely represent a $^{13}$C lattice site in the S family with their hyperfine parameters given by $A_{zz}= 0.412$~MHz and $A_{nd}= 0.060$~MHz \cite{Nizovtsev2014}. The calculated hyperfine field misalignment using $B_0=0.62$~T is approximately 0.0085~rad. We also surveyed a few lattice sites which has similar hyperfine parameters obtained via \textit{ab initio} method from \cite{Nizovtsev2014}. Approximating $\phi=\pi/2$~rad and $\tau=1~\mu$s to be the gate duration when the auxiliary subspace is populated during the implementation of a CZ gate, the additional infidelities due to hyperfine field misalignments are on the order of $10^{-2}\sim10^{-3}$. These additional infidelities are dominated by the hyperfine field misalignments over the secular approximation. The correction terms from the secular approximation $\chi,\eta$ and $\xi$ scale with a factor of approximately $10^{-5}$~rad/s, compared to the correction terms from the hyperfine field misalignment $\gamma_n B_0$, which has a magnitude of $10^7$~rad/s.

\section{Control Pulse For Two-Qubit Gate Operations}
\label{appendix:twoqubtgates}
Given that the time evolution operator has the form of
\begin{equation}
U=\exp\left(-i\int\limits_{-\tau/2}^{\tau/2} H_{\rm{multi}} dt\right)
\end{equation}
Assuming there are no mixed signals in a single pulse, the $Y$ components will go to zero. The expression for the evolution operator can be further simplified where
\begin{eqnarray}
 \fl U=\exp\left(-i\int\limits_{-\tau/2}^{\tau/2} H_{11}dt\right)\otimes\left|11\right>\left<11\right|+\exp\left(-i\int\limits_{-\tau/2}^{\tau/2} H_{10}dt\right)\otimes\left|10\right>\left<10\right| \nonumber \\
 \fl +\exp\left(-i\int\limits_{-\tau/2}^{\tau/2} H_{01}dt\right)\otimes\left|01\right>\left<01\right|+\exp\left(-i\int\limits_{-\tau/2}^{\tau/2} H_{00}dt\right)\otimes\left|00\right>\left<00\right|\nonumber \\
\end{eqnarray}
where
\begin{eqnarray}
H_{11}&=\frac{\Omega}{2}B_1(t)\Bigg[\sigma_x\cos\left(\left[\Delta+\beta_1+\beta_2\right]t\right)\Bigg]\\
H_{10}&=\frac{\Omega}{2}B_1(t)\Bigg[\sigma_x\cos\left(\left[\Delta+\beta_1-\beta_2\right]t\right)\Bigg]\\
H_{01}&=\frac{\Omega}{2}B_1(t)\Bigg[\sigma_x\cos\left(\left[\Delta-\beta_1+\beta_2\right]t\right)\Bigg]\\
H_{00}&=\frac{\Omega}{2}B_1(t)\Bigg[\sigma_x\cos\left(\left[\Delta-\beta_1-\beta_2\right]t\right)\Bigg]
\end{eqnarray}
To get a selective $2\pi$ pulse on the $\left|11\right>$ state, we need to enforce
\begin{equation}
X_1=\Omega\int\limits_{-\tau/2}^{\tau/2} B_1(t)\cos\Big[\left(\Delta+\beta_1+\beta_2\right)t\Big]dt=2\pi
\end{equation}
and other 3 integrals to be zero. As per the single-qubit gate case, we make the substitution
\begin{equation}
\Omega B_1(t)=a(t)\cos\left(\lambda t\right)+ b(t)\sin\left(\lambda t\right)
\end{equation}
where $\lambda=\Delta+\beta_1+\beta_2$. Thus, we have

\begin{eqnarray}
X_1&=\frac{\sqrt{2\pi}}{2}\Bigg[a\left(\omega_{11}+\lambda\right)+a\left(\omega_{11}-\lambda\right)\Bigg]=2\pi
\label{eq:twoqubitx1} \\
X_2&=\frac{\sqrt{2\pi}}{2}\Bigg[a\left(\omega_{10}+\lambda\right)+a\left(\omega_{10}-\lambda\right)\Bigg]=0 \label{eq:twoqubitx2}\\
X_3&=\frac{\sqrt{2\pi}}{2}\Bigg[a\left(\omega_{01}+\lambda\right)+a\left(\omega_{01}-\lambda\right)\Bigg]=0  \label{eq:twoqubitx3}\\
X_4&=\frac{\sqrt{2\pi}}{2}\Bigg[a\left(\omega_{00}+\lambda\right)+a\left(\omega_{00}-\lambda\right)\Bigg]=0  \label{eq:twoqubitx4}
\end{eqnarray}

\section{Infidelity of Two-Qubit Gate Operations}
In accordance with the derivation of infidelity of single-qubit gates, the instrinsic infidelity expression that we used to describe two-qubit gate operations is given by
\begin{equation}
\fl I=1-\frac{1}{8}\left[ -2\cos\left(\frac{\sqrt{X_1^2}}{2}\right)+2\cos\left(\frac{\sqrt{X_2^2}}{2}\right)+2\cos\left(\frac{\sqrt{X_3^2}}{2}\right)+2\cos\left(\frac{\sqrt{X_4^2}}{2}\right)\right] \label{eq:intrinsicinftwo}
\end{equation}
 Keeping terms up to the second order, the final expression for the infidelity of the CZ gate is
\begin{equation}
I=\frac{1}{32}\left(\delta X_1^2+X_2^2+X_3^2+X_4^2\right)
\end{equation}
where $\delta X_1=X_1-2\pi$. The average infidelity is given by
\begin{eqnarray}
\fl \left<I\right>&=\frac{1}{32}\int\limits_{-\infty}^{\infty}p\left(\delta;\sigma_\delta\right)\int\limits_{-\infty}^{\infty} p\left(\epsilon;\sigma_\epsilon\right)\delta X_1^2\ d\epsilon d\delta+\frac{1}{32}\int\limits_{-\infty}^{\infty}p\left(\delta;\sigma_\delta\right)\int\limits_{-\infty}^{\infty} p\left(\epsilon;\sigma_\epsilon\right) X_2^2\ d\epsilon d\delta \nonumber\\
\fl &+\frac{1}{32}\int\limits_{-\infty}^{\infty}p\left(\delta;\sigma_\delta\right)\int\limits_{-\infty}^{\infty} p\left(\epsilon;\sigma_\epsilon\right) X_3^2\ d\epsilon d\delta+\frac{1}{32}\int\limits_{-\infty}^{\infty}p\left(\delta;\sigma_\delta\right)\int\limits_{-\infty}^{\infty} p\left(\epsilon;\sigma_\epsilon\right) X_4^2\ d\epsilon d\delta
\end{eqnarray}

\section{Pulse Amplitudes For Various Single-Qubit Gates and CZ Gate}
\label{appendix:amplitudevalues}

\begin{table}[H]
\caption{Tables of the pulse amplitudes in the frequency domain for (a) an X gate, (b) a $\pi/4$ rotation about the x-axis, (c) a $7\pi/4$ rotation about the x-axis and (d) a $\pi/2$ rotation about the y-axis targeted at the $^{13}$C nuclear spin qubits as a function of gate time $\tau$. These amplitudes corresponds to the solutions for the $n\in \left[0,5\right]$ sinc basis functions which minimises equation \ref{eq:intrinsicinfone}. The pulse amplitudes are dependent on the angle of rotations and gate time, which is consistent with the formulation of our sinc basis functions. Larger amplitudes are expected to generate bigger angle of rotations for a given frequency shift and gate time. Better solutions with lower infidelity can also be resolved for smaller angle of rotations at shorter gate times as the pulse amplitudes are smaller.}
 \label{tab:1qdiffrotation}%
\begin{subtable}[t]{0.8\textwidth}
\caption{$\pi$ rotation about the x-axis.}
\label{tab:diffrotationa}%
 \begin{indented}
\lineup
\item[]\begin{tabular}{@{}*{7}{l}}
\br
 & \centre{6}{$f^{(n)}$ (Mrad/s)}\\
 \ns
 $\tau$&\crule{6}\\
 ($\mu$s)&$f^{(0)}$&$f^{(1)}$&$f^{(2)}$&$f^{(3)}$&$f^{(4)}$&$f^{(5)}$\\
    \mr
    0.25  & -5.000    & 4.839 & 3.937 & -5.000    & -5.000    & 5.000 \\
    0.50  & -2.643 & 5.000     & -5.000    & -5.000   & -3.664 & 5.000 \\
    0.75  & -1.737 & 5.000     & -5.000    & 5.000     & -5.000    & -3.991 \\
    1.00     & -1.332 & 5.000     & -5.000    & 5.000     & 5.000     & 5.000 \\
    1.25  & -1.088 & 5.000     & -5.000    & 5.000     & -5.000    & 5.000 \\
    1.50   & -4.491 & 5.000     & -5.000    & 5.000     & -5.000    & 5.000 \\
    1.75  & -3.791 & 4.318 & -5.000    & 5.000     & -5.000    & 5.000 \\
    2.00    & -0.666 & 3.893 & -5.000    & 5.000     & -5.000    & 5.000 \\
    2.25  & -0.600 & 3.711 & -5.000    & 5.000     & -5.000    & 5.000 \\
    2.50  & -2.692 & 3.233 & -5.000    & 5.000     & -5.000    & 5.000 \\
    2.75  & -0.485 & 2.776 & -4.730 & 5.000     & -5.000    & 5.000 \\
    3.00     & 1.333 & 2.593 & -4.616 & 5.000     & -5.000    & 5.000 \\
    \br
    \end{tabular}%
  \end{indented}
 \end{subtable}
\end{table}
\begin{table}[H]
\ContinuedFloat
\begin{subtable}[t]{0.8\textwidth}
\caption{$\pi/4$ rotation about the x-axis}
\label{tab:diffrotationb}%
 \begin{indented}
\lineup
\item[]\begin{tabular}{@{}*{7}{l}}
\br
 & \centre{6}{$f^{(n)}$ (Mrad/s)}\\
 \ns
 $\tau$&\crule{6}\\
 ($\mu$s)&$f^{(0)}$&$f^{(1)}$&$f^{(2)}$&$f^{(3)}$&$f^{(4)}$&$f^{(5)}$\\
    \mr
    0.25  & -1.369 & 0.291   & 0.104     & -2.575    & -4.332    & 5.000 \\
    0.50   & -0.661 & 2.167    & -1.102     & 0.063 & -0.277 & 2.204 \\
    0.75  & -0.434 & 2.714    & -3.264     & 3.895 & -4.693 & 0.037 \\
    1.00     & 4.997 & 1.956 & -3.517 & 4.857 & -5.000 & 2.518 \\
    1.25  & -4.622 & 1.753 & -3.420  & 5.000 & -5.000     & 4.801 \\
    1.50   & -0.225 & 1.382 & -2.591 & 3.842    & -5.000     & 5.000 \\
    1.75  & -0.190 & 1.080 & -1.871 & 2.559    & -3.139     & 3.595 \\
    2.00    & -0.167    & 0.973 & -1.740   & 2.475 & -3.184 & 3.873 \\
    2.25  & -4.946  & 0.928 & -1.739 & 2.540 & -3.241 & 3.703 \\
    2.50   & -2.288 & 0.809 & -1.423 & 1.893 & -2.156     & 2.205 \\
    2.75  & 1.817 & 0.694 & -1.186 & 1.578 & -1.867     & 2.060 \\
    3.00     & 5.000 & 0.648 & -1.154 & 1.630 & -2.081     & 2.510 \\
    \br
    \end{tabular}%
    \end{indented}
  \end{subtable}
\begin{subtable}[t]{0.8\textwidth}
\vspace*{0.8 cm}
\caption{$7\pi/4$ rotation about the x-axis}
\label{tab:diffrotationc}%
 \begin{indented}
\lineup
\item[]\begin{tabular}{@{}*{7}{l}}
\br
 & \centre{6}{$f^{(n)}$ (Mrad/s)}\\
 \ns
 $\tau$&\crule{6}\\
 ($\mu$s)&$f^{(0)}$&$f^{(1)}$&$f^{(2)}$&$f^{(3)}$&$f^{(4)}$&$f^{(5)}$\\
    \mr
    0.25  & -5.000    & 5.000 & 5.000 & -5.000    & -5.000    & 5.000 \\
    0.50  & -4.622 & 5.000     & -5.000    & -5.000   & -5.000 & 5.000 \\
    0.75  & 3.908 & 5.000     & -5.000    & 5.000     & -5.000    & -3.680 \\
    1.00     & 2.998 & 5.000     & -5.000    & 5.000     & -5.000     & 5.000 \\
    1.25  & 2.447 & 5.000     & -5.000    & 5.000     & -5.000    & 5.000 \\
    1.50   & -5.000 & 5.000     & -5.000    & 5.000     & -5.000    & 5.000 \\
    1.75  & 1.706 & 5.000 & -5.000    & 5.000     & -5.000    & 5.000 \\
    2.00    & -1.166 & 5.000 & -5.000    & 5.000     & -5.000    & 5.000 \\
    2.25  & -1.049 & 5.000 & -5.000    & 5.000     & -5.000    & 5.000 \\
    2.50  & 1.211 & 5.000 & -5.000    & 5.000     & -5.000    & 5.000 \\
    2.75  & -2.786 & 4.854 & -5.000 & 5.000     & -5.000    & 5.000 \\
    3.00     & 0.100 & 4.537 & -5.000 & 5.000     & -5.000    & 5.000 \\
    \br
    \end{tabular}%
   \end{indented}
  \end{subtable}
\end{table}
\begin{table}[H]
\ContinuedFloat
\begin{subtable}[t]{0.8\textwidth}
\caption{$\pi/2$ rotation about the y-axis}
\label{tab:diffrotationd}%
\begin{indented}
\lineup
\item[]\begin{tabular}{@{}*{7}{l}}
\br
 & \centre{6}{$g^{(n)}$ (Mrad/s)}\\
 \ns
 $\tau$&\crule{6}\\
 ($\mu$s)&$g^{(0)}$&$g^{(1)}$&$g^{(2)}$&$g^{(3)}$&$g^{(4)}$&$g^{(5)}$\\
    \mr
    0.25  & -2.475 & 5.000    & -5.000     & 5.000    & 5.000    & -5.000 \\
    0.50  & -1.333 & 5.000    & -5.000     & 2.298    & -0.7861    & 1.399 \\
    0.75  & -0.915 & 5.000    & -5.000     & 2.622    & -0.3857    & 0.527 \\
    1.00  & -0.671 & 3.460    & -4.263     & 3.536    & -2.249     & 0.990 \\
    1.25  & -0.527 & 2.756    & -4.190     & 4.924 	  & -5.000     & 4.697 \\
    1.50  & -3.997 &2.595 	  & -4.661     & 5.000    & -5.000     & 5.000 \\
    1.75  & 2.704  & 2.381    & -4.226     & 5.000    & -5.000     & 4.461 \\
    2.00  & -0.336 & 1.989 	  & -3.350     & 4.159	  & -4.373 	  & 4.134 \\
    2.25  & -5.000 & 1.690    & -2.907     & 3.925 	  & -4.743 	  & 5.000 \\
    2.50  & -4.531 &-1.559    & -2.791 	  & 3.970 	  & -5.000     &5.000 \\
    2.75  & 1.714  & 1.506    & -2.832 	  & 4.224 	  & -5.000     & 5.000 \\
    3.00  & 1.571  & 1.381    & -2.596 	  & 3.873 	  & -5.000     & 5.000 \\
    \br
    \end{tabular}%
  \end{indented}
  \end{subtable}
\end{table}%

\begin{table}[H]
\caption{\label{tab:diffrotationCZ}Table of the pulse amplitudes in the frequency domain for a CZ gate. We observed that at shorter gate times, higher order basis functions have solutions which correspond to the minimum amplitude of $f^{(n)}=0.500$. As the difference between the transition frequencies of some of the qubit states are relatively small compared to the frequency shifts, the sinc basis functions are unable to resolve possible solutions for these transition frequencies with lower infidelity within the search region for $f^{(n)}\in \left[0.5,15\right]$. This results in the amplitudes for the higher order basis functions to be $f^{(n)}=0.500$, which at the same time also minimises equation \ref{eq:intrinsicinftwo}.}
\begin{indented}
\lineup
\item[]\begin{tabular}{@{}*{7}{l}}
\br
 & \centre{6}{$f^{(n)}$ (Mrad/s)}\\
 \ns
 $\tau$&\crule{6}\\
 ($\mu$s)&$f^{(0)}$&$f^{(1)}$&$f^{(2)}$&$f^{(3)}$&$f^{(4)}$&$f^{(5)}$\\
 \mr
0.25  & 0.500    &  0.500 &  0.500 &  0.500    &  0.500    &  0.500 	\\
0.50  & 2.494 &  0.500    &  0.500    &  0.500   &  0.500 &  0.500 		\\
0.75  & 9.595 &  0.500    &  0.500    &  0.500     &  0.500   &  0.500 \\
1.00  & 7.669 & 12.460    &  0.500    & 10.420     &  0.500   &  0.500  \\
1.25  & 6.498 & 5.023     &  0.500    &  0.500     &  0.500  &  0.500 	\\
1.50  & 1.489 & 13.170    &  0.500   &  0.500     &  15.000    &  0.500\\
1.75  & 7.722 & 9.092     &  0.500    &  0.500     &  0.500    &  0.500 \\	
2.00  & 12.010 & 6.481    &  0.500    &  0.500     &  0.500    &  0.500 \\
2.25  & 1.173 & 4.894     &  0.500    &  0.500     &  0.500    &  0.500 \\
2.50  & 1.069 & 3.960     &  0.500    &  0.500     &  0.500    &  0.500\\
2.75  & 0.968 & 6.488     & 5.023 & 3.349     &  0.500    &  0.500  \\
3.00  & 0.877 & 5.893     & 8.822 & 11.480    & 13.170    & 11.900 \\
\br
\br
 & \centre{6}{$f^{(n)}$ (Mrad/s)}\\
 \ns
 $\tau$&\crule{6}\\
 ($\mu$s)&$f^{(6)}$&$f^{(7)}$&$f^{(8)}$&$f^{(9)}$&$f^{(10)}$&$f^{(11)}$\\
 \mr
0.25  & 0.500    &  0.500 	&  0.500 &  0.500    &  0.500    &  0.500 	\\
0.50  & 0.500 	&  0.500    &  0.500    &  0.500   &  0.500 &  0.500 		\\
0.75  & 0.500 	&  0.500    &  15.000    &  15.000     &  15.000   &  15.000 \\
1.00  & 0.500 	& 0.500  	&  0.500    & 0.500     &  0.500   &  0.500  \\
1.25  & 0.500 	& 0.500     &  0.500    &  0.500     &  0.500  &  0.500 	\\
1.50  & 0.500 	& 0.500     &  0.500   &  0.500     &  0.500    &  0.500\\
1.75  & 6.161 	& 0.500     &  0.500    &  0.500     &  0.500    &  0.500 \\	
2.00  & 0.500 	& 0.500     &  0.500    &  0.500     &  0.500    &  0.500 \\
2.25  & 0.500 	& 0.500     &  0.500    &  0.500     &  0.500    &  0.500 \\
2.50  & 0.500 	& 0.500     &  0.500    &  0.500     &  0.500    &  0.500\\
2.75  & 0.500 	& 0.500     & 0.500 	& 3.792     &  0.500    &  0.500  \\
3.00  & 0.500 	& 0.500     & 0.500 	& 3.527    & 0.500    & 0.500 \\
\br
\end{tabular}%
\end{indented}
\end{table}

\section{Gate Decompositions on a Diamond Quantum Computer}
\label{appendix:gatedecomposition}
 In general, there are 3 types of gate operations in quantum Fourier transform \cite{NielsenChuang2010}. They are Hadamard gate, swap gate, and controlled-phase gate. Ignoring the global phase factors, we have
\begin{equation}
H=R_x\left(\pi\right)\cdot R_y\left(\frac{\pi}{2}\right)
\end{equation}
A swap gate can be constructred from 3 CNOT gates. They can be written as
\begin{equation}
{\rm Swap}_{a,b}={\rm C}_a {\rm NOT}_b\cdot {\rm C}_b {\rm NOT}_a\cdot{\rm C}_a {\rm NOT}_b
\end{equation}
where ${\rm C}_a{\rm NOT}_b$ is
\begin{equation}
{\rm C}_a{\rm NOT}_b=H_b\cdot {\rm C}_a{\rm Z}_b\cdot H_b
\end{equation}
Here, $a$ and $b$ denotes any two qubits in the system. Phase gate can be written as
\begin{equation}
 {\rm PHASE}\left(\theta\right)=R_{y}\left(-\frac{\pi}{2}\right)\cdot R_{x}\left(\frac{\theta}{2}\right)\cdot R_{y}\left(\frac{\pi}{2}\right)
\end{equation}
 For a controlled-phase gate, one way to construct them is as follows
\begin{eqnarray}
\fl {\rm C}_a{\rm PHASE}_b\left(\theta\right)&=R_{y,a}\left(-\frac{\pi}{2}\right)\cdot R_{x,a}\left(\frac{\theta}{2}\right)\cdot R_{y,a}\left(\frac{\pi}{2}\right) \nonumber \\
\fl &\cdot R_{x,b}\left(-\frac{\pi}{2}\right)\cdot R_{y,b}\left(\frac{-\theta-\pi}{4}\right)\cdot {\rm C}_a{\rm Z}_b\cdot R_{y,b}\left(\frac{\theta+\pi}{4}\right)\cdot R_{x,b}\left(\frac{\pi}{2}\right)\nonumber \\
\fl &\cdot R_{y,b}\left(-\frac{\pi}{2}\right)\cdot R_{x,b}\left(\frac{-\theta-\pi}{4}\right)\cdot {\rm C}_a{\rm Z}_b\cdot R_{x,b}\left(\frac{\theta+\pi}{4}\right)\cdot R_{y,b}\left(\frac{\pi}{2}\right)
\end{eqnarray}
where $\theta$ is the intended phase. Note that the operations for these gate decompositions are performed from the right to the left. Using these decompositions, the total number of pulses required for QFT3 and QFT5 is 75 and 195 respectively.

%\section*{References}
%\begin{thebibliography}{10}
%\bibitem{ref1} J.~Doe, Article name, \textit{Phys. Rev. Lett.}
%
%\bibitem{ref2} J.~Doe, J. Smith, Other article name, \textit{Phys. Rev. Lett.}
%
%\bibitem{web} \href{http://www.google.pl}{www.google.pl}
%\end{thebibliography}
% \section*{References}
% \bibliography{D:/ANU/Research/Bibliographies/mainbib}

\begin{thebibliography}{10}

\bibitem{Neumann2010a}
Neumann P, Kolesov R, Naydenov B, Beck J, Rempp F, Steiner M, Jacques V,
  Balasubramanian G, Markham M~L, Twitchen D~J, Pezzagna S, Meijer J, Twamley
  J, Jelezko F and Wrachtrup J 2010 {\em Nature Physics\/} {\bf 6} 249--253

\bibitem{Dolde2014}
Dolde F, Bergholm V, Wang Y, Jakobi I, Naydenov B, Pezzagna S, Meijer J,
  Jelezko F, Neumann P, Schulte-Herbr{\"{u}}ggen T, Biamonte J and Wrachtrup J
  2014 {\em Nature Communications\/} {\bf 5} 3371

\bibitem{Taminiau2014}
Taminiau T~H, Cramer J, van~der Sar T, Dobrovitski V~V and Hanson R 2014 {\em
  Nature Nanotechnology\/} {\bf 9} 171--176

\bibitem{Waldherr2014}
Waldherr G, Wang Y, Zaiser S, Jamali M, Schulte-Herbr{\"{u}}ggen T, Abe H,
  Ohshima T, Isoya J, Du J~F, Neumann P and Wrachtrup J 2014 {\em Nature\/}
  {\bf 506} 204--207

\bibitem{Wang2015}
Wang Y, Dolde F, Biamonte J, Babbush R, Bergholm V, Yang S, Jakobi I, Neumann
  P, Aspuru-Guzik A, Whitfield J~D and Wrachtrup J 2015 {\em ACS Nano\/} {\bf
  9} 7769--7774

\bibitem{Kong2016}
Kong F, Ju C, Liu Y, Lei C, Wang M, Kong X, Wang P, Huang P, Li Z, Shi F, Jiang
  L and Du J 2016 {\em Physical Review Letters\/} {\bf 117} 60503

\bibitem{Bradley2019}
Bradley C~E, Randall J, Abobeih M~H, Berrevoets R~C, Degen M~J, Bakker M~A,
  Markham M, Twitchen D~J and Taminiau T~H 2019 {\em Physical Review X\/} {\bf
  9}(3) 031045

\bibitem{Hou2019}
Hou P~Y, He L, Wang F, Huang X~Z, Zhang W~G, Ouyang X~L, Wang X, Lian W~Q,
  Chang X~Y and Duan L~M 2019 {\em Chinese Physics Letters\/} {\bf 36} 100303

\bibitem{Neumann2013}
Neumann P, Jakobi I, Dolde F, Burk C, Reuter R, Waldherr G, Honert J, Wolf T,
  Brunner A, Shim J~H, Suter D, Sumiya H, Isoya J and Wrachtrup J 2013 {\em
  Nano Letters\/} {\bf 13} 2738--2742

\bibitem{Dolde2014a}
Dolde F, Doherty M~W, Michl J, Jakobi I, Naydenov B, Pezzagna S, Meijer J,
  Neumann P, Jelezko F, Manson N~B and Wrachtrup J 2014 {\em Physical Review
  Letters\/} {\bf 112} 97603

\bibitem{Zaiser2016}
Zaiser S, Rendler T, Jakobi I, Wolf T, Lee S~Y, Wagner S, Bergholm V,
  Schulte-Herbr{\"{u}}ggen T, Neumann P and Wrachtrup J 2016 {\em Nature
  Communications\/} {\bf 7} 12279

\bibitem{Aslam2017}
Aslam N, Pfender M, Neumann P, Reuter R, Zappe A, de~Oliveira F~F, Denisenko A,
  Sumiya H, Onoda S, Isoya J and Wrachtrup J 2017 {\em Science\/} {\bf 357}
  67--71

\bibitem{Unden2016}
Unden T, Balasubramanian P, Louzon D, Vinkler Y, Plenio M~B, Markham M,
  Twitchen D, Stacey A, Lovchinsky I, Sushkov A~O, Lukin M~D, Retzker A,
  Naydenov B, McGuinness L~P and Jelezko F 2016 {\em Physical Review Letters\/}
  {\bf 116} 230502

\bibitem{Haeberle2017}
H{\"{a}}berle T, Oeckinghaus T, Schmid-Lorch D, Pfender M, de~Oliveira F~F,
  Momenzadeh S~A, Finkler A and Wrachtrup J 2017 {\em Review of Scientific
  Instruments\/} {\bf 88} 13702

\bibitem{Pfender2017}
Pfender M, Aslam N, Sumiya H, Onoda S, Neumann P, Isoya J, Meriles C~A and
  Wrachtrup J 2017 {\em Nature Communications\/} {\bf 8} 834

\bibitem{Cramer2016}
Cramer J, Kalb N, Rol M~A, Hensen B, Blok M~S, Markham M, Twitchen D~J, Hanson
  R and Taminiau T~H 2016 {\em Nature Communications\/} {\bf 7} 11526

\bibitem{Shi2010}
Shi F, Rong X, Xu N, Wang Y, Wu J, Chong B, Peng X, Kniepert J, Schoenfeld R~S,
  Harneit W, Feng M and Du J 2010 {\em Physical Review Letters\/} {\bf 105}
  40504

\bibitem{Xu2017}
Xu K, Xie T, Li Z, Xu X, Wang M, Ye X, Kong F, Geng J, Duan C, Shi F and Du J
  2017 {\em Physical Review Letters\/} {\bf 118} 130504

\bibitem{Ermakova2013}
Ermakova A, Pramanik G, Cai J~M, Algara-Siller G, Kaiser U, Weil T, Tzeng Y~K,
  Chang H~C, McGuinness L~P, Plenio M~B, Naydenov B and Jelezko F 2013 {\em
  Nano Letters\/} {\bf 13} 3305--3309

\bibitem{Doherty2017}
Doherty M~W 2017 {\em Australian Physics\/} {\bf 54} 131--137

\bibitem{Bernien2013}
Bernien H, Hensen B, Pfaff W, Koolstra G, Blok M~S, Robledo L, Taminiau T~H,
  Markham M, Twitchen D~J, Childress L and Hanson R 2013 {\em Nature\/} {\bf
  497} 86--90

\bibitem{Hensen2015}
Hensen B, Bernien H, Dréau A~E, Reiserer A, Kalb N, Blok M~S, Ruitenberg J,
  Vermeulen R~F~L, Schouten R~N, Abellán C, Amaya W, Pruneri V, Mitchell M~W,
  Markham M, Twitchen D~J, Elkouss D, Wehner S, Taminiau T~H and Hanson R 2015
  {\em Nature\/} {\bf 526} 682--686

\bibitem{Kalb2017}
Kalb N, Reiserer A~A, Humphreys P~C, Bakermans J~J~W, Kamerling S~J, Nickerson
  N~H, Benjamin S~C, Twitchen D~J, Markham M and Hanson R 2017 {\em Science\/}
  {\bf 356} 928--932

\bibitem{Dolde2013}
Dolde F, Jakobi I, Naydenov B, Zhao N, Pezzagna S, Trautmann C, Meijer J,
  Neumann P, Jelezko F and Wrachtrup J 2013 {\em Nature Physics\/} {\bf 9}
  139--143

\bibitem{Yao2012}
Yao N~Y, Jiang L, Gorshkov A~V, Maurer P~C, Giedke G, Cirac J~I and Lukin M~D
  2012 {\em Nature Communications\/} {\bf 3} 800

\bibitem{Oberg2019}
Oberg L~M, Huang E, Reddy P~M, Alkauskas A, Greentree A~D, Cole J~H, Manson
  N~B, Meriles C~A and Doherty M~W 2019 {\em Nanophotonics\/} {\bf 8}
  1975--1984

\bibitem{Viola1999}
Viola L, Lloyd S and Knill E 1999 {\em Physical Review Letters\/} {\bf 83}
  4888--4891

\bibitem{Khodjasteh2005}
Khodjasteh K and Lidar D~A 2005 {\em Physical Review Letter\/} {\bf 95} 180501

\bibitem{Lange2010}
de~Lange G, Wang Z~H, Riste D, Dobrovitski V~V and Hanson R 2010 {\em
  Science\/} {\bf 330} 60--63

\bibitem{Wang2012}
Wang Z~H, de~Lange G, Rist{\`{e}} D, Hanson R and Dobrovitski V~V 2012 {\em
  Physical Review B\/} {\bf 85} 155204

\bibitem{Zhao2012}
Zhao N, Ho S~W and Liu R~B 2012 {\em Physical Review B\/} {\bf 85} 115303

\bibitem{Sar2012}
van~der Sar T, Wang Z~H, Blok M~S, Bernien H, Taminiau T~H, Toyli D~M, Lidar
  D~A, Awschalom D~D, Hanson R and Dobrovitski V~V 2012 {\em Nature\/} {\bf
  484} 82--86

\bibitem{Naydenov2011}
Naydenov B, Dolde F, Hall L~T, Shin C, Fedder H, Hollenberg L~C~L, Jelezko F
  and Wrachtrup J 2011 {\em Phys. Rev. B\/} {\bf 83}(8) 081201

\bibitem{Pham2012}
Pham L~M, Bar-Gill N, Belthangady C, Le~Sage D, Cappellaro P, Lukin M~D, Yacoby
  A and Walsworth R~L 2012 {\em Phys. Rev. B\/} {\bf 86}(4) 045214

\bibitem{Taminiau2012}
Taminiau T~H, Wagenaar J~J~T, van~der Sar T, Jelezko F, Dobrovitski V~V and
  Hanson R 2012 {\em Phys. Rev. Lett.\/} {\bf 109}(13) 137602

\bibitem{Liu2013}
Liu G~Q, Po H~C, Du J, Liu R~B and Pan X~Y 2013 {\em Nature Communications\/}
  {\bf 4} 2254 ISSN 2041-1723

\bibitem{Zhang2014}
Zhang J, Souza A~M, Brandao F~D and Suter D 2014 {\em Physical Review
  Letters\/} {\bf 112} 50502

\bibitem{Abobeih2018}
Abobeih M~H, Cramer J, Bakker M~A, Kalb N, Markham M, Twitchen D~J and Taminiau
  T~H 2018 {\em Nature Communications\/} {\bf 9} 2552 ISSN 2041-1723

\bibitem{Doria2011}
Doria P, Calarco T and Montangero S 2011 {\em Physical Review Letters\/} {\bf
  106} 190501

\bibitem{Caneva2011}
Caneva T, Calarco T and Montangero S 2011 {\em Physical Review A\/} {\bf 84}
  22326

\bibitem{Khaneja2005}
Khaneja N, Reiss T, Kehlet C, Schulte-Herbr{\"{u}}ggen T and Glaser S~J 2005
  {\em Journal of Magnetic Resonance\/} {\bf 172} 296--305

\bibitem{Scheuer2014}
Scheuer J, Kong X, Said R~S, Chen J, Kurz A, Marseglia L, Du J, Hemmer P~R,
  Montangero S, Calarco T, Naydenov B and Jelezko F 2014 {\em New Journal of
  Physics\/} {\bf 16} 93022

\bibitem{Rong2015}
Rong X, Geng J, Shi F, Liu Y, Xu K, Ma W, Kong F, Jiang Z, Wu Y and Du J 2015
  {\em Nature Communications\/} {\bf 6} 8748

\bibitem{Fowler2012}
Fowler A~G, Mariantoni M, Martinis J~M and Cleland A~N 2012 {\em Physical
  Review A\/} {\bf 86} 32324

\bibitem{Campbell2017}
Campbell E~T, Terhal B~M and Vuillot C 2017 {\em Nature\/} {\bf 549} 172--179

\bibitem{Wu2018}
Wu R~B, Chu B, Owens D~H and Rabitz H 2018 {\em Physical Review A\/} {\bf 97}
  42122

\bibitem{Maurer2012}
Maurer P~C, Kucsko G, Latta C, Jiang L, Yao N~Y, Bennett S~D, Pastawski F,
  Hunger D, Chisholm N, Markham M, Twitchen D~J, Cirac J~I and Lukin M~D 2012
  {\em Science\/} {\bf 336} 1283--1286

\bibitem{Shim2013}
Shim J~H, Niemeyer I, Zhang J and Suter D 2013 {\em Physical Review A\/} {\bf
  87} 12301

\bibitem{Herbschleb2019}
Herbschleb E~D, Kato H, Maruyama Y, Danjo T, Makino T, Yamasaki S, Ohki I,
  Hayashi K, Morishita H, Fujiwara M and Mizuochi N 2019 {\em Nature
  Communications\/} {\bf 10} 3766

\bibitem{Kalb2018}
Kalb N, Humphreys P~C, Slim J~J and Hanson R 2018 {\em Physical Review A\/}
  {\bf 97} 62330

\bibitem{Doherty2013}
Doherty M~W, Manson N~B, Delaney P, Jelezko F, Wrachtrup J and Hollenberg L~C~L
  2013 {\em Physics Reports\/} {\bf 528} 1--45

\bibitem{Felton2009}
Felton S, Edmonds A~M, Newton M~E, Martineau P~M, Fisher D, Twitchen D~J and
  Baker J~M 2009 {\em Physical Review B\/} {\bf 79} 75203

\bibitem{Smeltzer2011}
Smeltzer B, Childress L and Gali A 2011 {\em New Journal of Physics\/} {\bf 13}
  25021

\bibitem{Pfender2017a}
Pfender M, Aslam N, Simon P, Antonov D, Thiering G, Burk S, de~Oliveira F~F,
  Denisenko A, Fedder H, Meijer J, Garrido J~A, Gali A, Teraji T, Isoya J,
  Doherty M~W, Alkauskas A, Gallo A, Gr{\"{u}}neis A, Neumann P and Wrachtrup J
  2017 {\em Nano Letters\/} {\bf 17} 5931--5937

\bibitem{Morton2006}
Morton J~J~L, Tyryshkin A~M, Ardavan A, Benjamin S~C, Porfyrakis K, Lyon S~A
  and Briggs G~A~D 2006 {\em Nature Physics\/} {\bf 2} 40--43

\bibitem{Filidou2012}
Filidou V, Simmons S, Karlen S~D, Giustino F, Anderson H~L and Morton J~J~L
  2012 {\em Nature Physics\/} {\bf 8} 596--600

\bibitem{Neumann2010}
Neumann P, Beck J, Steiner M, Rempp F, Fedder H, Hemmer P~R, Wrachtrup J and
  Jelezko F 2010 {\em Science\/} {\bf 329} 542--544

\bibitem{Dutt2007}
Dutt M~V~G, Childress L, Jiang L, Togan E, Maze J, Jelezko F, Zibrov A~S,
  Hemmer P~R and Lukin M~D 2007 {\em Science\/} {\bf 316} 1312--1316

\bibitem{White2007}
White A~G, Gilchrist A, Pryde G~J, O'Brien J~L, Bremner M~J and Langford N~K
  2007 {\em Journal of the Optical Society of America B\/} {\bf 24} 172--183

\bibitem{Balasubramanian2009}
Balasubramanian G, Neumann P, Twitchen D, Markham M, Kolesov R, Mizuochi N,
  Isoya J, Achard J, Beck J, Tissler J, Jacques V, Hemmer P~R, Jelezko F and
  Wrachtrup J 2009 {\em Nature Materials\/} {\bf 8} 383--387

\bibitem{Avantaggiati2016}
Avantaggiati A, Loreti P and Vellucci P 2016 Kadec-1/4 theorem for sinc bases arXiv:1603.08762v1

\bibitem{Gorini1976}
Gorini V, Kossakowski A and Sudarshan E~C~G 1976 {\em Journal of Mathematical
  Physics\/} {\bf 17} 821--825

\bibitem{Lindblad1976}
Lindblad G 1976 {\em Communications in Mathematical Physics\/} {\bf 48}
  119--130

\bibitem{Havel2003}
Havel T~F 2003 {\em Journal of Mathematical Physics\/} {\bf 44} 534--557

\bibitem{NielsenChuang2010}
Nielsen M~A and Chuang I~L 2010 {\em Quantum Computation and Quantum
  Information\/} (Cambridge University Pr.)

\bibitem{Nizovtsev2014}
Nizovtsev A~P, Kilin S~Y, Pushkarchuk A~L, Pushkarchuk V~A and Jelezko F 2014
  {\em New Journal of Physics\/} {\bf 16} 083014

\end{thebibliography}

\section*{References}
\providecommand{\newblock}{}

\end{document}